\PassOptionsToPackage{hyphens}{url}
\documentclass{pas}
\usepackage{multirow}
\usepackage{siunitx}
\usepackage{longtable}
\usepackage{float}  

\newcommand{\prov}[1]{\textcolor{red}{#1}}

\usepackage[normalem]{ulem}
\definecolor{notegreen}{rgb}{0,0.55,0}

\usepackage[nohyperlinks]{acronym}

\begin{document}
\acrodef{UEMR}{unintended electromagnetic radiation}
\acrodef{SKA}{Square Kilometre Array}
\acrodef{ATNF}{Australia Telescope National Facility}
\acrodef{ATCA}{Australia Telescope Compact Array}
\acrodef{DTC}{Direct-to-Cell}
\acrodef{DTD}{Direct-to-device}
\acrodef{DTH}{Direct-to-Handset}
\acrodef{DTM}{Direct-to-mobile}
\acrodef{SNIFFLES}{Satellite measuremeNt of Intended emission, unwanted emission, and radio radiation to develop, Follow up, and veriFy mitigation measures, regulatory compliance, and Lawful usE of the radio Spectrum}
\acrodef{NGSO}{non-geostationary satellite orbit}
\acrodef{LEO}{low-Earth orbit}
\acrodef{MEO}{medium-Earth orbit}
\acrodef{EOP}{Earth Orientation Parameters}
\acrodef{GNSS}{Global Navigation Satellite Systems}
\acrodef{EMI}{electromagnetic interference}
\acrodef{EMC}{electromagnetic compatibility}
\acrodef{OOBE}{out-of-band emission}
\acrodef{NRAO}{National Radio Astronomy Observatory}
\acrodef{CABB}{Compact Array Broadband Backend}
\acrodef{BIGCAT}{Broadband Integrated GPU Correlator for ATCA}
\acrodef{LOFAR}{LOw-Frequency ARray}
\acrodef{RAS}{radio astronomy service}
\acrodef{ITU-R}{International Telecommunication Union Radiocommunication Sector}
\acrodef{RR}{Radio Regulations}
\acrodef{ASKAP}{Australian SKA Pathfinder}
\acrodef{VLBI}{very long baseline interferometry}
\acrodef{VGOS}{VLBI Global Observing System}
\acrodef{FSS}{Fixed Satellite Service}
\acrodef{RFI}{radio-frequency interference}
\acrodef{e.i.r.p}{effective isotropic radiated power}
\acrodef{EPFD}{equivalent power flux density}
\acrodef{UT1}{Universal Time 1}
\acrodef{ICRF3}{International Celestial Reference Frame 3}
\acrodef{ITRF2020}{International Terrestrial Reference Frame 2020}
\acrodef{SLR}{satellite laser ranging}
\acrodef{DORIS}{Doppler Orbitography and Radiopositioning Integrated by Satellite}
\acrodef{IVS}{International VLBI Service for Geodesy and Astrometry}
\acrodef{GGOS}{Global Geodetic Observing System}
\acrodef{ODC}{orbital data centre}
\acrodef{LNA}{low-noise amplifier}
\acrodef{WRC}{World Radiocommunication Conference}
\acrodef{OMM}{Orbit Mean-Elements Message}
\acrodef{PFD}{power flux density}
\acrodef{LO}{local oscillator}
\acrodef{GSO}{geostationary satellite orbit}
\acrodef{CISPR}{Comité international spécial des perturbations radioélectriques}
\acrodef{FCC}{Federal Communications Commission}
\acrodef{AIAA}{American Institute of Aeronautics and Astronautics}
\acrodef{JAXA}{Japan Aerospace Exploration Agency}
\acrodef{ECSS}{European Cooperation for Space Standardization}
\acrodef{ISO}{International Organization for Standardization}
\acrodef{GPS}{Global Positioning System}
\acrodef{TDRSS}{Tracking and Data Relay Satellite System}
\acrodef{TTC}[TT\&C]{telemetry, tracking, and command}
\acrodef{Tsys}{system temperature}
\acrodef{NCRIS}{National Collaborative Research Infrastructure Strategy}
\acrodef{CSIRO}{Commonwealth Scientific and Industrial Research Organisation}
\acrodef{CDF}{cumulative distribution function}
\acrodef{ECC}{Electronic Communications Committee}
\acrodef{UTC}{Coordinated Universal Time}
\acrodef{CEPT}{European Conference of Postal and Telecommunications Administrations}
\acrodef{MSS}{Mobile-satellite service}

\lefttitle{Publications of the Astronomical Society of Australia}
\righttitle{RAS \& Satellite coexistence: Emission and radiation limits}

\jnlPage{1}{1}
\jnlDoiYr{2026}
\doival{10.1017/pasa.xxxx.xx}

\articletitt{Research Paper}

\title{Towards Genuine Coexistence: Per-Satellite Emission and Radiation Limits to Protect Radio Astronomy and Geodetic VLBI at 1--14\,GHz from Satellite Constellations}

\author{\gn{Balthasar} \sn{Indermuehle}$^{1}$,
        \gn{Lucia} \sn{McCallum}$^{2}$,
        \gn{Emma} \sn{van der Wateren}$^{3}$,
        \gn{Hayo} \sn{Hase}$^{4}$,
        \gn{Benjamin} \sn{Winkel}$^{5}$,
        \gn{Liroy} \sn{Louren\c{c}o}$^{1}$,
        \gn{Federico} \sn{Di Vruno}$^{6}$,
        \gn{Michael} \sn{Lindqvist}$^{7}$,
        \gn{Gregory} \sn{Hellbourg}$^{8}$
        and
        \gn{Gyula I. G.} \sn{J\'ozsa}$^{9, 10}$
        }
        
\affil{$^1$CSIRO Space \& Astronomy, PO Box 76, Epping NSW 1710, Australia\\
$^2$School of Natural Sciences, University of Tasmania, PO Box 807, Sandy Bay TAS 7006, Australia\\
$^3$ASTRON, the Netherlands Institute for Radio Astronomy, Oude Hoogeveensedijk 4, 7991 PD Dwingeloo, The Netherlands\\
$^{4}$BKG, Federal Agency for Cartography and Geodesy, Sackenrieder Str. 25, 93444 Bad K\"otzting, Germany\\
$^{5}$MPIfR, Max-Planck-Institut f\"{u}r Radioastronomie, Auf dem H\"{u}gel 69, 53121 Bonn, Germany\\
$^{6}$SKAO, Square Kilometre Array Observatory, Jodrell Bank, Cheshire, SK11 9FT, United Kingdom\\
$^{7}$Department of Physics and Astronomy, Chalmers University of Technology, Onsala Space Observatory, SE-439 92 Onsala, Sweden\\
$^{8}$Cahill Center for Astronomy and Astrophysics, California Institute of Technology, Pasadena, CA 91125, USA\\
$^{9}$Max-Planck-Institut für Radioastronomie Radioobservatorium Effelsberg, Max-Planck-Strasse 28, 53902 Bad Münstereifel, Germany\\
$^{10}$Centre for Radio Astronomy Techniques and Technologies (RATT), Department of Physics and Electronics, Rhodes University, Makhanda 6140, South Africa
}

\corresp{B. Indermuehle, Email: balt.indermuehle@csiro.au}

\history{(Received xx xx xxxx; revised xx xx xxxx; accepted xx xx xxxx)}

\begin{abstract}
To meet its science goals, the radio astronomy service (RAS) relies on opportunistic access to spectrum allocated mostly to other services, with few bands protected on a primary basis. The unprecedented proliferation of non-geostationary satellite orbit (NGSO) constellations threatens this mode of operation. Geodetic very long baseline interferometry (VLBI) is a vulnerable application of RAS: it provides the fundamental link between the celestial and terrestrial reference frames, and is the only technique that uniquely determines UT1$-$UTC. The next-generation geodetic VLBI Global Observing System (VGOS) achieves millimetre accuracy by synthesising group delay across \SIrange{3}{14}{\giga\hertz} using $32\times\SI{32}{MHz}$ channels, most of which lie outside RAS primary allocations. The SNIFFLES-I survey \citep{indermuehle2026sniffles} measured intended emissions, unwanted emissions (spurious emissions, notably harmonics), and unintended electromagnetic radiation (UEMR) of NGSO systems from \SIrange{1}{26}{\giga\hertz}. On this basis, we model the equivalent power flux density (EPFD) of current and future constellations and compare against protection criteria of ITU-R RA.769. The analysis extends to frequencies without radio astronomy allocations where SNIFFLES-I made detections. For geodetic VLBI, we run a Monte-Carlo EPFD model at the AuScope VGOS stations and scale the aggregate from the present catalogued fleet ($\sim$\num{12000} satellites) to the hundreds of thousands on file with a validated method. Inverting the EPFD analysis against the interpolated RA.769 thresholds yields maximum tolerable per-satellite levels for spurious emissions and for UEMR, expressed as a field-strength limit in dB($\mu$V/m) at \SI{10}{m} for standard-setting bodies. We treat proposed orbital data centres (ODCs) in Sun-synchronous orbit as a distinctively UEMR-dominated case. We find that already today the single-dish protection criteria are exceeded in two primary RAS bands at 2.7 and 5\,GHz. Outside RAS allocations the measured levels likewise exceed the interpolated thresholds, information relevant to administrations protecting their facilities. For geodetic VLBI, the dominant threat is spurious emission from the \SI{2620}{MHz} direct-to-device (DTD) downlink, whose second harmonic at \SI{5240}{MHz} already exceeds the RA.769 VLBI threshold in \SI{68}{\percent} of the observed sky and time, the data-loss metric of ITU-R RA.1513. For a projected fleet of \num{300000} satellites, that data loss reaches \SI{100}{\percent}. Narrow-band UEMR, spread across a VGOS \SI{32}{MHz} channel, remains below the calculated threshold. At that population, the tolerable per-satellite level is 65--70\,dB($\mu$V/m) at \SI{10}{m} in \SI{1}{MHz} under the VLBI criterion and 18--46\,dB($\mu$V/m) under the single-dish criterion; current spacecraft exceed the former by $\sim$\SI{20}{dB} at the \SI{7860}{MHz} harmonic and the latter by 20--75\,dB at every calibrated detection.
\end{abstract}


\maketitle
\acresetall
\section{Introduction}

\subsection{Radio astronomy, geodetic VLBI, and who depends on them}

The radio spectrum is regulated by the \ac{ITU-R} via the so-called \ac{RR}\citep{ituRR}, which have the status of an international treaty, and associated Recommendations. Each application falls under a radio service, to which one or more frequency bands are allocated. The \ac{RAS} is a passive service: it transmits nothing and measures cosmic signals of order a few Jansky ($1\,\mathrm{Jy}=10^{-26}\,\mathrm{W\,m^{-2}\,Hz^{-1}}$). CSIRO, Australia's national science agency, operates RAS stations in Australia\footnote{The ASKAP, ATCA, Ceduna, Hobart, Katherine, Mopra, Murriyang, and Yarragadee dishes also serve in the astronomical Long Baseline Array (LBA). We treat only the \emph{geodetic} VLBI case here, on the dedicated \ac{VGOS} array (Hobart, Katherine, Yarragadee). The single-dish continuum criterion applies to the true single dishes (Ceduna \mbox{30-m}, Hobart \mbox{26-m}, Mopra \mbox{22m}, Murriyang \mbox{64-m}). The ATCA is an interferometer, to which the VLBI criterion properly applies; it enters the single-dish analysis only through the single-dish mode used for the SNIFFLES-I absolute-flux measurement.}, as does the University of Tasmania \citep{lovell2013}, including the AuScope array, a research collaboration between the University of Tasmania and Geoscience Australia. This paper assesses the impact of satellite emission and radiation on \ac{RAS} across observing frequencies from \SIrange{1}{14}{\giga\hertz}, and on geodetic VLBI. The facilities considered are listed in Table~\ref{tab:facilities}. 

\begin{table}[t]
\caption{Australian radio astronomy facilities considered as observers, with geographic latitude, representative dish diameter, observing-band coverage used here, and role. Coordinates and bands are consistent with Australian Radiofrequency Spectrum Plan (ARSP) footnotes AUS87/AUS103 \citep{arsp2025}.}
\label{tab:facilities}
{\tablefont\begin{tabular}{@{\extracolsep{\fill}}llrcl}
\toprule
Facility & Site & Lat. & Dish & Coverage\\
 & & (deg) & (m) & (GHz)\\
\hline
ASKAP & Murchison & $-26.7$ & 12 x 36 & 0.7--1.8\\
ATCA & Narrabri & $-30.3$ & 22 x 6 & 1.1--12.2\\
Mopra & Coonabarabran & $-31.3$ & 22 & 1.1--8\\
Murriyang & Parkes & $-33.0$ & 64 & 0.7--14\\
Ceduna & Ceduna & $-31.9$ & 30 & 1.5--14\\
Hobart & Mt Pleasant & $-42.8$ & 26 & 1.2--12\\
\hline
AuScope & Hobart    & $-42.8$ & 12 & 2--14 (VGOS)\\
AuScope & Katherine & $-14.4$ & 12 & 2--14 (VGOS)\\
AuScope & Yarragadee & $-29.0$ & 12 & 2--14 (VGOS)
\botrule
\end{tabular}}
\end{table}

Geodetic very long baseline interferometry \citep[VLBI;][]{hinteregger1972} measurements underpin essential geodetic products on which much of modern infrastructure depends \citep{hiddenrisk2024}. By observing compact extragalactic radio sources (quasars) with a global network of radio telescopes, geodetic VLBI realises the International Celestial Reference Frame \citep[ICRF3;][]{charlot2020} and ties that quasi-inertial frame to the International Terrestrial Reference Frame \citep[ITRF2020;][]{altamimi2023}. It is the \emph{only} technique that determines UT1$-$UTC and the precession and nutation of the Earth's spin axis: the satellite techniques (GNSS, SLR, DORIS) orbit within the rotating system and cannot sense the absolute phase of Earth rotation \citep{ituRA2178, dostal2025policy}.

These products are not only of purely scientific interest. The \ac{EOP} and reference frames produced by geodetic VLBI are required to compute and predict the precise orbits of Global Navigation Satellite System (GNSS) satellites. An error in UT1 propagates directly into positioning, timing, and navigation errors that affect aviation, space, maritime and land navigation, precision agriculture, financial transaction time\-stamp\-ing, power-grid synchronisation, and defence \citep{hiddenrisk2024,stateofgeodesy2026}. UN General Assembly Resolution 69/266 recognises the geodetic infrastructure for the global geodetic data supply chain as critical \citep{unga69266}. Maintaining it requires continued, global VLBI observing through the International VLBI Service for Geodesy and Astrometry \citep[IVS;][]{nothnagel2017}, against a Global Geodetic Observing System target of \SI{1}{mm} position accuracy and \SI{0.1}{mm/yr} stability \citep{ggos}. The AuScope array broadband antennas at Hobart (Tasmania), Katherine (Northern Territory) and Yarragadee (Western Australia) operate across \mbox{3--14\,GHz} \citep{plankAUSTRAL,mccallum2022} and form part of the global IVS network of roughly 25 sites in some 20 countries \citep{ivs2025flyer}.

\subsection{Opportunistic observing and the spectrum environment}

Like radio astronomy in general, geodetic VLBI has operated \emph{opportunistically} for more than four decades: it observes in spectrum that is mostly not allocated to or protected for the \ac{RAS}, historically within 2.2--2.4~GHz (S-band) and 8.2--8.9~GHz (X-band) \citep{SCHUH201268}. VGOS itself was the geodetic community's response to increasing activity in the S-band of the legacy S/X system, which was developed to avoid the WiFi and mobile-phone (2G, 3G, and 4G) allocations \citep{petrachenko2009}. The broadband group-delay technique that gives modern VGOS its accuracy (Section~\ref{sec:vgos}), requires access to $\sim$\SI{1}{GHz} of combined spectrum between \mbox{3--14\,GHz}. The separated observing bands spanning a broad frequency range enable the VLBI measurement to reach its target precision, with the additional separated bands being necessary to correct for the frequency-dependent ionospheric delay. Only $\sim$\SI{160}{MHz} of that range carries a primary \ac{RAS} allocation (of \SI{345}{MHz} total including secondary; Table~\ref{tab:ras-alloc}, \citealt{arsp2025}). VGOS was designed to be agile, placing its observing channels in locally quiet parts of the band, under the constraint that all VGOS stations worldwide must use the same frequencies. This worked while the radio sky was comparatively empty: interference sources were terrestrial, in the telescope's near-field, and below receiver-chain saturation, where the receiver remains in its linear operating range, so decorrelation protected the observations. That assumption is now failing.

\begin{table}[t]
\caption{\Acf{RAS} allocations and recognition in the \mbox{1--14\,GHz} range in the Australian Radiofrequency Spectrum Plan \citep{arsp2025} (ITU Region~3, Australian column). $\bullet$ marks a primary or secondary \ac{RAS} allocation (all L-band \ac{RAS} allocations are primary). Footnote 5.340 prohibits \emph{all} emissions (the \mbox{1400--1427\,MHz} HI line and \mbox{2690--2700}, \mbox{10680--10700\,MHz}); AUS87 recognises highly interference-sensitive \ac{RAS} receivers (1.2--1.8, 2.2--2.7, 4.5--6.7, 8--10\,GHz) but does not itself allocate spectrum. Scientifically important ranges, notably \mbox{8000--10000\,MHz}, which contains the legacy X-band of the geodetic S/X system and the third \ac{DTD} harmonic, carry no \ac{RAS} allocation at all.}
\label{tab:ras-alloc}
{\tablefont\begin{tabular}{@{\extracolsep{\fill}}rrccl}
\toprule
From & To & Prim. & Sec. & Applicable footnotes\\
(MHz) & (MHz) & & &\\
\hline
1200 & 1330   &          &          & AUS87\\
1330 & 1400   &          &          & 5.149, AUS87\\
1400 & 1427   & $\bullet$ &          & \textbf{5.340}, 5.341, AUS87\\
1427 & 1610.6 &          &          & AUS87\\
1610.6 & 1613.8 & $\bullet$ &        & 5.149, AUS87\\
1613.8 & 1660 &          &          & 5.149, AUS87\\
1660 & 1670   & $\bullet$ &          & 5.149, 5.379, AUS87\\
1718.8 & 1722.2 &        &          & 5.149\\
2200 & 2655  &          &          & AUS87\\
2655 & 2670  &          & $\bullet$ & 5.149, 5.208B, 5.420, AUS87\\
2670 & 2690  &          & $\bullet$ & 5.149, AUS87\\
2690 & 2700  & $\bullet$ &          & \textbf{5.340}, AUS87\\
4500 & 4800  &          &          & AUS87\\
4800 & 4825  &          & $\bullet$ & 5.149, 5.443, AUS87\\
4825 & 4835  & $\bullet$ &          & 5.443, 5.149, AUS87\\
4835 & 4940  &          & $\bullet$ & 5.149, 5.443, AUS87\\
4940 & 4950  &          & $\bullet$ & 5.149, 5.339, AUS87\\
4950 & 4990  & $\bullet$ &          & 5.443, 5.149, AUS87\\
4990 & 5000  & $\bullet$ &          & 5.149, AUS87\\
5000 & 6700  &          &          & AUS87\\
8000 & 10000 &          &          & AUS87\\
10600& 10680 & $\bullet$ &          & 5.149, 5.482, 5.482A\\
10680& 10700 & $\bullet$ &          & \textbf{5.340}\\
\hline
\multicolumn{5}{@{}p{\linewidth}@{}}{Total primary \ac{RAS}: \SI{210}{MHz};\newline
Total secondary \ac{RAS}: \SI{175}{MHz};\newline
L-band contributes \SI{40}{MHz} primary}
\botrule
\end{tabular}}
\end{table}

The number of active satellites in non-geostationary-orbit (NGSO) systems has grown from a few thousand to over fifteen thousand within a few years. Filings lodged with the \ac{ITU-R} imply of order a million or more satellites over the coming decades \citep{falle2023papersats,indermuehle2026sniffles, hellbourg2026}. Even after heavy discounting for filings that will never be launched, deployments of several hundred thousand satellites are still plausible, and forecasts of $\sim$\num{100000} transmitting satellites by 2030 are routinely cited in the regulatory literature \citep{wrc31vgos, ivs2025flyer}. Unlike terrestrial transmitters, satellites radiate from above the horizon into the main beam and sidelobes of every radio telescope on the ground, and a constellation distributes interferers across the whole visible sky. Consequently, the IVS community has identified the inclusion of VLBI in spectrum management as necessary for the long-term sustainability of geodetic VLBI products \citep{hase2026}. It is therefore raising the issue with stakeholders and governments \citep{dostal2025policy} and advocating a new agenda item at ITU-R's World Radio Conference 2031 (WRC-31) \citep{ivs2025flyer}.

\subsection{Interference sources}

The observational foundation of this paper is the SNIFFLES-I survey \citep{indermuehle2026sniffles}: 4629 tracked observations of satellites from the major NGSO constellations across \SIrange{1}{26}{\giga\hertz}, made with the Mopra \mbox{22m} telescope and flux-calibrated against the ATCA, yielding directly measured per-satellite emission and radiation levels rather than modelled or filed values. Every per-satellite input used in the predictions that follow derives from these measurements. Three distinct classes of emission and radiation are relevant to our analysis:
\begin{itemize}
  \item \textbf{Intended emission} within the satellite's licensed allocation which can easily saturate \ac{RAS} receivers if a satellite is located in the main beam of a telescope;
  \item \textbf{Unwanted emission}, which the \ac{RR} divide into two  \emph{mutually exclusive} classes \citep{ituSM329}. Firstly, \textbf{out-of-band emission} (RR No.~1.144) which includes  modulation sidebands and spectral regrowth immediately outside the necessary bandwidth, where spectral regrowth is the broadening of the transmitted signal caused by nonlinearities in the transmitter's power amplifier. Secondly, \textbf{spurious emission} (RR No.~1.145), comprising harmonic emissions, parasitic emissions, intermodulation products and frequency conversion products. The main SNIFFLES-I detections falling into this category are the harmonics up to the fourth order of the Starlink \ac{DTD} downlink at \SI{2620}{MHz}, placing emission at 5240, 7860, and 10480\,MHz, all within the \mbox{3--14\,GHz} VGOS span (and near the legacy bands);
  \item \textbf{\Acf{UEMR}}\footnote{The term UEMR is widely used in scientific literature and the media, but the \ac{ITU-R} is still discussing the terminology, which would be considered most appropriate in relation to the \ac{RR}.}, consisting of broadband noise from onboard electronics and power converters (predominantly at frequencies below \SI{350}{MHz}; \citealt{divruno2023, bassa2024}), and narrowband radiation from digital backplanes, clocks, and networking equipment, unrelated to any transmitter and not subject to the emission masks that govern intended emissions. This is the vast majority of what SNIFFLES-I discovered.
\end{itemize}
Harmonic emissions from satellites are particularly insidious for VGOS: they are often broadband, they fall inside the \mbox{3--14\,GHz} window, and the applicable spurious-emission limits (the RR Appendix 3 \emph{space-station limits}) do not automatically protect a sensitive receiver (see discussion in Section~\ref{sec:global}), as they are referring to single transmitters. Hence, whether (opportunistic) RAS observations could be affected depends on many details, such as the number of satellites in a constellation.

The downlink harmonic emissions at 5240, 7860, and \SI{10480}{MHz} are whole multiples of the \SI{2620}{MHz} centre frequency, and their bandwidth scales with harmonic order from the \SI{10}{MHz} fundamental: \SI{20}{MHz} at 2\textsuperscript{nd} order, \SI{30}{MHz} at 3\textsuperscript{rd}, and so forth. \ac{UEMR} in this frequency range is generally narrowband, but in practice will scale with satellite population and constellation architecture, resulting in \ac{UEMR} radiated at many frequencies and powers. Whilst the Radio Regulations cleanly separate the first two classes (intended and unwanted emission), the \SI{5190}{MHz} detection is an instructive case.\footnote{The Radio Regulations resolve this case explicitly: Recommendation ITU-R SM.329 \citep{ituSM329}, \S1.1.4, defines \emph{frequency conversion products} as spurious emissions ``at the frequencies, or whole multiples thereof, \ldots\ of any oscillations generated to produce the carrier'', which is precisely the \ac{LO}: the \SI{2595}{MHz} leakage and its second multiple at \SI{5190}{MHz} fall under this single clause. Both are therefore spurious \emph{emissions}, not unintended radiation and its second-order harmonic, because the oscillator belongs to the transmitter chain. Equally, they are not \emph{harmonic emissions}, which \S1.1.1 defines as whole multiples of the \emph{centre} frequency (\SI{2620}{MHz}) and which \S1.1.4 excludes by construction. Taxonomy and physics agree: a frequency conversion product inherits the bandwidth of the oscillator that produced it (narrowband, $\sim$\SI{10}{kHz}), a harmonic emission that of the modulated carrier (broadband, 20--\SI{40}{MHz}).} It is the second harmonic of the narrowband \ac{LO} leakage at \SI{2595}{MHz}. Both are \emph{frequency conversion products}, hence spurious emissions of the transmitter rather than unintended radiation, and both are correspondingly narrowband. This distinction is important because their narrow bandwidth and origin in the transmitter chain set them apart from the broadband UEMR considered elsewhere in this work.

\subsection{Aims of this paper}

This paper extrapolates the directly measured emission and radiation of current NGSO constellations onto the \ac{RAS} in general, and broadband geodetic VLBI in particular, assessing the interference potential to both single-dish \ac{RAS} and the VLBI measurements. In bands allocated to the RAS, protection criteria apply. For information, we also provide results for bands not allocated to the RAS based on (interpolated) power level thresholds. Our aims are to:
\begin{enumerate}
  \item introduce the reader to the difference between geodetic VLBI and radio astronomy VLBI, specifically the VGOS broadband system (Section~\ref{sec:vgos});
  \item review the relevant interference thresholds: The ITU-R RA.769 single-dish (continuum) and VLBI criteria in the few RAS allocations available across \mbox{1--14\,GHz} (Section~\ref{sec:thresholds}), and the linearity/damage limits of real receivers, and;
  \item extrapolate the SNIFFLES-I per-satellite intended-emission, spurious-emission and \ac{UEMR} measurements from the current fleet to a conservative $\sim$\num{300000}-satellite future, using both an analytic aggregate-power argument and a Monte-Carlo \ac{EPFD} simulation, for the Australian \ac{RAS} facilities of Table~\ref{tab:facilities} (Section~\ref{sec:methods});
  \item predict the resulting data loss for radio astronomy in general (Section~\ref{sec:results}), and treat broadband geodetic VLBI as a special case (Section~\ref{sec:geovlbi});
  \item recast the result as per-satellite spurious-emission and radiation limits in the units standards bodies use, and assess current compliance (Section~\ref{sec:limits});
  \item provide an analysis of the main threat to geodetic VLBI posed by \ac{DTD} and their harmonics (Section \ref{sec:global});
  \item consider a different threat model posed by ODCs where \ac{UEMR} is assumed to be the dominant source of interference (Section \ref{sec:odc}).
\end{enumerate}

\section{Geodetic VLBI and VGOS}
\label{sec:vgos}

\subsection{What makes geodetic VLBI different from radio astronomy}

Astronomical VLBI seeks to image or measure the cosmic source. Geodetic VLBI instead treats the quasars as fixed points and solves for the \emph{geometry}: the difference in arrival time (the group delay) of a wavefront at two stations, measured against a geometric model, yields the baseline vector and the orientation of the Earth in space. The observable is the group delay, whose precision scales inversely with the bandwidth via bandwidth synthesis \citep{hinteregger1972}. This necessitates a system design fundamentally different from spectral-line or continuum radio astronomy:
\begin{itemize}
  \item Geodetic VLBI does not depend on any protected spectral line. Observing frequencies are chosen for compatibility across the global network, for technical feasibility of the receiving and recording systems, and for delay precision, which favours the widest possible frequency span \citep{ituRA2507};
  \item sensitivity to interference is therefore set by group-delay corruption and receiver saturation;
  \item every station in the network must observe the \emph{same} frequency bands simultaneously, so interference at one station affects every baseline to it. Interference uncorrelated between stations correlates out and acts only by raising the system temperature and thereby reducing the sensitivity. A strong emitter could even saturate the radio astronomy front end leading effectively to a total loss of that station within the network. \citep{ituRA2178, haseivtw2024}.
\end{itemize}

\subsection{The VGOS broadband system}

VGOS replaces the legacy dual-band (S/X) system with fast-slewing \mbox{12--13\,m} antennas and broadband feeds covering \mbox{2--14\,GHz}\footnote{Although the VGOS observation range is now \mbox{3--14\,GHz}, the technical specifications were initially set to \mbox{2--14\,GHz}. Meanwhile the range of \mbox{2--3\,GHz} is lost for geodetic VLBI due to its heavy use by active services \citep{hase2026}. VGOS sites without filters remain sensitive in the range of \mbox{2--14\,GHz}.} \citep{petrachenko2009, niell2018}. Recording the full band is infeasible, so VGOS synthesises the wide effective bandwidth from $32\times\SI{32}{MHz}$ channels (1024\,MHz total); grouped in four sub-bands, each with eight channels (256\,MHz total) spanning up to 992\,MHz. In 2026, the IVS published Resolution 2026-01 on the frequency sequence for future VGOS operations \citep{IVS-Res-2026-01}; Table~\ref{tab:vgos-bands} summarises the four sub-bands defined for the proposed WRC-31 agenda item. The resolution itself states the exposure this paper quantifies: the channels ``may be adjusted slightly to improve geodetic products; however, there is no guarantee to obtain frequency protection by spectrum management'' \citep{IVS-Res-2026-01}. Delay precision is set by the separation between the lowest and highest observed frequency, so the system must reach both ends of the \mbox{3--14\,GHz} range; retreat to a narrow protected band would sacrifice accuracy \citep{ituRA2178, wrc31vgos}. VGOS band C overlaps the X-band (8.2--8.95~GHz) of the legacy S/X system \citep[e.g.,][]{SCHUH201268}, ensuring continuity of long-term results such as the Celestial Reference Frame.

\begin{table}[t]
 \caption{VGOS sub-bands and the corresponding frequency ranges named in the proposed WRC-31 agenda item \citep{IVS-Res-2026-01,wrc31vgos}. Each sub-band carries 8 of the 32 channels of 32\,MHz.}\label{tab:vgos-bands}
 {\tablefont\begin{tabular}{@{\extracolsep{\fill}}clc}
   \toprule
   Sub-band & VGOS channel span & Regulatory range \\
   \hline
   A & 3.000--3.384\,GHz  & 2\,900--3\,400\,MHz \\
   B & 4.824--5.816\,GHz  & 4\,800--5\,850\,MHz \\
   C & 8.760--9.752\,GHz  & 8\,750--9\,800\,MHz \\
   D & 12.984--13.976\,GHz & 12.75--14.00\,GHz \\
   \botrule
 \end{tabular}}
\end{table}

The broadband feed \citep[e.g.,][]{akgiray2013} is, by design, receptive to any signal stronger than the cosmic background across the whole \mbox{3--14\,GHz} window, not only within the 32 observed channels. Filters inserted into the VGOS receiving chain can remove individual strong (mostly local) interferers \citep[e.g.,][]{turner2023,yebesfilters2024}, but they are not a general mitigation. Any filter degrades sensitivity over the whole band, while providing additional suppression of spurious signals over a large bandwidth \citep{yebesfilters2024}. Only a few can be installed, and each installation is station-specific work comparable to developing a new system. The AuScope receivers offer no access ahead of the first \ac{LNA} stage, which eliminates filtering as a mitigation technique entirely. Against cosmic signals of a few Jansky, even faint anthropogenic emission or radiation is overwhelming.

\section{Interference thresholds and protected bands}
\label{sec:thresholds}

No \ac{ITU-R} Recommendation yet prescribes a protection criterion specific to broadband geodetic VLBI outside RAS allocated bands. Report ITU-R RA.2507 \citep{ituRA2507}, prepared by the IVS community and cited by Recommendation ITU-R RA.2178 \citep{ituRA2178}, derives VGOS-specific thresholds across \mbox{2--14\,GHz} by applying the RA.769 methods with VGOS operating parameters (\SI{100}{s} calibration integrations, \SI{32}{MHz} channels), and concludes that the resulting VLBI levels are approximately those of RA.769-2 Table~3. As an ITU-R report it is informative, not normative. Its levels, like those of RA.769, cap the total interference power at the receiver. Neither document provides a mechanism that holds the combined emission of many independent systems\footnote{A system refers to a whole satellite constellation not to individual satellites.} below that cap. In the primary RAS bands, a registered station has to be protected from interference. Geodetic VLBI stations fall within that protection. Recommendation ITU-R RA.2178 states that geodetic VLBI stations operate under the radio astronomy service, and recommends that administrations consider registering their geodetic VLBI sites as RAS stations, in accordance with RR No.~11.12, in the bands allocated to the RAS. The RA.769 criteria, whose Table~3 gives the VLBI thresholds, and the RA.1513 data-loss limits therefore apply to a registered geodetic VLBI station in the same way as to any other radio astronomy station; neither Recommendation distinguishes the astronomical from the geodetic use of the observation. The de-facto standard for protection criteria is provided in Recommendation ITU-R RA.769. For VLBI measurements, its Table 3 is relevant (see our Table~\ref{tab:ra769-vlbi}), however, VLBI stations also need to perform calibration measurements in single dish mode for which the \emph{continuum} mode thresholds are the appropriate limits. While the tables in RA.769 contain entries for selected frequency bands only, the set of equations in the recommendation also allow calculation of levels for any other frequency and receiver properties such as different measurement bandwidths and antenna/noise temperatures. The levels in Recommendation RA.769 do not specify how to address aggregate emissions, but this is contained in Recommendation ITU-R RA.1513, which sets maximum levels of acceptable data loss applicable for primary RAS allocations. Hence, there is currently no criterion tailored to broadband geodetic VLBI across the full \mbox{3--14\,GHz} span for cases of aggregate emissions (nor are there any means to consider the impact of these emissions/radiations at non-allocated RAS frequencies within the ITU-R Radio Regulations framework). We therefore assess the predicted emission against a ladder of criteria of increasing severity and decreasing regulatory standing.

\subsection{The RA.769 VLBI criterion}

Recommendation ITU-R RA.769-2 \citep{ituRA769} defines detrimental interference for single-dish operations as interference that perturbs the measured radiometric noise \emph{fluctuation} by more than 10\%. The reference quantity is not the system noise power $P$ but the far smaller rms fluctuation $\Delta P = P/\sqrt{\Delta f\, t}$ that remains after averaging over the \SI{2000}{s} reference integration, about five orders of magnitude below $P$ itself. For VLBI it adopts a separate, less stringent criterion: the interfering power must not exceed 1\% of the receiver noise \emph{power}, on the grounds that interference is rarely correlated between widely separated stations. Table~\ref{tab:ra769-vlbi} reproduces the resulting VLBI thresholds (RA.769-2, Table~3) at the frequencies relevant to VGOS. Report ITU-R RA.2507 derives its VGOS-specific VLBI thresholds from these values. Because the 1\%-of-noise criterion is independent of integration time, the shorter VGOS integrations leave them unchanged; its Table A-2 agrees with the thresholds we propose in Table~\ref{tab:vlbi-threshold} to better than \SI{1}{dB} at every band centre.

The two percentages reference different quantities, which resolves an apparent paradox: 10\% of the post-integration fluctuation is a far smaller power than 1\% of the receiver noise. This is why the single-dish thresholds sit some \SI{40}{dB} \emph{below} the VLBI thresholds despite the larger percentage.

\begin{table}[t]
 \caption{ITU-R RA.769-2 Table~3 threshold spectral power flux density for VLBI observations \citep{ituRA769}, at the frequencies most relevant to VGOS.}
 \label{tab:ra769-vlbi}
 {\tablefont\begin{tabular}{@{\extracolsep{\fill}}cc}
   \toprule
   Centre frequency (MHz) & Threshold (dB(W\,m$^{-2}$\,Hz$^{-1}$)) \\
   \hline
   2\,695  & $-205$ \\
   4\,995  & $-200$ \\
   10\,650 & $-193$ \\
   15\,375 & $-189$ \\
   \botrule
 \end{tabular}}
\end{table}

Constellation-scale aggregate interference at the receiver does not arise via main-beam coupling alone. The main beam of a \mbox{12-m} antenna spans only $\sim$\SI{0.6}{\degree} at \SI{3}{GHz} ($\sim$\SI{0.2}{\degree} at \SI{8}{GHz}). Even for a \num{300000}-satellite fleet, with roughly \num{10000} above the horizon at any moment\footnote{Corroborated by the operator side: SpaceX's Gen3 filing computes a maximum of \num{4054} satellites visible above \SI{0}{\degree} elevation for its \num{100000}-satellite system \citep{spacex2026gen3}, the same $\sim$3--4\% visibility fraction.}, the expected number inside the main beam is $\sim$0.2, i.e.\ usually none. The aggregate enters through the sidelobes. An antenna retains finite gain in every direction above the horizon; for the radio astronomy reference antenna pattern of Recommendation ITU-R RA.1631, the far-sidelobe floor lies at $-12$ to $-7$\,dBi. The received powers of all satellites in view, several hundred for the present fleet, add incoherently in the receiver. RA.769 is built on this premise: it evaluates its thresholds for a \SI{0}{dBi} sidelobe\footnote{The \SI{15}{dB} originates as the near-sidelobe gain (\SI{15}{dBi}) of the SA.509 reference pattern at \SI{5}{\degree} from boresight, the antenna's pointing direction, the minimum \ac{GSO} spacing RA.769 recommends; \S2.1 then requires the \emph{summed} received power of all such transmitters to stay \SI{15}{dB} below the tabulated level. The \SI{0}{dBi} reference against which the thresholds are tabulated corresponds to \SI{19.05}{\degree} from boresight, a cone of \SI{0.344}{sr}, or \SI{5.5}{\percent} of the visible \mbox{$2\pi$\,sr} sky, a fraction RA.769 (\S1.3) notes coincides with the percentage-of-time data loss apportioned in Recommendation ITU-R RA.1513.} and notes that ``most interference [\ldots] is received through the far side lobes of the telescope''. Rare main-beam transits produce brief, strong increases in the received power but the persistent sensitivity loss quantified in this paper is the sidelobe sum, which we refer to as the \emph{statistical sensitivity loss}. How the RA.769 criterion is exercised is a central point of this paper. The threshold caps the \emph{total} interference power at the receiver, and RA.769 already requires aggregate interference from multiple transmitters to be considered: for \ac{GSO} networks the \emph{sum} of all interfering signals must sit \SI{15}{dB} below the detrimental level (\S 2.1), and for NGSO systems the combined, \ac{EPFD}-style response to all satellites within a system applies (\S 2.2).

\subsection{The single-dish continuum criterion}

Where the more conservative 10\%-of-fluctuation continuum criterion of RA.769-2 (Table~1) is applicable, for example for continuum or spectral line work with single-dish telescopes, or when considering correlated emission, the thresholds are some 40\,dB more stringent than the VLBI values (e.g. $-247$ vs $-205$\,dB(W\,m$^{-2}$\,Hz$^{-1}$) at 2700\,MHz). We carry this as a conservative bound.

\subsection{Extending the thresholds to frequencies not allocated to RAS}
\label{sec:interp}

\begin{figure}[t]
\centerline{\includegraphics[width=\columnwidth]{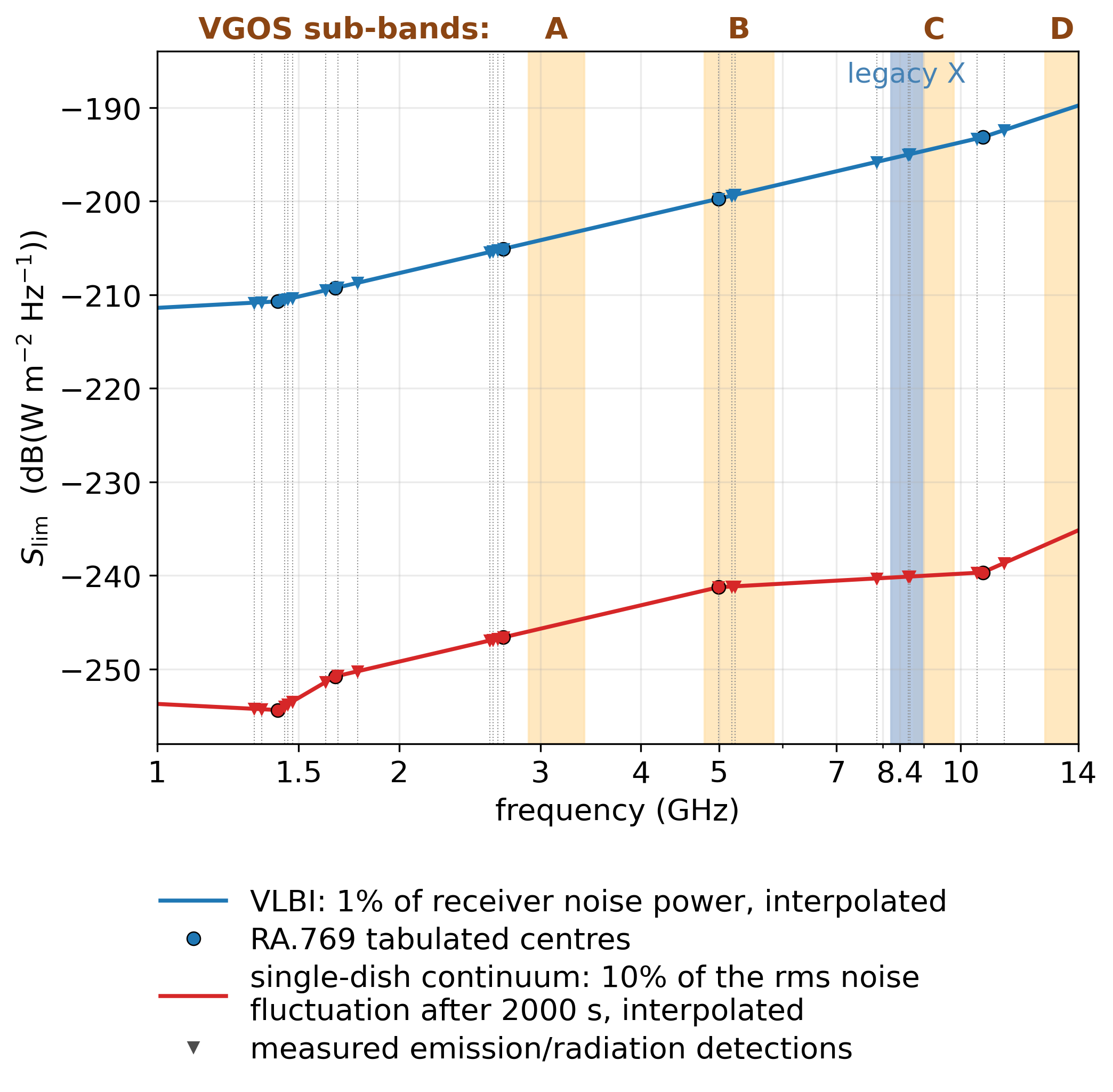}}
\caption{The RA.769 protection thresholds interpolated across \mbox{1--14\,GHz} (linear in dB against $\log_{10}\nu$, hence straight segments on the logarithmic frequency axis): the VLBI criterion (interference below 1\% of the receiver noise power) above and the single-dish continuum criterion (interference perturbing the far smaller post-integration noise fluctuation by no more than 10\%), some \SI{40}{dB} more stringent, below. Circles mark the discrete band centres RA.769 tabulates; triangles mark the measured satellite emission or radiation features of Table~\ref{tab:inputs} at the threshold applying to each. Tan bands mark the four VGOS sub-bands, the blue band the legacy X-band.}
\label{fig:sliminterp}
\end{figure}

RA.769 itself already applies its criterion beyond primary allocations. Its spectral-line table states thresholds at the \SI{4830}{MHz} formaldehyde line, in a band carrying no RAS allocation at all (footnote RR No.~5.149 applies), and at \SI{14488}{MHz}, where the allocation is secondary. The threshold is thus a property of the observation, rather than allocation status. RA.769 specifies the threshold spectral power flux density $S_{\rm lim}$ only at discrete band centres, some twenty values from \SI{13}{MHz} to \SI{270}{GHz} per criterion; Table~\ref{tab:ra769-vlbi} lists those relevant to VGOS. Most SNIFFLES-I emission and radiation detections fall \emph{between} these centres, in spectrum with no RAS allocation and no prescribed limit. We therefore interpolate the threshold to each emission or radiation frequency. $S_{\rm lim}$ is set by the assumed receiver noise and the reference-antenna gain, both smooth and slowly varying with frequency, so we interpolate it (in dB) linearly against $\log_{10}\nu$ between the two nearest tabulated centres,
\begin{equation}
S_{\rm lim}(\nu) = S_1 + (S_2-S_1)\,\frac{\log_{10}(\nu/\nu_1)}{\log_{10}(\nu_2/\nu_1)},
\end{equation}
where $(\nu_1,S_1)$ and $(\nu_2,S_2)$ are the bracketing centres (for example \SI{1665}{} and \SI{2695}{MHz} for the \SI{2620}{} and \SI{2656}{MHz} detections, and \SI{4995}{} and \SI{10650}{MHz} for the \SI{5240}{}, \SI{7860}{} and \SI{10480}{MHz} harmonics). Figure~\ref{fig:sliminterp} shows both interpolated criteria across \mbox{1--14\,GHz}: the tabulated RA.769 centres they pass through, the measured emission or radiation detections at which they are evaluated, and the VGOS sub-bands and legacy X-band for orientation.

Every measured satellite feature lies within RA.769's tabulated range, so this is interpolation between adjacent centres, not extrapolation beyond the data. The same procedure is applied to both the VLBI and the single-dish continuum criteria. It yields a defensible estimate of the protection level that \emph{would} apply at each satellite feature, against which the aggregate \ac{EPFD} is compared (Section~\ref{sec:methods}). Report ITU-R RA.2507 offers further precedent for stating protection levels in unallocated spectrum: it tabulates its VGOS thresholds continuously across \mbox{2--14\,GHz}, in every \SI{32}{MHz} channel, irrespective of the allocation.

\subsection{Receiver linearity and damage}

Neither RA.769 criterion captures the failure mode that is of greatest concern to VGOS operators: \emph{saturation} of the broadband front-end, where a signal is strong enough to introduce non-linearities that affect the entire observing band. 
For the VGOS case, we adopt typical threshold levels \citep{haseivtw2024, lopezperez2019} of an input power below \SI{-40}{dBm} ($10^{-7}$\,W) to keep the low-noise amplifier in its linear regime, noting a damage level between \SI{10}-\SI{12}{dBm} (\SI{10}-\SI{15}{mW}). Both regimes have ITU-R precedent. Report ITU-R RA.2188 \citep{ituRA2188} tabulates the power flux-density and \ac{e.i.r.p}.\ levels potentially damaging to radio astronomy receivers, and Recommendation ITU-R RS.2066 \citep{ituRS2066} already constrains the strongest spaceborne X-band emitters, the synthetic-aperture radars operating in \mbox{9.3--9.9\,GHz}, with respect to RAS stations in the \mbox{10.6--10.7\,GHz} band. Both instruments predate the broadband \mbox{2--14\,GHz} VGOS front-ends and the sensitivity of their \acp{LNA}; a revision of Report RA.2188 could prepare their update to VGOS-class receivers.

A strong in-band emitter can corrupt all 32 channels at once, at either of two stages. At the analogue front-end, the broadband \ac{LNA} responds to the integrated in-band power, so a single strong carrier compresses the whole \mbox{2--14\,GHz} chain before it is channelised; the \SI{-40}{dBm} figure above is this limit. At the digital back-end, the VLBI sampler has limited dynamic range. VGOS specifies 8--10\,bit sampling providing \SI{6}{dB} of RFI headroom per bit, before re-quantisation to 2\,bit for transmission to the correlator; \citealt{petrachenko2009}; this is implemented in current wide-band backends such as the DBBC3, \citealt{tuccari2019dbbc3}. Switchable attenuation keeps the analogue-to-digital converter at its optimal loading. A strong carrier can therefore force the use of heavy attenuation that, where it acts on the whole receiver rather than per sub-band, drives the faint astronomical signal in every channel towards the quantisation floor. Either way, the result is a total outage rather than a graceful loss of sensitivity. SNIFFLES-I used the most resilient receiver in ATNF's fleet, the ATCA broadband backend BIGCAT, to measure flux densities up to $\sim$\SI{e6}{Jy} for the strongest detections, levels capable of saturating sensitive receivers \citep{indermuehle2026sniffles}. We evaluate the predicted aggregate and peak power flux against these hardware limits as a third criterion.

\subsection{Towards a geodetic VLBI-specific threshold}

Combining the above, Section~\ref{sec:limits} proposes an interference threshold tailored to broadband geodetic VLBI. It has four defining properties.
\begin{enumerate}
\renewcommand{\labelenumi}{(\roman{enumi})}
\item The threshold caps the summed emission\footnote{Translating an aggregate cap into per-system or per-spacecraft limits requires an agreed \emph{apportionment} of the interference budget among the contributing systems. No such mechanism is in prospect at ITU-R even for dividing the budget between the \ac{GSO} and NGSO services, let alone across the many NGSO systems now on file.} and radiation of the entire simultaneously radiating population, assuming isotropic radiation, verified against a defined compliance statistic. The RA.769 levels equally cap the total power at the receiver, but they are exercised in coordination (entry-by-entry), and no defined path (regulatory or methodological) leads from a constellation's population to the permissible levels of its individual spacecraft.
\item The threshold is stated per VGOS sub-band across \mbox{3--14\,GHz}, not only in the few narrow RAS bands.
\item The threshold bounds the receiver-saturation regime in addition to the statistical sensitivity loss regime. The latter is the progressive loss of sensitivity that arises while the front-end stays linear, when the aggregate of many simultaneous emitters exceeds the RA.769 detrimental threshold without reaching the saturation limit, quantified against the RA.769 criteria by the Monte-Carlo \ac{EPFD} simulations of Section~\ref{sec:methods}.
\item The threshold is inverted into \emph{per-satellite} emission and radiation ceilings that fall as the constellations grow: the form a regulator or a standards body can apply to an individual spacecraft.
\end{enumerate}

\section{From SNIFFLES-I to the future sky: methodology}
\label{sec:methods}

Our approach extrapolates the directly measured emission and radiation of current constellations \citep{indermuehle2026sniffles} to a projected future fleet by two complementary methods: an analytic aggregate-power scaling, which gives a transparent upper bound, and a Monte-Carlo \ac{EPFD} simulation, which captures the spatial and temporal statistics of the interference. The geodetic VLBI analysis uses the AuScope sites; the continuum case uses all Australian single-dish sites.

\subsection{Per-satellite emission/radiation inputs}

SNIFFLES-I provides detection rates and the \ac{e.i.r.p} spectral densities (detection floor $\sim$\SI{-120}{dB(W/Hz)} at 850\,km slant range), together with absolute flux densities calibrated against the ATCA, for every emission/radiation detection and per constellation/version. These are \emph{apparent} \ac{e.i.r.p} densities, referenced to the direction of the telescope at the time of measurement; no intrinsic isotropy is assumed of any transmission. Section~\ref{sec:vgos-methods-mc} sets out how each class enters the aggregate: platform radiation is taken to be physically isotropic, \ac{DTD} harmonics have been measured to be near-isotropic, and for the steered \ac{DTD} downlink, we measured the serving-beam peak power, i.e.\ the maximum power associated with the main beam, and use this peak as a conservative upper bound on the transmitted power in the aggregate, applied isotropically (Section~\ref{sec:vgos-methods-mc}). The case of a downlink beam pointed directly at a station is treated separately in Section~\ref{sec:limits}; operator-side boresight avoidance, which suppresses precisely this geometry, has been trialled at the ATCA and Murriyang telescopes \citep{indermuehle2026odsba}, with implications taken up below. We construct per-satellite inputs separately for each class (intended emission, spurious emission, and unintended electromagnetic radiation, UEMR) because the classes differ in level, bandwidth, and the fraction of the fleet that exhibits them. We adopt \emph{two brackets} per detection: a conservative lower bound at the SNIFFLES-I detection-floor \ac{e.i.r.p} density, and a nominal level from the ATCA-calibrated absolute flux density. Table~\ref{tab:inputs} lists the adopted inputs. The measured ATCA flux densities are converted to \ac{e.i.r.p} spectral density via
\begin{equation}
\mathrm{EIRP}_\nu = 4\pi d^2\, S_\nu
\end{equation}
where $S_\nu$ is the ATCA-calibrated spectral flux density ($1\,\mathrm{Jy}=10^{-26}$\,W\,m$^{-2}$\,Hz$^{-1}$) received at the measured slant range $d$. The subscript $\nu$ denotes a per-hertz spectral density throughout; $\nu$ itself always denotes frequency, and bandwidths are written $\Delta\nu_e$ (Section~\ref{sec:bwcorr}). All emissions and radiations are taken to be continuous (duty cycle unity: the \ac{DTD} service and platform electronics are effectively always on). The radiating fraction of each source population is set to the SNIFFLES-I detection rate of that feature, the fraction of the observed satellites that showed it.

At these frequencies the measurements show a simple pattern: a given emission or radiation feature is either present in a satellite at close to a characteristic level or absent altogether, and when present its level scatters by no more than about an order of magnitude. This contrasts with the low-frequency regime. At \SIrange{110}{188}{MHz}, \citet{divruno2023} detected \ac{UEMR} from 47 of 68 satellites, a comparable presence fraction, but with levels spanning two orders of magnitude (\SIrange{0.1}{10}{Jy} broadband, \SIrange{10}{500}{Jy} in the narrowband features), \citet{bassa2024} find an order-of-magnitude generation dependence, and EDA2 observations of the \ac{DTD} fleet find the radiation heterogeneously expressed across the fleet and modulated by operational state, brighter in eclipse than in sunlight \citep{dong2026eda2}. This contrast can be explained mechanistically: the narrowband GHz detections are clock and digital harmonics whose amplitude is fixed by the hardware design of a satellite version, whereas the low-frequency broadband radiation is dominated by power-electronics switching noise that varies with load and thermal state. The single-level, present-or-absent model adopted here is therefore appropriate at these frequencies, and is not claimed for the low-frequency regime.

\begin{table}[t]
\caption{Adopted per-satellite inputs from SNIFFLES-I. E.i.r.p spectral densities in dB(W/Hz); ``floor'' is the detection-limited lower bound, ``ATCA'' the flux density-calibrated nominal level. Class: Intended, Spurious (unwanted emission: harmonic emissions of the downlink, and frequency conversion products of the carrier-generating oscillator), UEMR. DTD $=$ SpaceX \ac{DTD}. Every detection carrying an ATCA value is a solid measurement; values shown in red are placeholders for the two L-band detections not yet measured at the ATCA (\SI{1474}{} and \SI{1680}{MHz}), quoted at the detection floor as a conservative lower bound. The \SI{5240}{}, \SI{7860}{}, \SI{8599}{} and \SI{10480}{MHz} ATCA values are Y-factor calibrated (peak power flux density (PFD) $\geq$\SI{321740}, \SI{21413}, \SI{2786}, and \SI{740}{Jy} respectively). The three harmonic values are the polarisation-mean peaks of the brightest tracked pass (STARLINK-11117, 2026 May 31), range-corrected to \SI{890}{km}. The \SI{2595}{}, \SI{2620}{}, and \SI{5240}{MHz} values are \emph{lower bounds}, since those bright detections drive the calibrating receiver into compression. The last column shows the \emph{active fraction}, indicating how many of the observed satellites showed that emission or radiation feature.}
\label{tab:inputs}
{\tablefont\setlength{\tabcolsep}{5pt}\begin{tabular}{@{\extracolsep{\fill}}rllrrr}
\toprule
Freq & Class & Source & Floor & ATCA & Act.\\
(MHz) & & & \multicolumn{2}{c}{dB(W/Hz)} & frac.\\
\hline
1320 & UEMR & v2-Mini          & $-118.4$ & $-100.2$ & 0.83\\
1350 & UEMR & v2-Mini          & $-118.4$ & $-99.8$  & 0.96\\
1440 & UEMR & v2-Mini          & $-118.4$ & $-102.0$ & 1.00\\
1454 & UEMR & GuoWang (comb)   & $-118.4$ & $-89.7$  & 0.25\\
1474 & UEMR & DTD              & \prov{$-118.4$} & ---     & \prov{0.88}\\
1620 & UEMR & v2-Mini          & $-118.4$ & $-98.7$  & 0.83\\
1680 & UEMR & v2-Mini          & \prov{$-118.4$} & ---     & \prov{0.89}\\
1777 & UEMR & GuoWang (comb)   & $-118.4$ & $-74.7$  & 0.12\\
2595 & Spurious & DTD \ac{LO} (conv.\ prod.) & $-118.4$ & $\geq-68.1$ & 0.87\\
2620 & Intended & DTD (downlink) & $-118.4$ & $\geq-64.1$ & 0.87\\
2656 & UEMR & DTD              & $-118.4$ & $-85.4$ & 0.80\\
2700 & UEMR & v2-Mini Ku/std   & $-118.4$ & $-86.5$ & 0.77\\
4995 & UEMR & Starlink         & $-120.5$ & $-101.0$ & ---\\
5190 & Spurious & DTD \ac{LO} 2nd mult.      & $-120.5$ & $-93.6$     & 0.87\\
5240 & Spurious & DTD 2nd harmonic            & $-120.5$ & $\geq-74.9$ & 0.87\\
7860 & Spurious & DTD 3rd harmonic            & $-120.0$ & $-86.7$     & 0.87\\
8599 & UEMR & OneWeb           & $-120.0$ & $-91.3$ & 1.00\\
8640 & UEMR & v2-Mini          & $-120.0$ & $-93.7$ & 0.94\\
10480& Spurious & DTD 4th harmonic            & $-119.0$ & $-101.3$    & 0.87\\
\botrule
\end{tabular}}
\end{table}

\subsection{Analytic aggregate scaling}

For uncorrelated emitters distributed over the visible sky, the aggregate spectral power flux density is linear in the number of simultaneously radiating satellites. We therefore run the full Monte-Carlo \ac{EPFD} at the \emph{present} radiating population (real orbits; Section~\ref{sec:vgos-methods-mc}) and scale the resulting distribution to larger constellations by
\begin{equation}
+10\log_{10}(N/N_0),
\label{eq:scaling}
\end{equation}
where $N_0$ is the present emitting NGSO total. This scaling assumes that future constellations have altitude and inclination distributions similar to those of the present fleet. That need not hold: the Sun-synchronous orbital data centres of Section~\ref{sec:odc} are a case in point, and are therefore modelled on their own orbits rather than by this scaling. Our Orbit Mean-Elements Message (OMM) snapshot contains \num{11981} satellites of the five measured systems (Starlink, OneWeb, Amazon Leo, GuoWang, Qianfan). Every present-day \ac{EPFD} result in this paper is computed on their catalogued orbits; only the growth scenarios are modelled. This measured emitting population is the normalisation $N_0$ for the per-satellite limits and the growth scenarios, within the $\sim$\num{15000} NGSO satellites currently active. The \ac{DTD} downlink and its harmonics, which drive the VLBI impact, are produced only by satellites carrying that downlink; the uncharacterised remainder of the fleet does not add to those detections. The radiating subset differs by detection, and the DTD detections are the narrowest. They are carried only by the \mbox{STARLINK-11xxx} series: \num{642} of the \num{11981} satellites of $N_0$, or \SI{5.4}{\percent} of the modelled fleet, of which the measured active fraction of \num{0.869} leaves \num{558} radiating. The remaining \SI{94.6}{\percent} of the fleet contributes nothing to them. We model this fraction of active satellites as a subset which we assume to be continuously radiating rather than assuming a duty cycle applied per satellite. This overestimates the number of simultaneously active transmitters and so leans conservative, potentially overestimating the impact. The same is true of the peak-power assumption for the steered downlink, where the transmit beam is directed towards a served user: every radiating satellite is given the strongest serving level observed, treated as isotropic, and assumed to transmit continuously. A satellite can carry dozens of co-frequency beams formed by its phased array, each operating at the same frequency, but steered towards its own serving cell (a defined area of the ground grid). When the main beam is not directed at a telescope, the array presents only the incoherent sum of its beams' sidelobes, far below any serving-beam level. Assigning every satellite its peak serving level isotropically therefore gives an upper limit on the aggregate emission from all beams. Replacing this convention with modelled beam patterns \citep{divruno2025} could only lower the predicted aggregate at a random pointing, never raise it. The isotropic assumption is therefore conservative: it cannot underestimate the aggregate interference from the beam-formed downlinks. The convention touches only the intended \SI{2620}{MHz} downlink; the harmonics that drive the VLBI result are measured near-isotropic (Section~\ref{sec:vgos-methods-mc}). A transmit beam pattern changes the outcome in exactly one geometry, a serving beam steered onto the station itself; that case is excluded from the statistical aggregate and treated separately, as the single-satellite saturation worst case of Section~\ref{sec:limits}. The scaling above multiplies the whole aggregate by $N/N_0$ and holds each detection's radiating share of the fleet \emph{constant}: the \num{300000}-satellite sky therefore assumes some \num{16000} DTD satellites, today's \SI{5.4}{\percent} share extrapolated. The aggregate is linear in the number of radiators, so any larger share scales every DTD detection directly. A fleet of \num{30000} such satellites, the \mbox{Gen2} filing taken as wholly DTD capable, would add \SI{2.7}{dB} to each of them; half of a \num{300000}-satellite fleet would add \SI{9.7}{dB}, and an entirely DTD fleet \SI{12.7}{dB}. The composition of the future constellation is thus a larger source of uncertainty in the DTD detections than the present-day beam state and is the one aspect of our assumptions that could underestimate the interference. Separately, we have not measured whether the other operators' satellites radiate platform UEMR of their own. If they do, the present aggregate at the unintended-radiation detections is a lower bound, and the single-dish limits derived from it are to that extent optimistic. The \num{100000}- and \num{300000}-satellite scenarios correspond to factors of $\times 8.4$ ($+9.2$\,dB) and $\times 25.0$ ($+14.0$\,dB) respectively. We verified that the \ac{EPFD} aggregate scales linearly with population over this range: the analytic aggregate scaling of Equation \ref{eq:scaling} and a direct \ac{EPFD} run agree to within the Monte-Carlo scatter.

These scenarios are no longer hypothetical: in July 2026 SpaceX applied
for a third-generation system of \num{100000} satellites in orbital shells, i.e. groups of satellites at similar altitudes, at
\mbox{323--327.5} and \mbox{473--477.5\,km} \citep{spacex2026gen3}, so the
\num{100000}-satellite column corresponds to a single filed constellation.
The proposed shells also validate our scaled geometry: the upper set lies
within the current v2-Mini shell range (\mbox{448--482\,km}), and the lower set sits within \SI{0.8}{dB} of zenith flux of the \ac{DTD} median
altitude of \SI{358}{km}, so scaling today's catalogued orbits reproduces the proposed architecture to $<$\SI{1}{dB}.

\subsection{Monte-Carlo \ac{EPFD} simulation}
\label{sec:vgos-methods-mc}

We use a Monte-Carlo \ac{EPFD} calculation, built on the \texttt{pycraf} spectrum-management package \citep{winkel2018pycraf} and the \texttt{cysgp4} satellite-propagation wrapper \citep{cysgp4_software}. Orbital states are taken from a frozen OMM snapshot of the full catalogue. Satellites are propagated over the RA.769 \SI{2000}{s} integration window, and telescope pointings are sampled on an equal-area sky-cell grid (Recommendation ITU-R M.1583) above the minimum elevation of the radio telescope. The elevation limit restricts the pointings only: satellites contribute to every pointing whenever they are above the horizon. Each satellite's spectral power flux density, weighted by the receiving antenna's gain towards it, is summed into an aggregate \emph{equivalent} power flux density (the \ac{EPFD} concept of RR No.~22.5C, which is explained in more detail in Recommendations ITU-R M.1583 and S.1586).

For full consistency with the \ac{EPFD} method used within the ITU-R framework, the simplified single-dish antenna gain model defined in Recommendation ITU-R RA.1631 is employed, noting that real antenna patterns are usually more complicated, especially in the far side lobes. The receiver physically accumulates spectral \emph{power}; the gain weighting expresses that power in the flux-density units in which the RA.769 thresholds are stated (for a \SI{0}{dBi} side lobe), so the aggregate and the threshold compare directly. We use the RA.1631 \texttt{pycraf} implementation and evaluate it for each facility's dish diameter (Table~\ref{tab:facilities}), with the aperture efficiency term $\eta_a$ at unity, as specified in the Recommendation. A realistic efficiency would lower only the main-beam and nearby sidelobe gain, by $10\log_{10}\eta_a$, leaving the far-sidelobe envelope, and with it every continuum-criterion result, unchanged. The near-beam VLBI statistics might overestimate the interference, because the unity-efficiency assumption gives higher antenna gain than a realistic antenna in these directions.

On the transmit side, each satellite is modelled as a $0$\,dBi isotropic radiator. This is physical for the non-directional UEMR as we demonstrate: ATCA tracking of the second, third and fourth \ac{DTD} harmonics (19, 8 and 5 satellite passes respectively) finds that, after correcting for the changing satellite range, their measured signal level varies by less than $\sim$\SI{2}{dB} over the accessible elevation.  This is expected because the phased array is designed for the fundamental frequency, not for its harmonics, and so forms no coherent beam at them. Therefore, we take the isotropic assumption to hold for the harmonics too. From all features investigated in this study, only the intended \SI{2620}{MHz} downlink is subject to beamforming on the transmitter side, noting that no beam was ever pointing to the radio telescope owing to the established exclusion area. The assumed power levels used in our calculations refer to the maximum of the measured \ac{PFD} values along the tracked passes, i.e. the footprint served by this satellite may have been relatively closely aligned towards telescope boresight, but not serving the telescope site itself, so cannot have been boresight aligned. This produces a (compressed) lower bound on a beam (Section~\ref{sec:limits}). The resulting model gives an upper bound on the interference from satellites whose beams are not directed at the telescope. The worst case, a satellite beam pointed directly at the telescope (we call this case beam-coupled), is treated separately in Section~\ref{sec:limits}.

Each Monte-Carlo iteration draws an independent realisation of the observing geometry: a random start epoch for the \SI{2000}{s} window (uniform over one day, which fully samples orbital phase, as low-Earth-orbit satellites complete $\sim$15 revolutions per day) and a random telescope pointing within each sky cell. The frozen snapshot fixes the shell architecture (altitudes, inclinations, plane populations); what is iterated is where the satellites lie on those orbits at observation time, and where within each cell the beam points. Each iteration yields one \SI{2000}{s}-averaged aggregate \ac{PFD} per cell; the ensemble builds the distribution from which the data loss is read.

Because the aggregate spectral \ac{EPFD} is a single physical quantity, we compare it cell-by-cell against \emph{both} relevant RA.769 criteria, each interpolated to the emission or radiation frequency as in Section~\ref{sec:interp}: the single-dish radio astronomy (continuum) limits of Table~1 and the VLBI limits of Table~3 \citep{ituRA769}. The percentage data loss under each criterion is the fraction of aggregate-PFD samples that exceed the interpolated limit $S_{\rm lim}(\nu)$. The samples combine every sky cell above the minimum elevation and every Monte-Carlo iteration, so each sample represents one accessible (pointing,\,epoch) realisation. The data loss is therefore the fraction of realisations in which the aggregate is detrimental: the fraction of observable sky and time lost. Recommendation ITU-R RA.1513 reads the same criterion slightly differently, comparing the 98th percentile of the aggregate distribution with the threshold; that is the statistic to which the per-satellite limits of Section~\ref{sec:limits} are anchored. The two criteria yield almost identical data-loss values, as independent tests conducted for ECC Report~363 confirm \citep{ecc363}. The sky-coverage maps show this fraction per cell, taken over iterations; the single quoted figure combines all cells.

\begin{figure}[t]
\centerline{\includegraphics[width=\columnwidth]{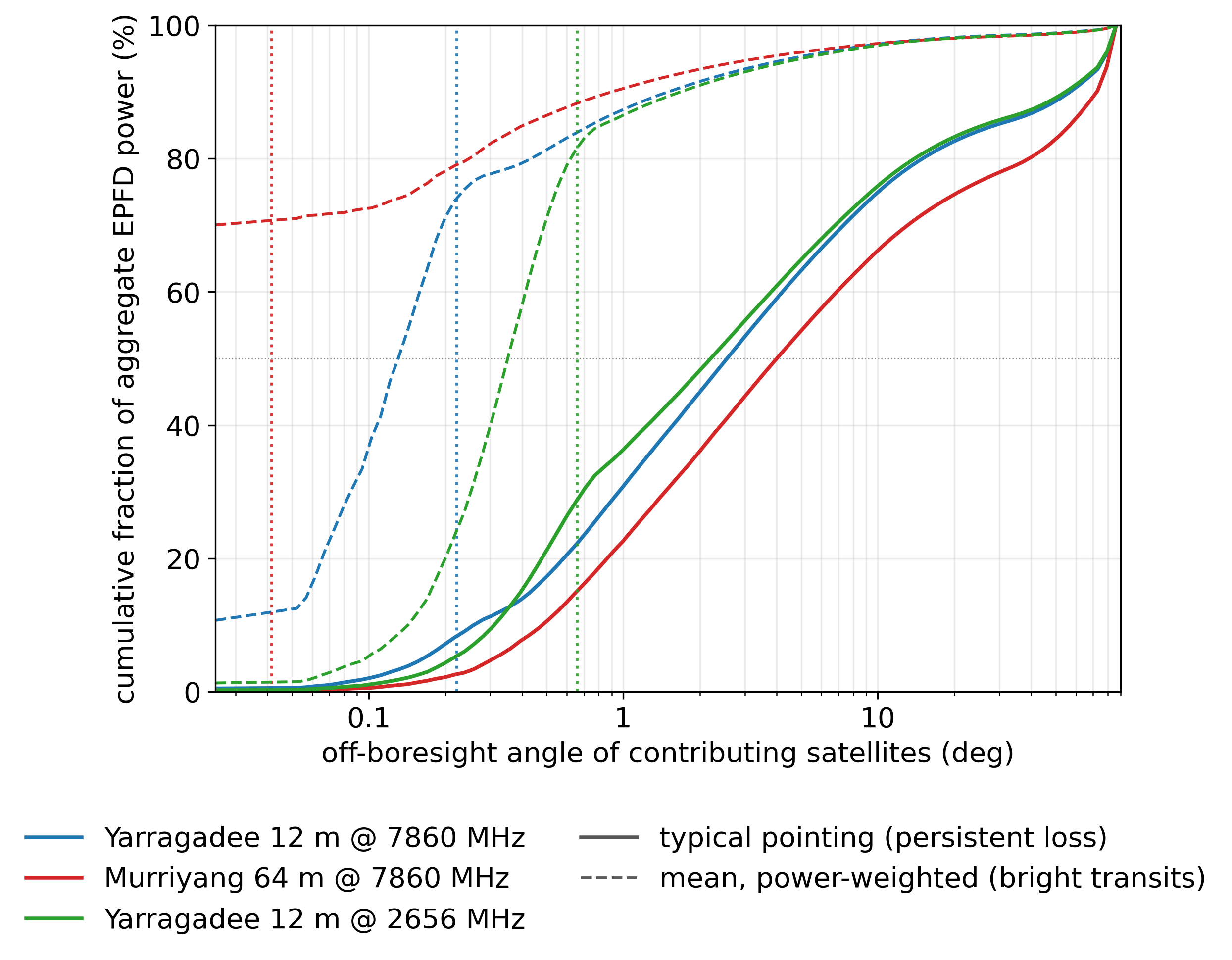}}
\caption{Cumulative fraction of the aggregate \ac{EPFD} power versus the off-boresight angle of the contributing satellites, from the RA.1631 receiving pattern. Solid curves show the \emph{typical pointing} (mean of the per-pointing split, which drives the persistent data-loss statistic); dashed curves the power-weighted \emph{mean} (set by rare bright main-beam transits). Dotted verticals mark each dish's half-power beamwidth. The mean is concentrated near boresight, while the typical pointing is a near sidelobe sum. A count-based breach attribution (Section~\ref{sec:methods}) shows single-dish continuum losses are driven by the far-sidelobe floor, and VLBI losses by contributions within $\sim$\SI{5}{\degree} of boresight. Equation~\ref{eq:angdecomp} defines the two curves.}
\label{fig:angdecomp}
\end{figure}

We verified where the aggregate interference comes from by decomposing it as a function of off-boresight angle in the \ac{EPFD} engine (Figure~\ref{fig:angdecomp}): against the single-dish continuum limit the loss is carried by the far-sidelobe floor. Removing every contribution within \SI{5}{\degree} of boresight does not change this. Against the higher VLBI limit however, the far floor never reaches the threshold, and the realisations exceeding the threshold are driven by the main beam and nearby sidelobes within $\sim$\SI{5}{\degree}, a region that widens with population. This criterion dependence enables a potential mitigation: operator-side boresight avoidance, in which the operator suppresses its downlink while a satellite crosses a telescope's near-boresight cone. This can relieve the VLBI criterion (at \num{300000} satellites every realisation that exceeds the threshold at \SI{7860}{MHz} falls below it if the contributions within \SI{5}{\degree} are removed) but offers no relief against the single-dish continuum criterion. Trial observations at the ATCA and Murriyang measure exactly this division: near-boresight transits are suppressed by factors of 12--40 while the aggregate occupancy is unchanged \citep{indermuehle2026odsba}. Per-satellite emission control therefore remains the only mitigation that reaches the single-dish criterion, including the 6.5 dB relaxed geodetic VLBI single-dish calibration limit. The continuum data-loss figures therefore rest on the far-sidelobe level specified by the reference pattern rather than on any free sidelobe parameter.

To make these statements quantitative we decompose the aggregate in the production \ac{EPFD} engine itself: in every (pointing,\,epoch) realisation each satellite's contribution is retained together with its angle from boresight, instead of only their sum, so every per-cell aggregate is identical to the production result. Figure~\ref{fig:angdecomp} shows two cumulative readings of the same contributions, which are the unweighted and the power-weighted averages of one quantity. For realisation $r$, let $p_{r,s}$ be the window-averaged power received from satellite $s$, $P_r=\sum_s p_{r,s}$ the realisation's aggregate, and $f_r(\theta)$ the fraction of $P_r$ arriving from within an angle $\theta$ of boresight. Then
\begin{equation}
F_{\rm typ}(\theta) = \frac{1}{R}\sum_{r=1}^{R} f_r(\theta), \qquad
F_{\rm mean}(\theta) = \frac{\sum_{r} P_r\, f_r(\theta)}{\sum_{r} P_r}.
\label{eq:angdecomp}
\end{equation}
The \emph{typical-pointing} curve $F_{\rm typ}$ gives every realisation equal weight: it describes where the power comes from at the typical pointing, unaffected by rare bright transits, and is the reading relevant to the persistent data-loss statistic. The \emph{mean} curve $F_{\rm mean}$ weights each realisation by its own aggregate; the rare realisations containing a near-beam transit carry aggregates tens of decibels above the typical, so those few terms dominate both sums, and the curve describes the composition of the bright transits, concentrated near boresight. Equivalently, $F_{\rm mean}$ is the fraction of the ensemble-total power arriving within $\theta$.

We evaluate each AuScope station (Hobart, Katherine, Yarragadee) at the SNIFFLES-I emission/radiation frequencies of Table~\ref{tab:inputs}. For each we propagate the present radiating population and apply the linear-in-$N$ scaling described above, stepping from the present fleet to \num{2000000} satellites in nine stages and producing a sky-coverage map and a cumulative-distribution data-loss plot. For each detection, the radiating population is the source constellation restricted to the relevant version (e.g. the \ac{DTD} subset for harmonics) with only the fraction of satellites corresponding to the detection rate treated as radiating.

The Monte-Carlo ensemble averaging applies to the two sensitivity criteria only. The saturation bound of Section~\ref{sec:thresholds} is neither a data-loss statistic nor time-averaged: front-end compression responds to the instantaneous total power across the full receiver passband, on the timescale of the transit itself. It is therefore assessed against peak, band-integrated power flux, dominated by the brightest single transit (Section~\ref{sec:limits}). The data-loss statistics use \num{50} Monte-Carlo iterations of a \SI{2000}{s} window (converged). The plotted cases that produce the per-cell sky maps use \num{200} iterations, reduced towards \num{30} for the largest radiating populations corresponding to the orbital data centre scenario. The window is sampled at \SI{2}{s} on a \SI{5}{deg} sky grid above a \SI{5}{deg} minimum elevation. The \SI{2}{s} step is verified by convergence: rerunning the \SI{5240}{MHz} case at \SI{1}{}, \SI{0.5}{} and \SI{0.25}{s} steps with identical random draws leaves the data-loss figures unchanged and shifts the 98th-percentile aggregate by only $\sim$\SI{1}{dB}. The \SI{2}{s} step tends to overestimate the contribution: when a main-beam transit is shorter than \SI{2}{s}, the step is over-weighted. Each station, detection, and level bracket is propagated once at the present fleet; the larger constellations follow by analytic scaling. Orbital elements are from the OMM snapshot of 2026 June 20.

\subsection{Emission bandwidth and the recording channel}
\label{sec:bwcorr}

The \ac{EPFD} above yields the aggregate \emph{spectral} \ac{PFD} (per Hz), but a radiometer or VLBI correlator responds to the interference \emph{power} integrated over a frequency channel of width $\Delta f_{\rm ch}$. For VGOS, this is the \SI{32}{MHz} recorded channel. Throughout, the ``\SI{32}{MHz} channel'' means this integration bandwidth, the width over which the correlator accumulates power within each of its 32 channels, not the full broadband span synthesised for the geodetic group delay. The RA.769 thresholds are the spectral \ac{PFD} at which that in-channel power becomes detrimental, derived assuming the interference fills the channel. An emission or radiation detection of bandwidth $\Delta\nu_e$ at aggregate spectral density $S_{\rm agg}$ deposits an in-channel power $S_{\rm agg}\,\Delta\nu_e\,A_{\rm eff}$. Doppler drift over the \SI{2000}{s} integration redistributes this within the channel but conserves the total, and the sum over satellites remains linear in $N$. The detrimental condition $S_{\rm agg}\,\Delta\nu_e\ge S_{\rm lim}\,\Delta f_{\rm ch}$ is therefore equivalent to comparing the aggregate against an effective threshold
\begin{equation}
S_{\rm lim}^{\rm eff} = S_{\rm lim}\,\max\!\left(1,\ \frac{\Delta f_{\rm ch}}{\Delta\nu_e}\right).
\end{equation}
Emission or radiation that fills or overfills the channel ($\Delta\nu_e\ge\Delta f_{\rm ch}$) is unaffected; narrower emission or radiation is held to a proportionally raised threshold. The downlink harmonic emissions are broadband (20--40\,MHz, comparable to a VGOS channel) and fill the channel, so no correction applies. The UEMR detections and the frequency conversion products at \SI{2595}{} and \SI{5190}{MHz} are narrow, of order 1--10\,kHz, so against a $\Delta f_{\rm ch}=\SI{32}{MHz}$ VGOS channel their threshold is raised by 35--45\,dB. We adopt per-line emission/radiation bandwidths, applied throughout: \SI{5}{kHz} default for the narrowband UEMR detections, \SI{10}{kHz} for the OneWeb \SI{8599}{MHz} detection, and the measured broadband widths for the GuoWang L-band combs and the downlink harmonics at 5240, 7860 and \SI{10480}{MHz}, with $\Delta f_{\rm ch}=\SI{32}{MHz}$. The aggregate still scales linearly with $N$; only the threshold it must cross changes. A channel is counted as lost when its aggregate in-channel power exceeds the effective threshold $S_{\rm lim}^{\rm eff}$, the RA.769 statistical sensitivity criterion of Section~\ref{sec:vgos-methods-mc} applied per channel. This is distinct from the finite dynamic range of the digitiser, where strong in-band power consumes the sampler's headroom ahead of the correlator. That is a hardware effect of the receiver-saturation regime bounded in Section~\ref{sec:limits}, mitigable by e.g. filtering, and it does not enter the sensitivity loss statistic used here. For the same reason the saturation regime is not evaluated with the Monte-Carlo machinery at all. A hard threshold on instantaneous total power is governed by the peak power reached during the window, not by a window-averaged ensemble statistic; and because the per-satellite spread between a serving beam pointed at the station and far-sidelobe coupling ($\sim$\SI{110}{dB}) far exceeds the gain from summing all other visible satellites ($\lesssim$\SI{30}{dB} even at \num{300000} satellites), that peak is set by the single brightest transit, with the combined contribution from the other visible satellites remaining $\sim$\SI{55}{dB} below the linearity limit for the beam counts and per-satellite levels measured in the present DTD fleet. A payload with substantially more co-frequency beams or greater bandwidth would raise the aggregate contribution only as $10\log_{10}$ of the multiplicity. A wide margin would therefore remain before the combined emission would drive the front-end into compression. It is therefore bounded analytically by assuming the worst-case transit, with the satellite beam directed towards the telescope. Section~\ref{sec:limits} shows this single beam coupling case already exceeds the front-end linearity limit, which no statistical refinement could soften.

\subsection{Observational validation of the \ac{EPFD} model}
\label{sec:epfd-validation}

We validated the \ac{EPFD} machinery end-to-end against a dedicated observation, using the identical code path as every result in this paper: the same orbit propagation, the RA.1631 receiving pattern, the \SI{0}{dBi} transmit convention, and \SI{2}{s} sampling of a \SI{2000}{s} window. We forecast the aggregate apparent flux density over the whole visible sky for a case bright enough to measure directly: the Ku-band downlink of the present Starlink and OneWeb fleets at \SI{11.3}{GHz}, received by an ATCA \mbox{22m} antenna, with the per-satellite level calibrated by tracking a satellite through the main beam. From that all-sky forecast, we selected the pointing where the predicted \SI{2000}{s} aggregate is both strongest and most repeatable across Monte-Carlo realisations: azimuth \SI{180}{\degree}, elevation \SI{15}{\degree}, in the satellite-dense low southern sky, with a forecast mean aggregate some thirty times the receiver noise floor. The aggregate there is carried by many satellites at once, so it does not depend on any particular satellite steering its service beam towards the telescope. We then parked the antenna at that pointing for \SI{2098}{s} on 2026 July 2. The comparison is independent of gain normalisation, since the antenna's peak gain cancels in the model-to-measurement ratio. The receiver noise floor is measured in the adjacent \SI{10.6}{}--\SI{10.7}{GHz} radio astronomy band, which carries no downlink emission.

\begin{figure*}[t]
\centerline{\includegraphics[width=\textwidth]{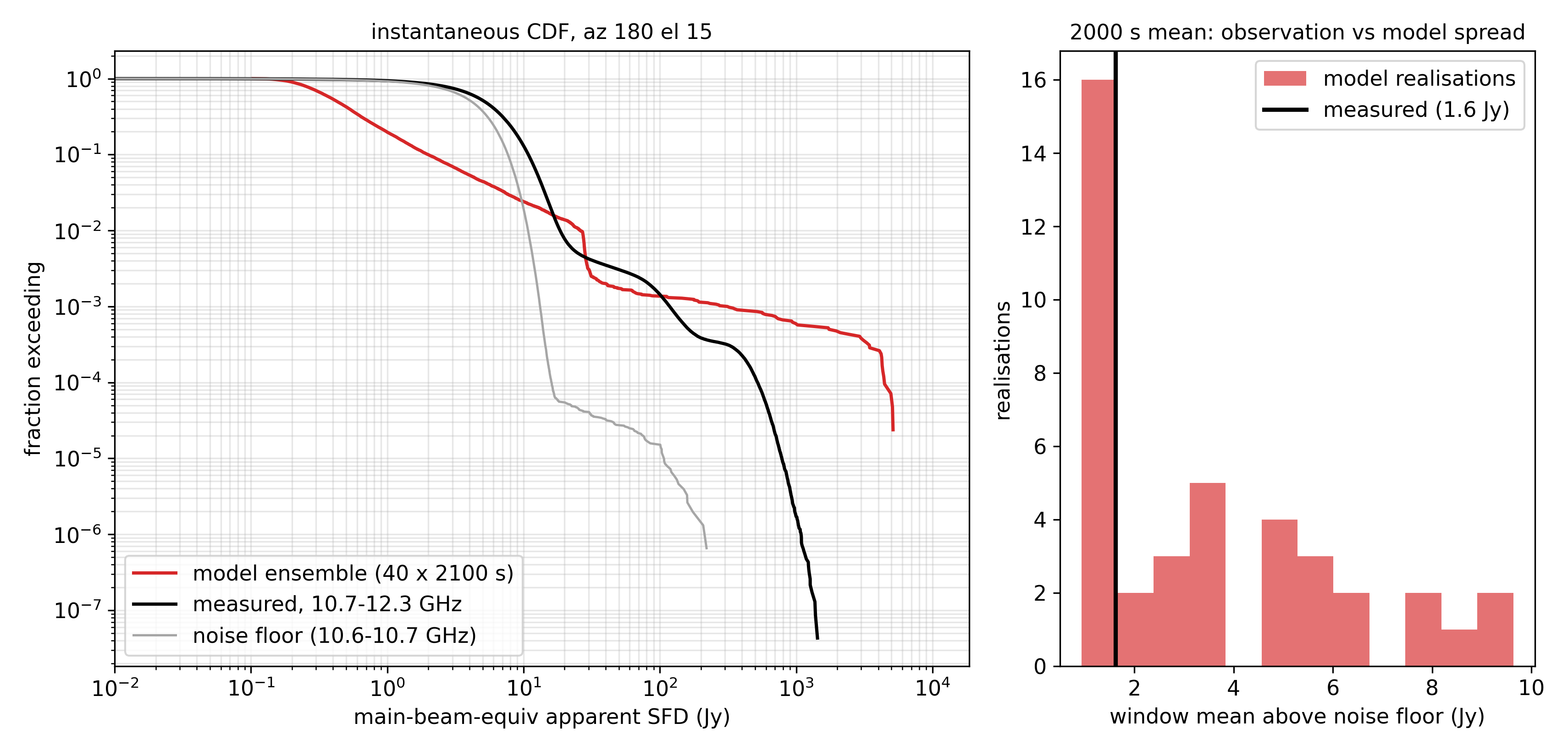}}
\caption{Observational validation of the \ac{EPFD} model: a \SI{2098}{s} parked stare at azimuth \SI{180}{\degree}, elevation \SI{15}{\degree} (ATCA, \SI{10.7}{}--\SI{12.3}{GHz}, 2026 July 2) against the model forecast for the same pointing, fleet and epoch (40 Monte-Carlo realisations of the \SI{2000}{s} window). \emph{Left:} exceedance distribution of the instantaneous aggregate; the measurement (black) tracks the forecast (red) through the full observable range above the receiver noise floor (grey, measured in the emission-free \SI{10.6}{}--\SI{10.7}{GHz} RAS band); below the noise floor, the aggregate is not measurable. The model exceeds the measurement only in the extreme tail, where each satellite is modelled at its beam-on worst case. \emph{Right:} the \SI{2000}{s} mean aggregate, the quantity the RA.769 data-loss statistics are built from: the histogram is the forecast spread over realisations (a tail-dominated statistic, hence broad), and the measured value (black line) falls at its 32nd percentile.}
\label{fig:epfdval}
\end{figure*}

Figure~\ref{fig:epfdval} shows the outcome: every statistic the observation can constrain agrees with the forecast: the 98th-percentile instantaneous aggregate to within $20\%$, the fraction of time above \SI{100}{Jy} to within $10\%$ ($0.129\%$ measured against $0.136\%$ forecast), the number of bright transits in the window (2 measured against a forecast median of 1 and range 0--5), and the \SI{2000}{s} mean aggregate, which lands at the 32nd percentile of the distribution across the forecast realisations. The one deliberate departure is the extreme tail, which the model over-predicts because each satellite is modelled at its measured beam-coupled worst case; this conservatism acts in the protective direction.

\section{Results: impact on radio astronomy (\texorpdfstring{1--14\,GHz}{1-14 GHz})}
\label{sec:results}

We assess the impact on the Australian RAS facilities of Table~\ref{tab:facilities} against the single-dish (continuum) RA.769 criterion across \mbox{1--14\,GHz}; the geodetic VLBI special case (the looser 1\%-of-noise criterion and the VGOS broadband bands) is treated separately in Section~\ref{sec:geovlbi}. The result is uniform at the bright detections: the intended downlink, its \ac{LO} fundamental, the harmonics, and the \SI{2700}{} and \SI{8640}{MHz} unintended radiation cause $100\%$ data loss at every facility, already at the present fleet, so the single-dish radio astronomy threshold is exceeded for all emissions and radiations we found. These findings describe the impact on astronomical observations across the whole spectrum, not interference in the regulatory sense, since most of the detections lie outside the RAS allocated bands.

The facility-to-facility differences appear only in the marginal detections. The \SI{2656}{MHz} radiation, for instance, ranges from $44\%$ at Murriyang to $71\%$ at the Hobart \mbox{26-m} dish. Table~\ref{tab:facilities-cont} gives the present-fleet continuum data loss at each facility for every measured detection, using the ATCA-calibrated level where available and the detection floor otherwise. The detection floor constrained entries are detected but not yet flux-calibrated, so the floor bounds the level from below and, the loss being monotonic in the level, the tabulated data loss is likewise a lower bound. Because the data loss is a percentile count, it is highly non-linear and saturates at \SI{100}{\percent} once the whole distribution clears the threshold; Table~\ref{tab:facilities-cont-margin} therefore gives the companion 98th-percentile margins, which remain linear in per-satellite level and population and state directly the additional attenuation a compliant aggregate would require. The scale of these margins is itself a finding. The spurious-emission detections sit 27--62\,dB above the continuum criterion, and suppression of that depth at the source, harmonic filtering at the output of every element of a transmitting phased array, is intrusive engineering: inexpensive at design time, costly to prohibitive as a retrofit. Margins of tens of decibels will not be closed by incremental mitigation.

To show how a single entry of Table~\ref{tab:facilities-cont} is built, and why a strong detection need not reach $100\%$ loss, take the brightest S-band unintended-radiation detection, the \SI{2656}{MHz} Starlink detection, at the Hobart \mbox{26-m} dish. The applicable protection level at this frequency is $S_{\rm lim}=\SI{-208.7}{dB(W\,m^{-2}\,Hz^{-1})}$: the RA.769 single-dish continuum level interpolated between the tabulated bands (Section~\ref{sec:interp}), raised \SI{38.1}{dB} by the emission/radiation-bandwidth correction for this narrowband detection (Section~\ref{sec:bwcorr}). The Monte-Carlo \ac{EPFD} run returns an aggregate spectral \ac{EPFD} for every sky cell of the M.1583 grid and every \SI{2000}{s} realisation. Pooled, these samples form a distribution with median \SI{-206.0}{}, linear mean \SI{-198.5}{} and 98th percentile \SI{-188.5}{dB(W\,m^{-2}\,Hz^{-1})}. The data loss uses none of these statistics directly: it is the fraction of samples that exceed $S_{\rm lim}$, here \SI{71.3}{\percent}, and this becomes the entry in Table~\ref{tab:facilities-cont}. The statistics instead locate the distribution for what follows: the sky maps of Figure~\ref{fig:facuemrsky} colour the per-cell mean, the median is the typical sample, and the 98th percentile sets the margin (Table~\ref{tab:facilities-cont-margin}) and the per-satellite limits of Section~\ref{sec:limits}.

This also resolves an apparent paradox in the sky maps of Figure~\ref{fig:facuemrsky}: their cells sit near \SI{-200}{dB(W\,m^{-2}\,Hz^{-1})}, above the \SI{-208.7}{} threshold, yet the loss is not \SI{100}{\percent}. The colour-coded maps show the per-cell \emph{mean} aggregate \ac{EPFD}, whereas the loss is a property of the full distribution, which is strongly right-skewed. A few epochs, when a satellite crosses a bright rise or set azimuth close to the beam, dominate the linear mean and lift it about \SI{10}{dB} above the threshold; the \emph{typical} instantaneous sample lies far lower (the median is only \SI{2.7}{dB} above threshold). The loss counts samples above $S_{\rm lim}$ rather than asking whether the mean clears it, so the upper \SI{71}{\percent} of the distribution qualifies, not all of it. The same detection at the \mbox{64-m} Murriyang gives only \SI{43.5}{\percent}. Its \SI{30}{deg} elevation limit removes the low-elevation pointings whose cells are the brightest (boresight and the near sidelobes aimed into the satellite-dense sky just above the horizon). The satellites below the limit remain in the aggregate of every accessible pointing, but couple only through the far-sidelobe floor, shifting the whole distribution down until its sample median falls just \emph{below} the threshold. A brighter detection behaves differently again: for the \SI{2700}{MHz} radiation, and for every harmonic, the \emph{entire} distribution clears the threshold and the loss saturates at \SI{100}{\percent}, the regime drawn for the \SI{5240}{MHz} harmonic in Figure~\ref{fig:cdf5240}. The southernmost Hobart antenna is the most exposed on these marginal detections. The \mbox{64-m} Murriyang is the \emph{least} exposed despite its collecting area, because its \SI{30}{deg} elevation limit excludes the brightest pointings from its accessible sky; the aperture instead dominates the saturation risk of Section~\ref{sec:limits}. ASKAP, observing only \mbox{0.7--1.8\,GHz}, carries the L-band detections, now ATCA-calibrated apart from two (\SI{1474}{} and \SI{1680}{MHz}) still held at the detection floor. The calibrated levels exceed the continuum criterion across ASKAP's whole accessible sky at several of them; the \SI{1320}{}, \SI{1350}{}, \SI{1440}{} and \SI{1620}{MHz}\footnote{The \SI{1620}{MHz} detection lies just above the \SIrange{1610.6}{1613.8}{MHz} band used for OH observations, where the spectral-line criterion (RA.769 Table~2) applies rather than the continuum criterion used here. For kHz-wide radiation the spectral-line criterion is effectively the more stringent by $\sim$\SI{18}{dB}: its \SI{20}{kHz} channel admits almost no bandwidth dilution (Section~\ref{sec:bwcorr}), whereas the same detection dilutes by \SI{38}{dB} in the broadband detection bandwidth adopted here. The continuum figures tabulated near spectral-line bands are to that extent conservative, and SNIFFLES-I detected radiation inside the protected band itself, at \SI{1613.19}{MHz} \citep{indermuehle2026sniffles}.} radiation all reach $100\%$ loss (Appendix~\ref{app:facplots}, Figure~\ref{fig:appASK}). The full data-loss grid (all stations, scenarios and criteria) is in Appendix~\ref{app:full}.

\begin{table*}[t]
\caption{Present-fleet single-dish continuum data loss (\%) at each Australian RAS facility, per emission or radiation detection (ATCA-calibrated level where available, else the detection floor as a lower bound). Dish diameter in the column head; a dash marks a detection outside the facility's observing band. Values in red are floor-only lower bounds (detections not yet measured at ATCA). The Hobart \mbox{26-m} column assumes an adopted \SI{6}{deg} minimum elevation (matching the comparable Ceduna dish), the antenna's operational limit not being separately catalogued here. Class as in Table~\ref{tab:inputs}. $^\dagger$The \SI{2620}{MHz} row is an intended, licensed in-band emission: within its authorised band an operator may transmit at its licensed power, so this data loss measures the incompatibility of a co-frequency radio-astronomy receiver with an authorised downlink, a question of spectrum allocation, and is not a compliance failure. Data loss is rounded to whole percent; Table~\ref{tab:appendix} in the appendix gives the same values to one decimal, together with both level brackets and the companion margins.}
\label{tab:facilities-cont}
{\tablefont 
\begin{tabular}{@{\extracolsep{\fill}}rlrrrrrr}
\toprule
Freq & Class & ATCA & Mopra & Murriyang & Ceduna & Hobart & ASKAP\\
(MHz) & & 22\,m & 22\,m & 64\,m & 30\,m & 26\,m & 12\,m\\
\hline
1320 & UEMR & 100 & 100 & 100 & --- & 100 & 100\\
1350 & UEMR & 100 & 100 & 100 & --- & 100 & 100\\
1440 & UEMR & 100 & 100 & 100 & --- & 100 & 100\\
1454 & UEMR & 10 & 9 & 8 & --- & 10 & 8\\
1474 & UEMR & \textcolor{red}{0} & \textcolor{red}{0} & \textcolor{red}{0} & --- & \textcolor{red}{0} & \textcolor{red}{0}\\
1620 & UEMR & 100 & 100 & 100 & 100 & 100 & 100\\
1680 & UEMR & \textcolor{red}{0} & \textcolor{red}{0} & \textcolor{red}{0} & \textcolor{red}{0} & \textcolor{red}{0} & \textcolor{red}{0}\\
1777 & UEMR & 36 & 37 & 33 & 37 & 34 & 25\\
2595 & Spurious & 100 & 100 & 100 & 100 & 100 & ---\\
2620$^\dagger$ & Intended & 100 & 100 & 100 & 100 & 100 & ---\\
2656 & UEMR & 51 & 53 & 44 & 57 & 71 & ---\\
2700 & UEMR & 100 & 100 & 100 & 100 & 100 & ---\\
4995 & UEMR & 32 & 33 & 7 & 32 & 42 & ---\\
5190 & Spurious & 5 & 5 & 4 & 5 & 8 & ---\\
5240 & Spurious & 100 & 100 & 100 & 100 & 100 & ---\\
7860 & Spurious & 100 & 100 & 100 & 100 & 100 & ---\\
8599 & UEMR & 5 & 5 & 2 & 4 & 5 & ---\\
8640 & UEMR & 100 & 100 & 100 & 100 & 100 & ---\\
10480 & Spurious & 100 & 100 & 100 & 100 & 100 & ---\\
\botrule
\end{tabular}
}
\end{table*}

\begin{table*}[t]
\caption{Present-fleet 98th-percentile margin $M = S_{\rm lim}-\mathrm{pfd}_{98}$ (dB) against the single-dish continuum criterion at each Australian RAS facility, per emission or radiation detection: the companion to Table~\ref{tab:facilities-cont} in the linear quantity. Where the data loss saturates at $100\%$ and flattens, the margin still resolves the depth of the exceedance: negative $M$ gives directly the aggregate attenuation required to restore RA.769 compliance, and positive $M$ the headroom in hand. Margins at the \num{100000} and \num{300000}-satellite scenarios follow by subtracting \SI{9.2}{} and \SI{14.0}{dB}. For the red floor-only detections the tabulated $M$ is an upper bound (the true margin is at most this large), while the corresponding data loss in Table~\ref{tab:facilities-cont} is a lower bound. Their margin entries carry an explicit $<$. The Monte-Carlo sampling uncertainty on the margins (bootstrap standard error of the pooled 98th percentile, resampled over iterations) has median \SI{0.1}{dB} and lies below \SI{0.3}{dB} for \SI{90}{\percent} of entries, reaching \SI{1.4}{dB} only for the sparsest radiating population (the 21-satellite GuoWang detection at \SI{1777}{MHz}); the uncertainty budget is instead dominated by the input levels, bracketed by the floor and ATCA rows and by the compression-limited $\geq$ bounds.}
\label{tab:facilities-cont-margin}
{\tablefont 
\begin{tabular}{@{\extracolsep{\fill}}rlrrrrrr}
\toprule
Freq & Class & ATCA & Mopra & Murriyang & Ceduna & Hobart & ASKAP\\
(MHz) & & 22\,m & 22\,m & 64\,m & 30\,m & 26\,m & 12\,m\\
\hline
1320 & UEMR & $-$16.0 & $-$16.3 & $-$18.9 & --- & $-$18.6 & $-$14.5\\
1350 & UEMR & $-$17.0 & $-$17.3 & $-$19.7 & --- & $-$19.6 & $-$15.5\\
1440 & UEMR & $-$14.7 & $-$15.0 & $-$17.6 & --- & $-$17.3 & $-$13.2\\
1454 & UEMR & $-$15.1 & $-$15.1 & $-$7.5 & --- & $-$15.6 & $-$13.6\\
1474 & UEMR & \textcolor{red}{$<$9.2} & \textcolor{red}{$<$9.2} & \textcolor{red}{$<$9.4} & --- & \textcolor{red}{$<$7.2} & \textcolor{red}{$<$11.3}\\
1620 & UEMR & $-$15.0 & $-$15.2 & $-$18.5 & $-$16.1 & $-$17.4 & $-$13.3\\
1680 & UEMR & \textcolor{red}{$<$5.1} & \textcolor{red}{$<$4.9} & \textcolor{red}{$<$1.6} & \textcolor{red}{$<$4.0} & \textcolor{red}{$<$2.7} & \textcolor{red}{$<$6.8}\\
1777 & UEMR & $-$21.9 & $-$21.3 & $-$17.1 & $-$17.8 & $-$17.2 & $-$21.9\\
2595 & Spurious & $-$36.0 & $-$36.4 & $-$28.3 & $-$37.0 & $-$37.9 & ---\\
2620$^\dagger$ & Intended & $-$73.2 & $-$73.4 & $-$65.4 & $-$74.0 & $-$74.8 & ---\\
2656 & UEMR & $-$18.5 & $-$18.8 & $-$10.7 & $-$19.3 & $-$20.1 & ---\\
2700 & UEMR & $-$23.2 & $-$23.4 & $-$28.0 & $-$24.5 & $-$25.3 & ---\\
4995 & UEMR & $-$6.1 & $-$6.4 & $-$10.6 & $-$7.4 & $-$8.0 & ---\\
5190 & Spurious & $-$8.3 & $-$8.8 & $-$2.0 & $-$8.6 & $-$10.1 & ---\\
5240 & Spurious & $-$59.9 & $-$60.4 & $-$53.6 & $-$60.2 & $-$61.6 & ---\\
7860 & Spurious & $-$44.8 & $-$45.9 & $-$44.0 & $-$44.7 & $-$48.8 & ---\\
8599 & UEMR & $-$6.8 & $-$7.1 & 0.1 & $-$6.7 & $-$7.7 & ---\\
8640 & UEMR & $-$19.2 & $-$19.4 & $-$16.2 & $-$20.4 & $-$20.8 & ---\\
10480 & Spurious & $-$27.3 & $-$27.8 & $-$29.3 & $-$28.6 & $-$30.5 & ---\\
\botrule
\end{tabular}
}
\end{table*}

\subsection{Single-dish radio astronomy is already exceeded across the whole sky}

Every result in this section is judged against the single-dish continuum criterion; the (Geodetic) VLBI case is discussed in Section~\ref{sec:geovlbi}. The continuum criterion is exceeded so readily because it is the stringent one of the pair. Its threshold protects the radiometric fluctuation that remains after averaging over the \SI{2000}{s} integration (Section~\ref{sec:thresholds}) and therefore lies some \SI{40}{dB} below the VLBI threshold. The emission/radiation-bandwidth dilution that spares VLBI from the narrowband detections (Section~\ref{sec:bwcorr}) does \emph{not} apply here: even the kHz-wide UEMR detections, integrated over the wide continuum channel, exceed it. 

\begin{figure}[t]
\centerline{\includegraphics[width=\columnwidth]{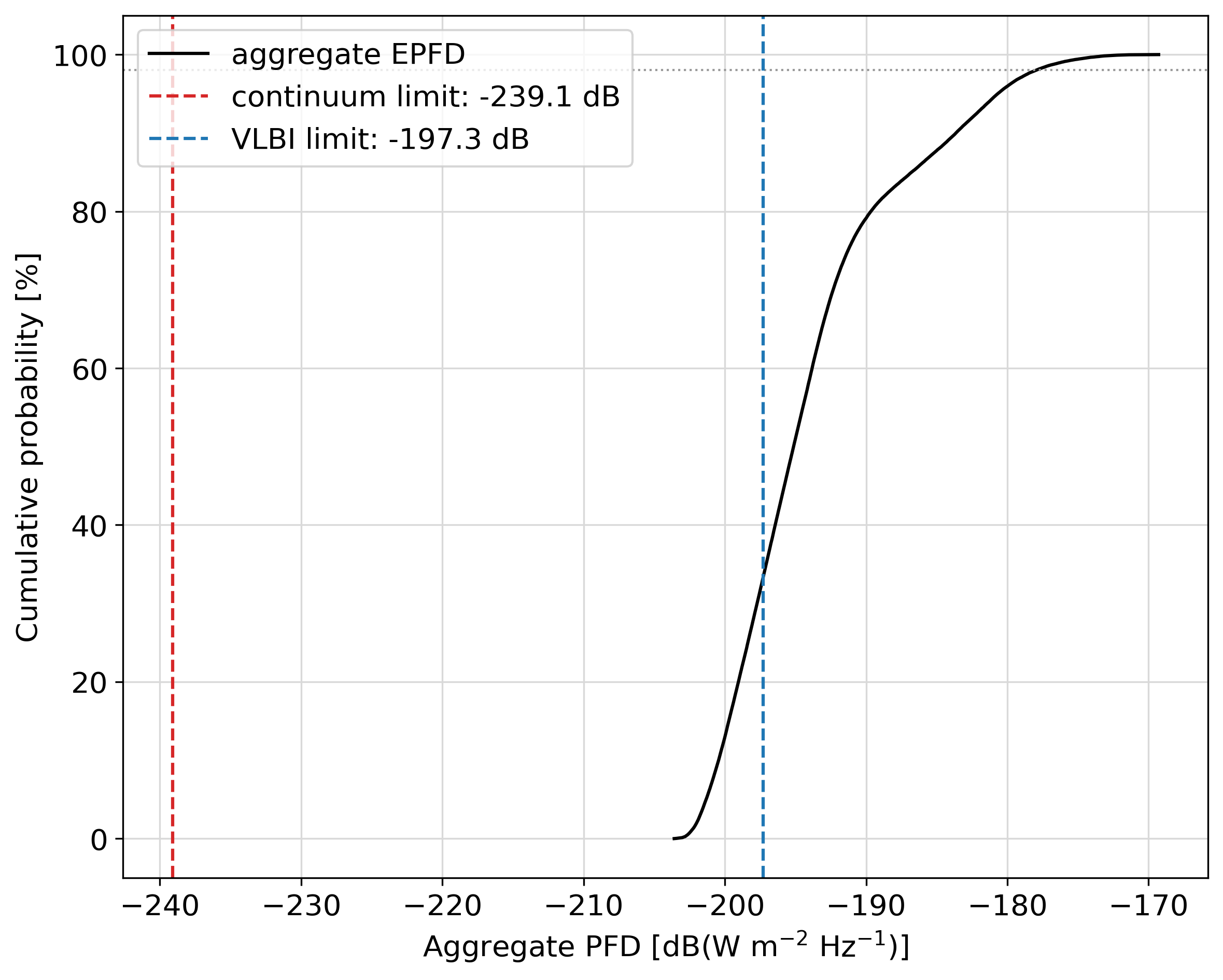}}
\caption{Cumulative distribution of the aggregate spectral \ac{EPFD} at Hobart for the \SI{5240}{MHz} Starlink \ac{DTD} second harmonic (present fleet, ATCA level). The single-dish (Continuum) RA.769 limit is exceeded by the whole distribution ($100\%$ loss); the VLBI limit is exceeded over $68\%$ of the distribution. (Y-factor calibrated; a lower bound, as this harmonic compresses the calibrating receiver.)}
\label{fig:cdf5240}
\end{figure}

Figure~\ref{fig:cdf5240} shows the \SI{5240}{MHz} \ac{DTD} second harmonic as an example. The entire cumulative distribution of the aggregate \ac{EPFD} lies above the continuum limit (left, red; $100\%$ loss), while the VLBI limit (right, blue) is crossed at the $\sim$32nd percentile, i.e.\ $68\%$ data loss. This harmonic drives the calibrating receiver into compression, so its flux density, and hence this loss, is a lower bound. Section~\ref{sec:saturation} locates the compression in the digital back-end of the calibrating system and bounds how far the measured levels understate the true ones.

\begin{figure*}[t]
\centerline{\includegraphics[width=\textwidth]{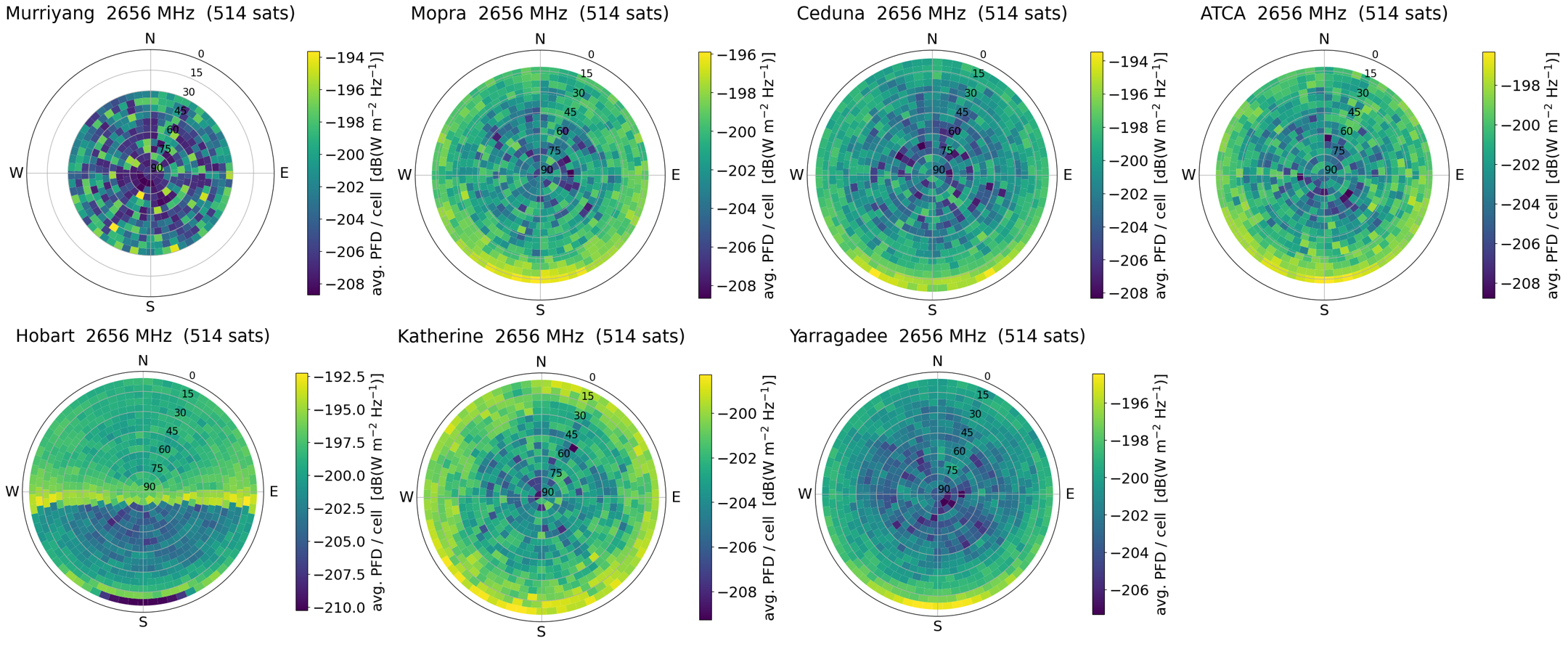}}
\caption{Modelled aggregate-\ac{EPFD} sky coverage (present fleet, ATCA level; a simulation output, not a measured map of the sky) of the \SI{2656}{MHz} Starlink unintended-radiation detection, the brightest S-band UEMR forest detection, at each tested facility (polar all-sky maps: zenith at centre, horizon at rim, north at top; each panel labelled with the facility and radiating-satellite count). Murriyang is sparser because its \SI{30}{deg} elevation limit excludes the low-elevation pointings. ASKAP, observing only \mbox{0.7--1.8\,GHz}, does not see this detection.}
\label{fig:facuemrsky}
\end{figure*}

\begin{figure*}[t]
\centerline{\includegraphics[width=\textwidth]{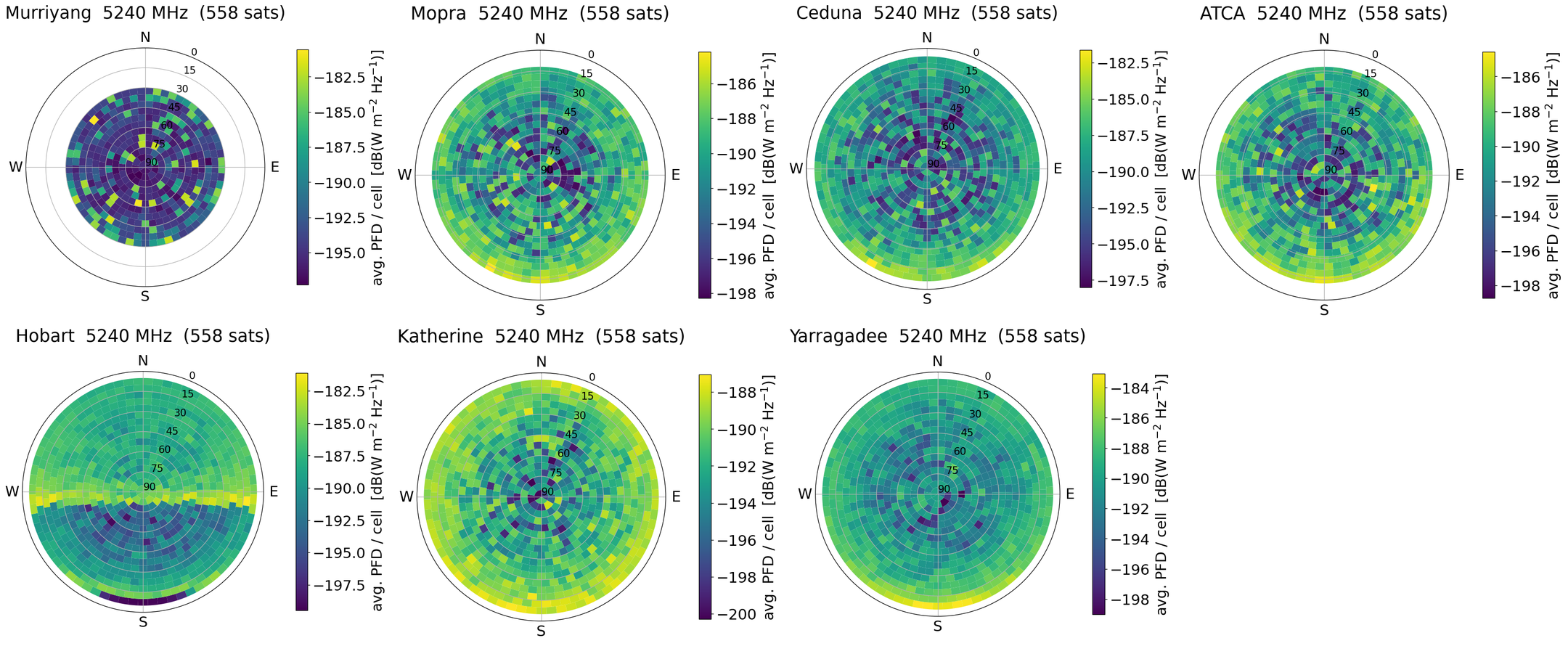}}
\caption{As Figure~\ref{fig:facuemrsky}, but for the \SI{5240}{MHz} \ac{DTD} second harmonic, whose single-dish data loss reaches $100\%$ at every facility.}
\label{fig:facharmsky}
\end{figure*}

Figures~\ref{fig:facuemrsky} and~\ref{fig:facharmsky} give the per-facility detail behind Table~\ref{tab:facilities-cont}, mapping the aggregate-\ac{EPFD} sky coverage of a strong unintended-radiation detection (the \SI{2656}{MHz} Starlink forest detection) and of the \ac{DTD} second harmonic (\SI{5240}{MHz}) at each facility. The \ac{EPFD} concentrates along the satellite rise and set azimuths and thins towards the southern sky as seen from these southern-hemisphere sites; Murriyang is markedly sparser because its \SI{30}{deg} elevation limit excludes the low-elevation pointings. For these bright detections the single-dish data loss saturates to near-total across the accessible sky and carries little spatial detail (Table~\ref{tab:facilities-cont}); we map it cell-by-cell only for the one illustrative case of Figure~\ref{fig:loss7860}. Appendix~\ref{app:facplots} collects the per-facility \ac{EPFD} sky coverage for the remaining non-saturated detections.

\section{Geodetic VLBI: the special case}
\label{sec:geovlbi}

Geodetic VLBI is the scientifically and societally distinct case introduced in Section~\ref{sec:vgos}: a global interferometer whose broadband group-delay technique (32 channels of \SI{32}{MHz} synthesised across \mbox{3--14\,GHz}) realises the reference frames and Earth-orientation parameters. It is judged against the looser VLBI protection criterion (interference $<1\%$ of receiver noise) because interference rarely correlates across long baselines. Here we assess the VLBI-specific impact for the Australian AuScope array (Hobart, Katherine, Yarragadee). Hobart is the most severely affected station: a shell's orbital density peaks at latitudes equal to its inclination, and Hobart's latitude (\SI{-42.8}{\degree}) lies directly under the density cusp of the \SI{43}{\degree}-inclined \ac{DTD} shell (330 of the 642 \ac{DTD} satellites; the remaining 312, at \SI{53}{\degree}, culminate only \SI{10}{\degree} further south). It therefore sees the most \ac{DTD} satellites at any moment (25 on average, against 23 at Yarragadee and 17 at Katherine), at higher elevations and correspondingly shorter ranges; the time-averaged $1/d^{2}$ sum over the visible fleet reproduces the station ordering of the \ac{EPFD} results; we therefore quote Hobart values. Table~\ref{tab:key} summarises the VLBI data loss and the constellation size $N_{\rm breach}$ at which the 98th-percentile aggregate first reaches the VLBI threshold. $N_{\rm breach}$ is the total number of satellites of that type in orbit, not a number near the telescope, at which the sidelobe-received aggregate becomes detrimental in $2\%$ of pointings and integration windows. At $N_{\rm breach}$ the data loss is therefore $2\%$, not $100\%$, and no satellite need cross the main beam.

\begin{table}[t]
\caption{VLBI data loss (\%) at Hobart for the measured (ATCA) emission/radiation levels, for the present fleet and the \num{100000}/\num{300000}-satellite scenarios, and the constellation size $N_{\rm breach}$ at which the 98th-percentile aggregate reaches the RA.769 VLBI threshold. $N_{\rm breach}$ is a total in-orbit population; the aggregate is received through the sidelobes, and at $N_{\rm breach}$ the data loss is $2\%$ by construction. Class as in Table~\ref{tab:inputs}. The $^\dagger$ on the \SI{2620}{MHz} row carries the same meaning as in Table~\ref{tab:facilities-cont}; its $N_{\rm breach}$ is not a threshold that nine spacecraft violate, but the population at which a co-frequency receiver can no longer share the band.}
\label{tab:key}
{\tablefont 
\begin{tabular}{@{\extracolsep{\fill}}rlrrrr}
\toprule
Freq & Class & \multicolumn{3}{c}{VLBI data loss (\%)} & $N_{\rm breach}$\\
(MHz) & & now & 100k & 300k & (sats)\\
\hline
2595 & Spurious & 0.0 & 11.0 & 29.8 & \num{44100}\\
2620$^\dagger$ & Intended & 100.0 & 100.0 & 100.0 & \num{9}\\
2656 & UEMR & 0.0 & 0.0 & 0.0 & \num{2.5e6}\\
2700 & UEMR & 0.0 & 0.0 & 0.1 & \num{6.4e5}\\
4995 & UEMR & 0.0 & 0.0 & 0.0 & \num{4.2e7}\\
5190 & Spurious & 0.0 & 0.0 & 0.0 & \num{2.0e7}\\
5240 & Spurious & 68.2 & 100.0 & 100.0 & \num{144}\\
7860 & Spurious & 7.6 & 34.6 & 71.2 & \num{2730}\\
8599 & UEMR & 0.0 & 0.0 & 0.0 & \num{8.5e7}\\
8640 & UEMR & 0.0 & 0.0 & 0.0 & \num{5.8e6}\\
10480 & Spurious & 0.0 & 1.6 & 4.2 & \num{1.2e5}\\
\botrule
\end{tabular}
}
\end{table}

\begin{figure}[t]
\centerline{\includegraphics[width=\columnwidth]{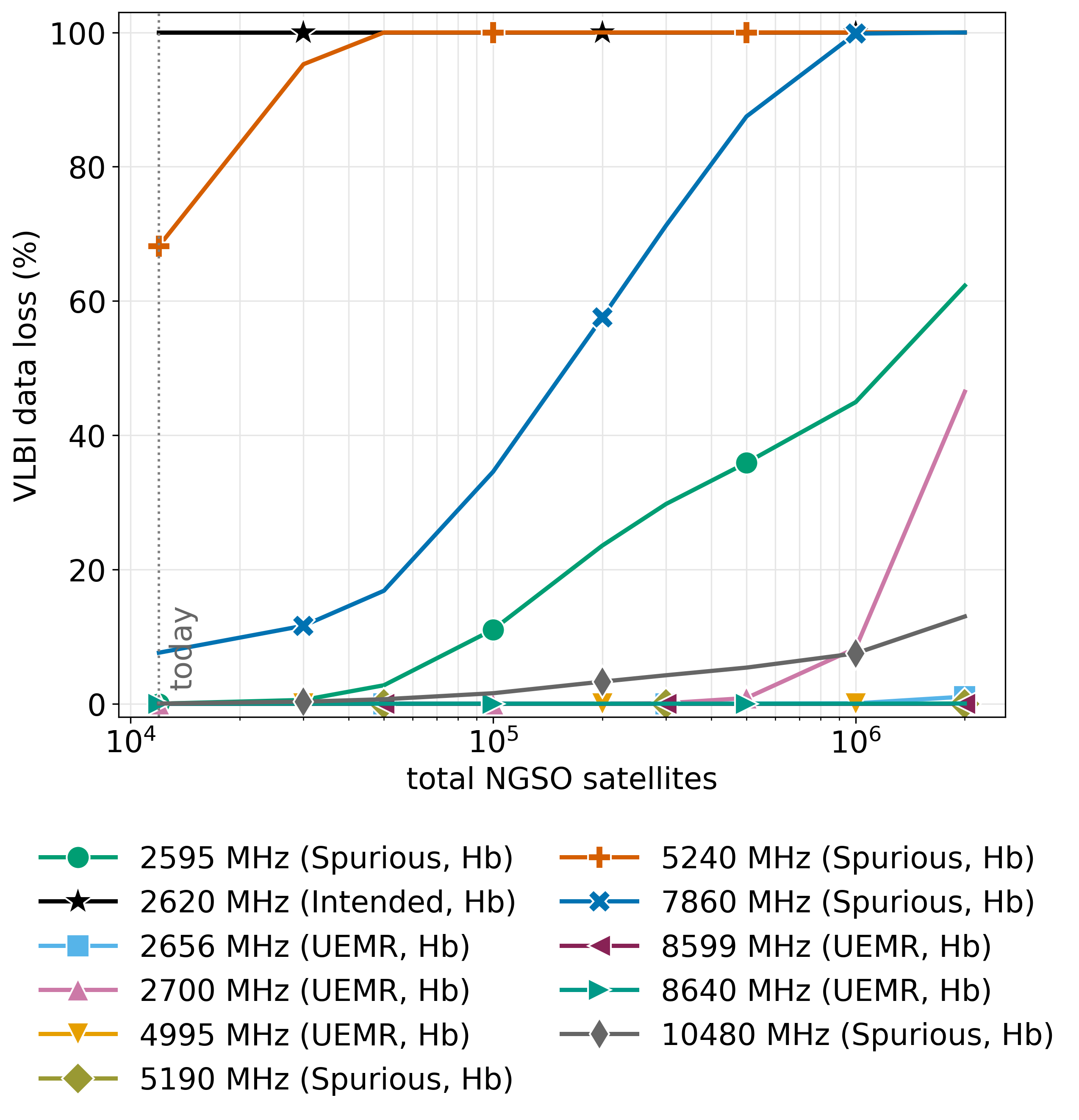}}
\caption{Predicted VLBI data loss versus total NGSO constellation size (ATCA measured levels, worst AuScope station) for each SNIFFLES-I emission or radiation detection, swept from the present fleet to \num{2000000} satellites, with the emission/radiation-bandwidth correction of Section~\ref{sec:bwcorr} applied. The vertical dotted line marks today's modelled population ($\sim$\num{12000}). The bright \ac{DTD} payload dominates: the \SI{2620}{MHz} intended downlink (\SI{4.3}{MJy}, compressed) is at $100\%$ loss at the present fleet, the \SI{5240}{} and \SI{7860}{MHz} harmonics exceed the threshold at the present fleet, the \SI{5240}{MHz} second harmonic reaching $100\%$ loss by $\sim$\num{100000} satellites, and the \SI{2595}{MHz} \ac{LO} fundamental (\SI{1.7}{MJy}, compressed) breaches by $\sim$\num{44000}. The incidental narrowband platform UEMR (\SI{2656}{}, \SI{2700}{}, \SI{8599}{MHz}), diluted against the \SI{32}{MHz} channel, stays near zero across the modelled range; only \SI{2700}{MHz} rises appreciably, and only beyond $\sim$\num{640000} satellites. All detections shown are ATCA-measured.}
\label{fig:breach}
\end{figure}

The VLBI criterion is $\sim$40\,dB less stringent than the single-dish continuum criterion (for the interferometric observable; the calibration tier of the geodetic chain is held to continuum-class levels, Section~\ref{sec:limits}), and the emission/radiation-bandwidth correction (Section~\ref{sec:bwcorr}) shows why: what breaches VLBI is the bright \ac{DTD} payload, not the incidental platform radiation. The intended DTD downlink at \SI{2620}{MHz}, measured at \SI{4.3}{MJy} (a compressed lower bound), already exceeds the VLBI criterion with $100\%$ data loss for the present fleet, crossing the threshold at $N_{\rm breach}\approx9$ satellites. A single satellite at the measured level, received at \SI{1500}{km} through a \SI{0}{dBi} sidelobe, delivers $\approx-199$\,dB(W\,m$^{-2}$\,Hz$^{-1}$), some \SI{7}{dB} above the VLBI threshold, so nine satellites suffice for the aggregate to be detrimental in $2\%$ of all realisations. The \ac{DTD} harmonics follow. The \SI{5240}{MHz} second harmonic is lost $68\%$ of the time today (a lower bound, as it too compresses the calibrating receiver, Section~\ref{sec:saturation}) and reaches $100\%$ loss by \num{100000} satellites. The \SI{7860}{MHz} third harmonic, just below the legacy X-band, rises from $7.6\%$ data loss today to $71\%$ at \num{300000}. The $7.6\%$ present-fleet value is at the \SI{2000}{s} reference window; at the \SI{10}{}--\SI{30}{s} VGOS scan length the third harmonic clears the present fleet (Section~\ref{sec:window}), though its growth to $71\%$ holds at every window. The DTD local-oscillator fundamental at \SI{2595}{MHz}, also a compressed lower bound (\SI{1.7}{MJy}), breaches as well, reaching $30\%$ at \num{300000}: it is narrowband (a frequency conversion product of the carrier-generating oscillator) yet bright enough to overcome the bandwidth dilution. The remaining narrowband UEMR, the incidental platform radiation at \SI{2656}{}, \SI{2700}{} and \SI{8599}{MHz}, does \emph{not} reach the VLBI threshold at any plausible fleet size. Its \SI{5}{kHz}-class bandwidth raises the effective threshold by 38--45\,dB, so the brightest detection (\SI{2700}{MHz}) crosses only near \num{640000} satellites and the rest beyond several million. This does not make that radiation harmless to VLBI: a strong narrow detection still saturates the front-end (Section~\ref{sec:limits}) and corrupts the phase of the \SI{32}{MHz} channel it falls in. Its effect is simply not the aggregate sensitivity loss that the bright detections inflict. The $N_{\rm breach}$ column of Table~\ref{tab:key} and Figure~\ref{fig:breach} make the split clear: $\approx9$ satellites for the \SI{2620}{MHz} downlink, \num{144} for the \SI{5240}{MHz} harmonic, $\approx$\num{2700} for the \SI{7860}{MHz} harmonic and \num{44100} for the \SI{2595}{MHz} fundamental, against $>$\num{600000} for the incidental platform detections.

\begin{figure}[t]
\centerline{\includegraphics[width=0.92\columnwidth]{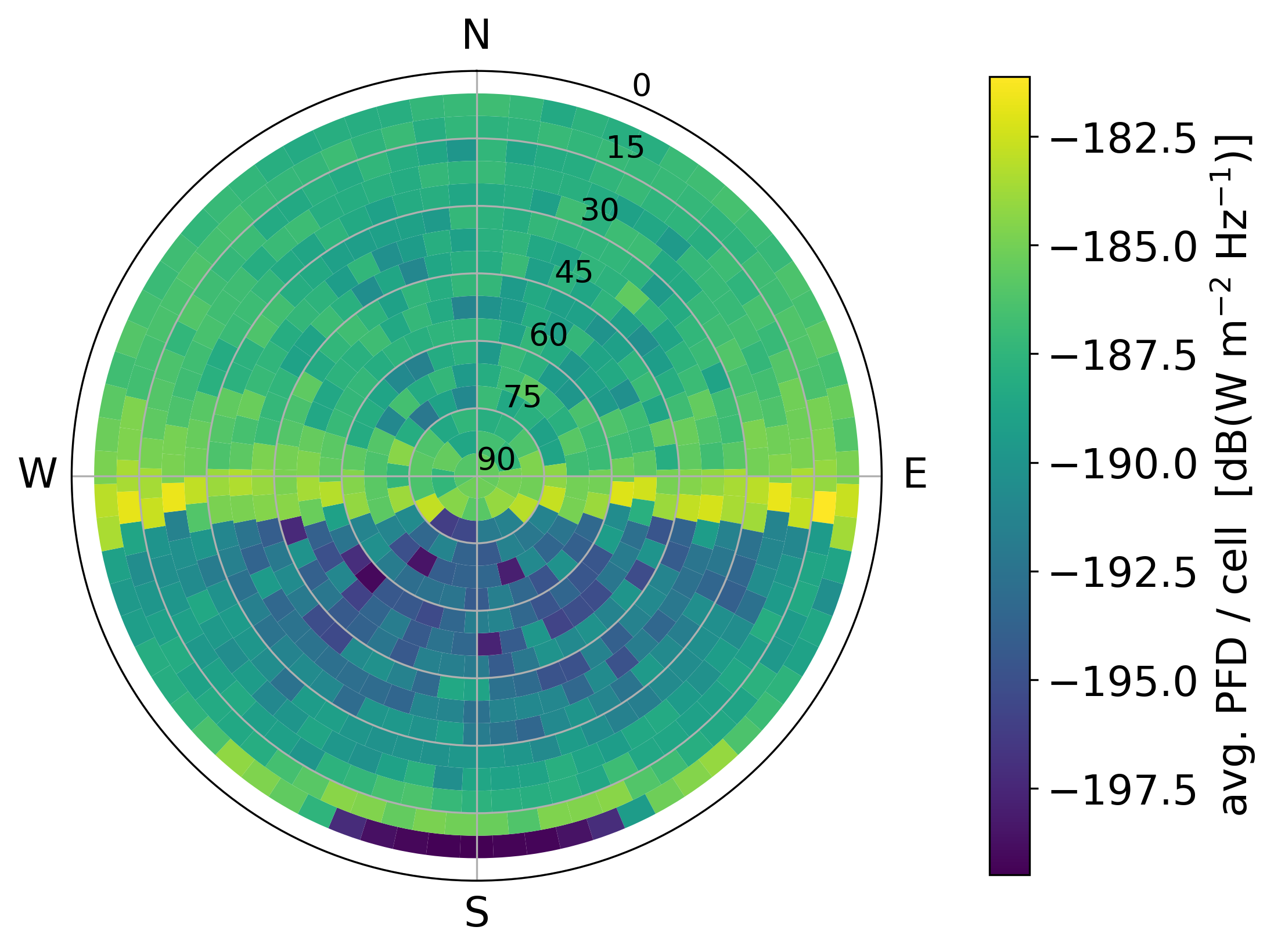}}
\caption{Polar all-sky map (zenith at centre, horizon at rim; N at top) of the average aggregate \ac{EPFD} at Hobart for the \SI{5240}{MHz} \ac{DTD} second harmonic at the present fleet.}
\label{fig:sky5240}
\end{figure}

\begin{figure*}[t]
\centerline{%
\includegraphics[width=0.49\textwidth]{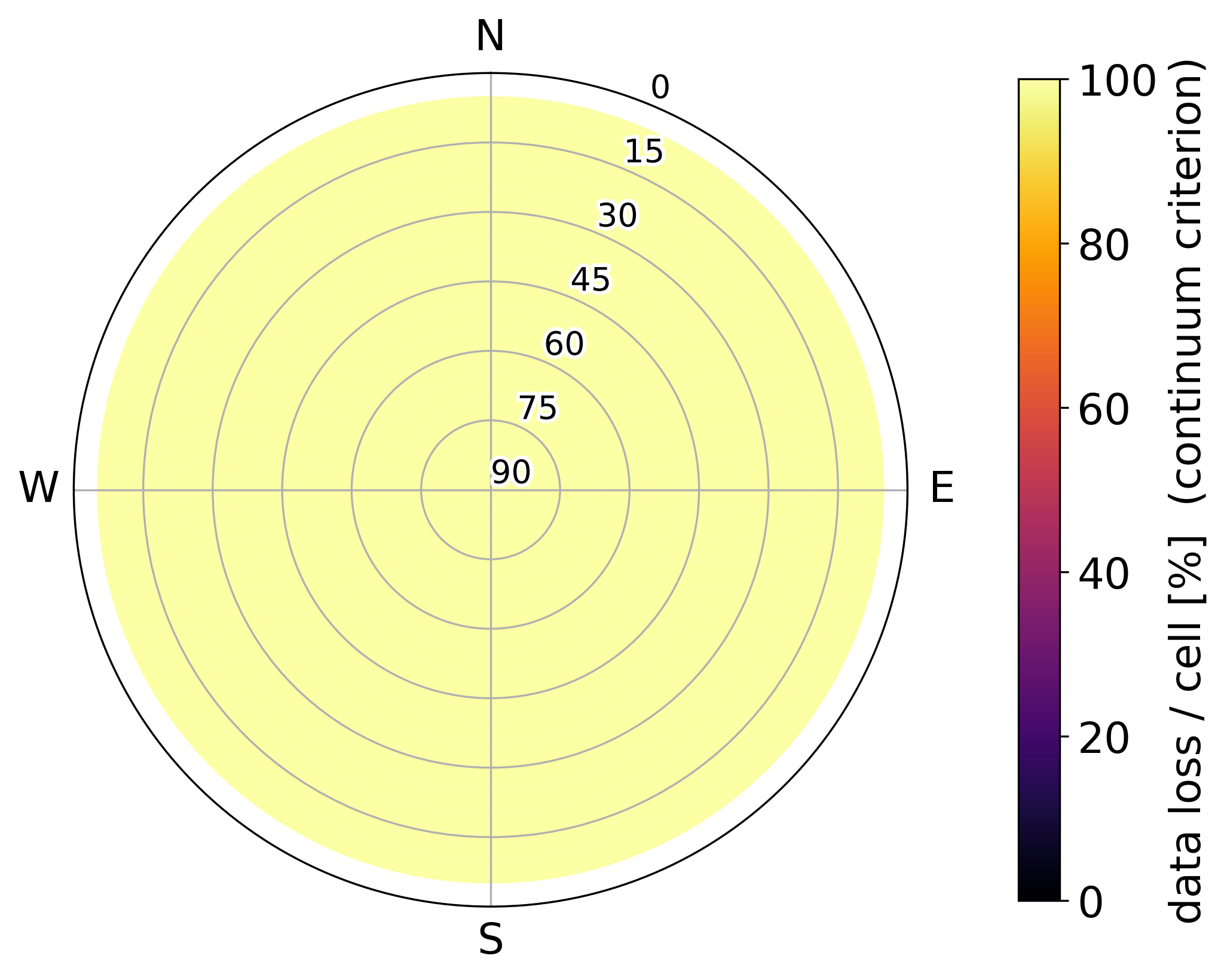}\hfill
\includegraphics[width=0.49\textwidth]{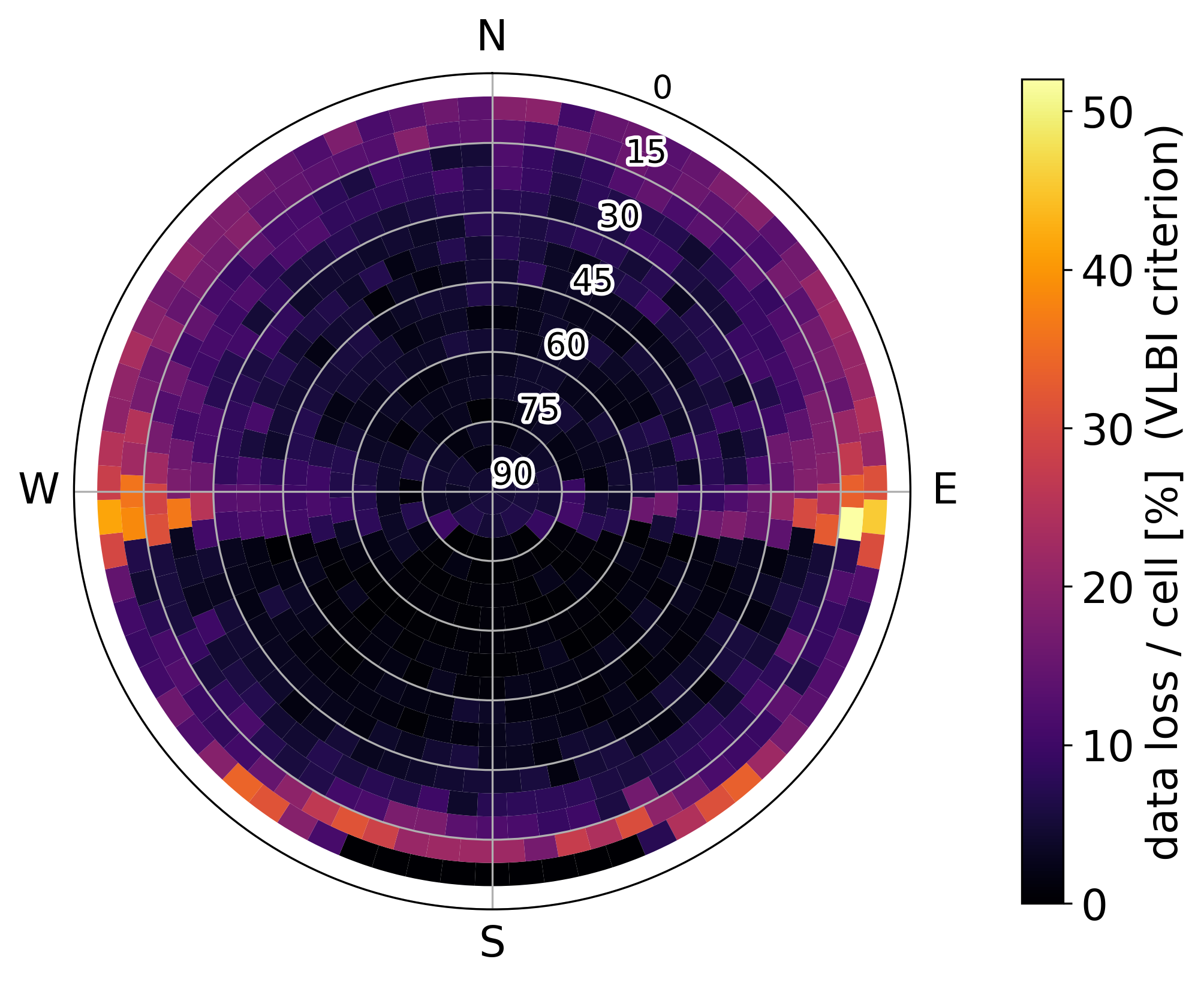}}
\caption{Per-cell data loss at Hobart for the \SI{7860}{MHz} \ac{DTD} third harmonic (present fleet, ATCA level) against the single-dish continuum criterion (left; near-total across the sky) and the VLBI criterion (right; $7.6\%$ at the \SI{2000}{s} reference window, concentrated towards the satellite-dense azimuths). The same aggregate emission is catastrophic for single-dish RAS and already eroding VLBI at the edge of the legacy X-band. (This detection is Y-factor calibrated.)}
\label{fig:loss7860}
\end{figure*}

The spatial structure of the loss is shown in Figures~\ref{fig:sky5240} and~\ref{fig:loss7860}. The aggregate \ac{EPFD} concentrates along the satellite rise/set azimuths and away from the southern sky, where the inclined shells leave a relative gap as seen from Tasmania; the data loss correspondingly fills most of the accessible sky for the brightest detections while remaining patchier (and criterion-dependent) for the weaker ones.

Which band an administration licenses for DTD largely determines its potential impact on VGOS measurements, as this indirectly determines where the harmonics of the downlink will appear (Section~\ref{sec:global}, Figure~\ref{fig:dtdharm}). Sweeping the authorised space-to-Earth bands against the four VGOS sub-bands and the legacy X-band, two regions are benign and two are the worst possible. Downlinks in \mbox{852--960\,MHz} are clear: none of their 2nd to 5th harmonics fall in any geodetic band. The lower \mbox{617--821\,MHz} bands reach only VGOS A, and only through their weak 4th and 5th harmonics. At the other extreme, the upper S-band is the most problematic. The \mbox{1930--2000} and \mbox{2620--2690\,MHz} bands each contaminate two VGOS sub-bands at once through strong low-order harmonics and land a further harmonic just below the legacy X-band: the 2nd harmonic of \mbox{2620--2690\,MHz} is the \SI{5240}{MHz} detection in VGOS B, and its 3rd the \SI{7860}{MHz} detection immediately below the legacy X-band. The \mbox{1475--1518\,MHz} band is also not well suited, its 2nd harmonic landing in VGOS A. From a geodetic-protection standpoint, DTD downlinks therefore belong below \mbox{$\sim$1\,GHz}, ideally in \mbox{852--960\,MHz}, and the \mbox{1.9--2.7\,GHz} bands are the ones to avoid. It is notable that the first band in Australia used for DTD, \mbox{2620--2690\,MHz}, is the single worst choice in the set, whereas the second Australian \mbox{870--890\,MHz} DTD band sits in the cleanest part of the spectrum.

This optimum for geodetic VLBI is, however, in direct tension with the protection of radio astronomy in the same low band, and it contradicts CSIRO's protection requirements for our own facilities. The reason is ASKAP, which observes between roughly \mbox{0.7} and \mbox{1.8\,GHz}, squarely across the candidate DTD low band. A downlink at \mbox{852--960\,MHz} strikes ASKAP twice over: its bright \emph{fundamental} lands directly in band, and its \emph{second harmonic} (\mbox{1.70--1.92\,GHz}) folds into the upper ASKAP band, so the strongest and next-strongest emission products both corrupt large portions of its observed science. For geodetic VLBI, by contrast, only the far weaker high-order harmonics of that band matter, and those fall between the VGOS sub-bands. The two requirements pull in opposite directions: the band that best protects geodetic VLBI is among the worst for ASKAP, while the band that spares ASKAP, by pushing DTD up into \mbox{1.9--2.7\,GHz}, is the worst for geodetic VLBI. It is a Sophie's choice, and a clear illustration of how the radio astronomy service is increasingly squeezed into sacrificing one science case to protect another as the spectrum fills.

\subsection{Validation of the linear extrapolation}

\begin{figure*}[t]
\centerline{\includegraphics[width=\textwidth]{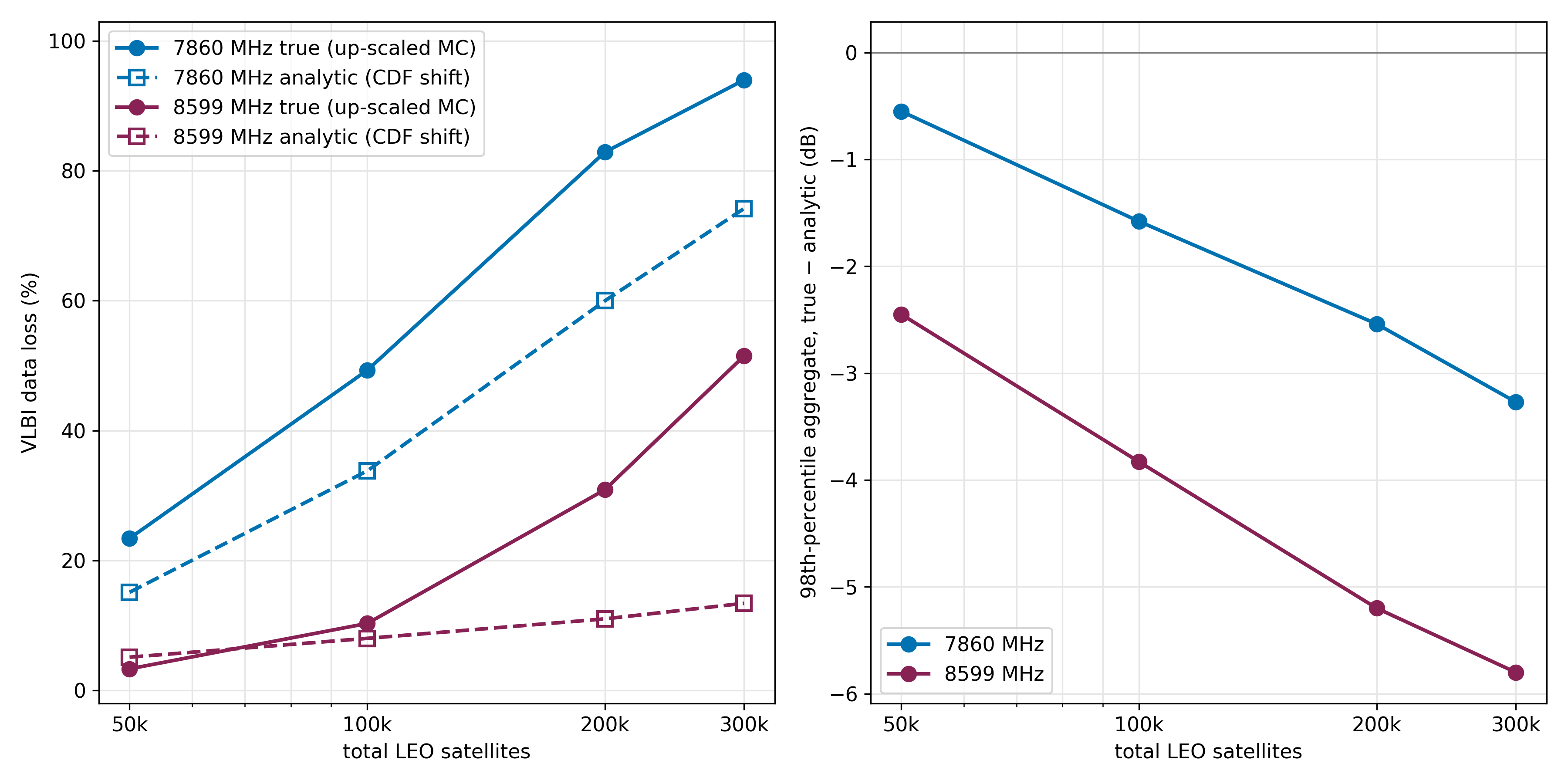}}
\caption{Validation of the linear-in-$N$ extrapolation against direct Monte-Carlo of up-scaled fleets (present radiating population cloned with randomised node and mean anomaly, preserving the altitude/inclination distribution), at Hobart. \emph{Left:} VLBI data loss from the true up-scaled simulation (solid) versus the analytic $10\log_{10}(N/N_0)$ CDF-shift (dashed) for the \SI{7860}{} and \SI{8599}{MHz} detections; the analytic estimate is a conservative lower bound for the not-yet-saturated detections. \emph{Right:} the analytic method overestimates the 98th-percentile aggregate \ac{EPFD} (true $-$ analytic is negative and grows with $N$), because the spatial distribution narrows as the sky fills, so the per-satellite limits and breach populations are, if anything, conservative.}
\label{fig:linval}
\end{figure*}

We verified the linear-aggregate extrapolation against a direct Monte-Carlo simulation of genuinely up-scaled fleets, cloning the present radiating population to the target count with randomised right ascension of the ascending node and mean anomaly, preserving the measured altitude and inclination distribution (Figure~\ref{fig:linval}). The analytic $10\log_{10}(N/N_0)$ scaling proves conservative in both relevant senses. It \emph{over}-estimates the 98th-percentile aggregate \ac{EPFD} by 2--6\,dB at $N\sim$\num{300000}, because the sky distribution narrows as it fills and the spiky present-day spatial variance is not retained; the derived per-satellite limits (Table~\ref{tab:cispr}) and breach populations are therefore conservative. At the same time it \emph{under}-estimates the fractional data loss for the detections not yet saturated: the \SI{7860}{MHz} third harmonic reaches a true $94\%$ VLBI data loss at \num{300000} satellites, against the $74\%$ of the linear estimate (Figure~\ref{fig:breach}). The conclusions of this paper therefore hold \emph{a fortiori}. The deviation is correctable: a fleet scaled by $G$ with randomised node and anomaly is statistically $G$ independent copies of the present one, and received powers add, so the up-scaled aggregate in each sky cell is the sum of $G$ independent draws from that cell's Monte-Carlo ensemble. The corrected model builds each up-scaled sample exactly so, as a per-cell sum of $G$ independently resampled iteration samples rather than one sample multiplied by $G$: the mean still scales as $G$, but the fluctuations grow only as $\sqrt{G}$, reproducing the narrowing; the resampled samples form a synthetic ensemble of the same form as a simulation's output (one aggregate per sky cell per iteration), from which the pooled distribution, its 98th percentile and the data loss are read in the same way as a standard \ac{EPFD} simulation run. Validated against the direct up-scaled runs (Appendix~\ref{app:full}, Figure~\ref{fig:sqrtncorr}), this compound resampling reproduces the true 98th percentile to within \SI{1}{dB} and the true data loss to within a few points at every tested population, where the rigid shift deviates by up to \SI{5.8}{dB}. At \num{300000} satellites the rigid shift overstates the 98th-percentile aggregate by \SI{4.1}{dB} (\SI{7860}{MHz}) and \SI{5.7}{dB} (\SI{8599}{MHz}): the per-satellite ceilings of Section~\ref{sec:limits} are conservative by that quantified margin, while the corresponding data-loss figures are understatements. We retain the rigid shift for the tabulated limits, as a verifiable single-parameter convention whose bias direction protects the service.

\subsection{Sensitivity to the integration window}
\label{sec:window}

\begin{figure}[t]
\centerline{\includegraphics[width=\columnwidth]{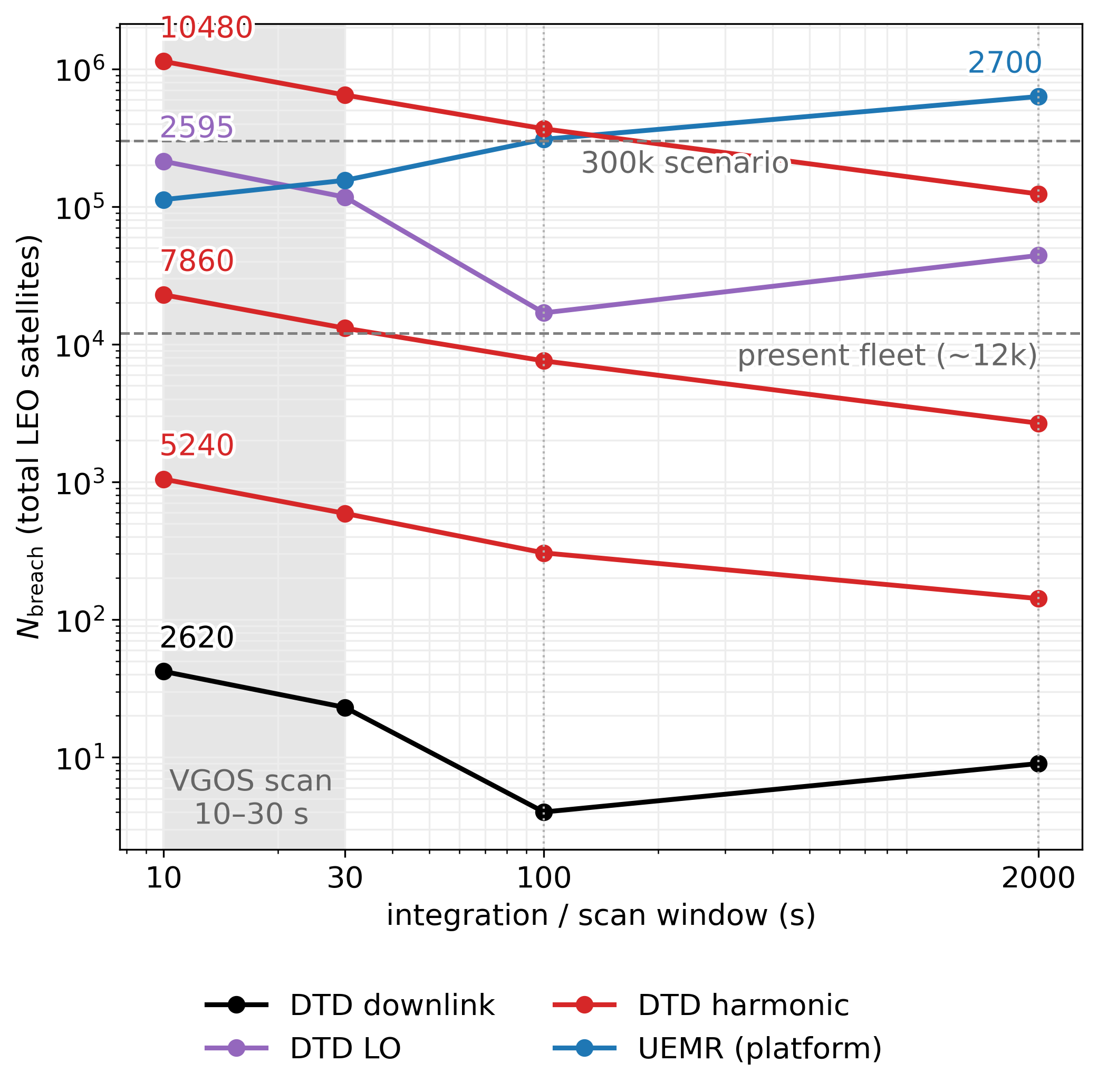}}
\caption{VLBI breach population $N_{\rm breach}$ versus scan/integration window (Hobart, the worst affected AuScope station). The sparse \ac{DTD} harmonics (red) have $N_{\rm breach}$ \emph{rising} as the window shortens, so the \SI{2000}{s} RA.769 reference is conservative for them, while the dense \SI{2700}{MHz} platform UEMR (blue) \emph{falls}, mildly anti-conservative. The \SI{5240}{MHz} second harmonic breaches below the present fleet at every window; the \SI{7860}{MHz} third harmonic crosses the present-fleet detection between the \SI{10}{}--\SI{30}{s} VGOS scan length and the \SI{2000}{s} reference. Dotted verticals mark the RA.2507 \SI{100}{s} and RA.769 \SI{2000}{s} windows. The \SI{2595}{MHz} \SI{100}{s} point carries visible Monte-Carlo scatter.}
\label{fig:window}
\end{figure}

The VLBI results above adopt the RA.769 \SI{2000}{s} reference integration. A geodetic VGOS scan is far shorter, typically \SI{10}{}--\SI{30}{s}, and the VLBI threshold is \emph{independent} of integration time $\tau$: it is set at $1\%$ of the receiver noise power, not by a sensitivity that improves as $\sqrt{\tau}$ (Report ITU-R RA.2507 \citep{ituRA2507}, whose Table~A-2 VGOS thresholds agree with our Table~\ref{tab:vlbi-threshold} to better than \SI{1}{dB}). We therefore re-ran the aggregate at scan windows of \SI{10}{}, \SI{30}{}, \SI{100}{} (the value RA.2507 adopts for VGOS continuum) and \SI{2000}{s}. In the model the telescope pointing is held fixed for the window while the satellites move, so a run of length $W$ is a $W$-second scan. The \SI{2000}{s} case reproduces the breach populations of Table~\ref{tab:key} to within the Monte-Carlo scatter (Figure~\ref{fig:window}).

The effect is detection-dependent and governed by the density of the radiating population. For the sparse \ac{DTD} harmonics (a few hundred radiators), a short scan most often falls \emph{between} transits: the short-scan 98th-percentile aggregate is lower than the \SI{2000}{s} average, and the breach population $N_{\rm breach}$ \emph{rises} as the window shortens. The \SI{2000}{s} reference is the conservative choice for these detections. For the dense platform UEMR (the \SI{2700}{MHz} detection, $\sim$\num{8000} radiators), many emitters are always present, so reduced averaging \emph{raises} the short-scan 98th percentile and $N_{\rm breach}$ \emph{falls}; here \SI{2000}{s} is mildly anti-conservative. Neither shift overturns the conclusions. The \SI{5240}{MHz} second harmonic breaches at the present fleet at every window ($N_{\rm breach}=\num{140}$--\num{1050}, all below the present $\sim$\num{12000}), and the \SI{2700}{MHz} detection remains a future-fleet concern at every window ($N_{\rm breach}>\num{100000}$). The one materially window-dependent case is the \SI{7860}{MHz} third harmonic, whose present-fleet status is marginal. It breaches at \SI{2000}{s} ($N_{\rm breach}\approx\num{2700}$), while at a \SI{10}{}--\SI{30}{s} scan length $N_{\rm breach}$ rises to \numrange{13000}{23000}, just above the present fleet; it still breaches firmly by the \num{100000}-satellite scenario. We retain the \SI{2000}{s} reference throughout for comparability with RA.769, noting that for the binding \ac{DTD} harmonics it is the conservative choice.

\section{Per-satellite emission and radiation limits}
\label{sec:limits}

Section~\ref{sec:thresholds} argued that the entry-by-entry application of the RA.769 criterion does not by itself protect geodetic VLBI against a constellation of many simultaneous emitters. We therefore state first the \emph{aggregate} threshold that does, and only then invert it. Table~\ref{tab:vlbi-threshold} gives, for each VGOS sub-band and the legacy X-band, the maximum tolerable aggregate spectral power flux density: the RA.769 VLBI (1\%-of-noise) value interpolated continuously across the \mbox{3--14\,GHz} span (Section~\ref{sec:interp}), to be met by the \emph{sum} over the whole simultaneously-radiating constellation rather than by any single satellite. The threshold is a composite, mirroring the operational requirements of the geodetic chain: the VLBI (1\%-of-noise) column protects the interferometric observable, while the $\sim$\SI{40}{dB} more stringent continuum column protects the single-dish-mode calibration on which that observable rests (the noise-diode, \ac{Tsys}, measurements against strong continuum sources), applied at the \SI{100}{s} calibration integration of Report ITU-R RA.2507, which relaxes it by \SI{6.5}{dB}; the same column doubles as the conservative single-dish bound. The proposal differs from the RA.769 limits as tabulated in two further respects. It is stated across the entire VGOS span, not only at the few narrow RAS allocations. And it is complemented by a receiver-saturation bound: a total in-beam power flux of $\approx-88.7$\,dB(W\,m$^{-2}$) drives a \mbox{12m} VGOS front-end to its \SI{-40}{dBm} linearity limit, roughly independent of sub-band\footnote{It should be noted that certain detrimental effects, such as receiver-induced intermodulation products, can already occur below the nominal linearity limit as intermodulation products can scale with higher orders; compare Annex 4 of ECC Report 271 \citep{ecc271}.}. These are the aggregate, per-sub-band and saturation-bounding properties anticipated in Section~\ref{sec:thresholds}. The VLBI column is independently corroborated by Report ITU-R RA.2507: its Table A-2, computed by the IVS from VGOS operating parameters, agrees with our interpolated values to better than \SI{1}{dB} at every band centre. What Table~\ref{tab:vlbi-threshold} adds is the aggregate interpretation, the per-sub-band statement, and the saturation bound.

\begin{table}[t]
\caption{Proposed aggregate protection threshold for geodetic VLBI, per VGOS sub-band and the legacy X-band (\mbox{8.2--8.95\,GHz}). $S_{\rm lim}$ is the maximum tolerable \emph{aggregate} spectral power flux density, summed over the whole simultaneously-radiating constellation, from the RA.769-2 VLBI (1\%-of-noise) criterion interpolated to each band centre and for the legacy X-band to the 8.4 GHz ICRF reference frequency in dB against $\log_{10}\nu$ (Section~\ref{sec:interp}); the continuum (10\%-of-fluctuation) column protects the single-dish-mode calibration measurements of the geodetic chain, applied at the \SI{100}{s} calibration integration of Report ITU-R RA.2507 (add \SI{6.5}{dB} to the tabulated \SI{2000}{s} values), and doubles as the conservative single-dish bound. Beyond the sensitivity regime, a total in-beam power flux of $\approx-88.7$\,dB(W\,m$^{-2}$) drives a \mbox{12m} VGOS aperture to its \SI{-40}{dBm} front-end linearity limit (the saturation regime), roughly independent of sub-band. Sub-band ranges are listed in Table~\ref{tab:vgos-bands}.}
\label{tab:vlbi-threshold}
{\tablefont\begin{tabular}{@{\extracolsep{\fill}}lcrr}
\toprule
Band & Centre & VLBI & Continuum \\
 & (GHz) & \multicolumn{2}{c}{$S_{\rm lim}$ (dB(W\,m$^{-2}$\,Hz$^{-1}$))} \\
\hline
VGOS A             & 3.15  & $-203.8$ & $-245.3$ \\
VGOS B             & 5.32  & $-199.2$ & $-241.1$ \\
legacy X           & 8.4  & $-195.2$ & $-240.2$ \\
VGOS C             & 9.28  & $-194.4$ & $-240.0$ \\
VGOS D             & 13.38 & $-190.4$ & $-235.9$ \\
\botrule
\end{tabular}}
\end{table}

\begin{figure}[t]
\centerline{\includegraphics[width=\columnwidth]{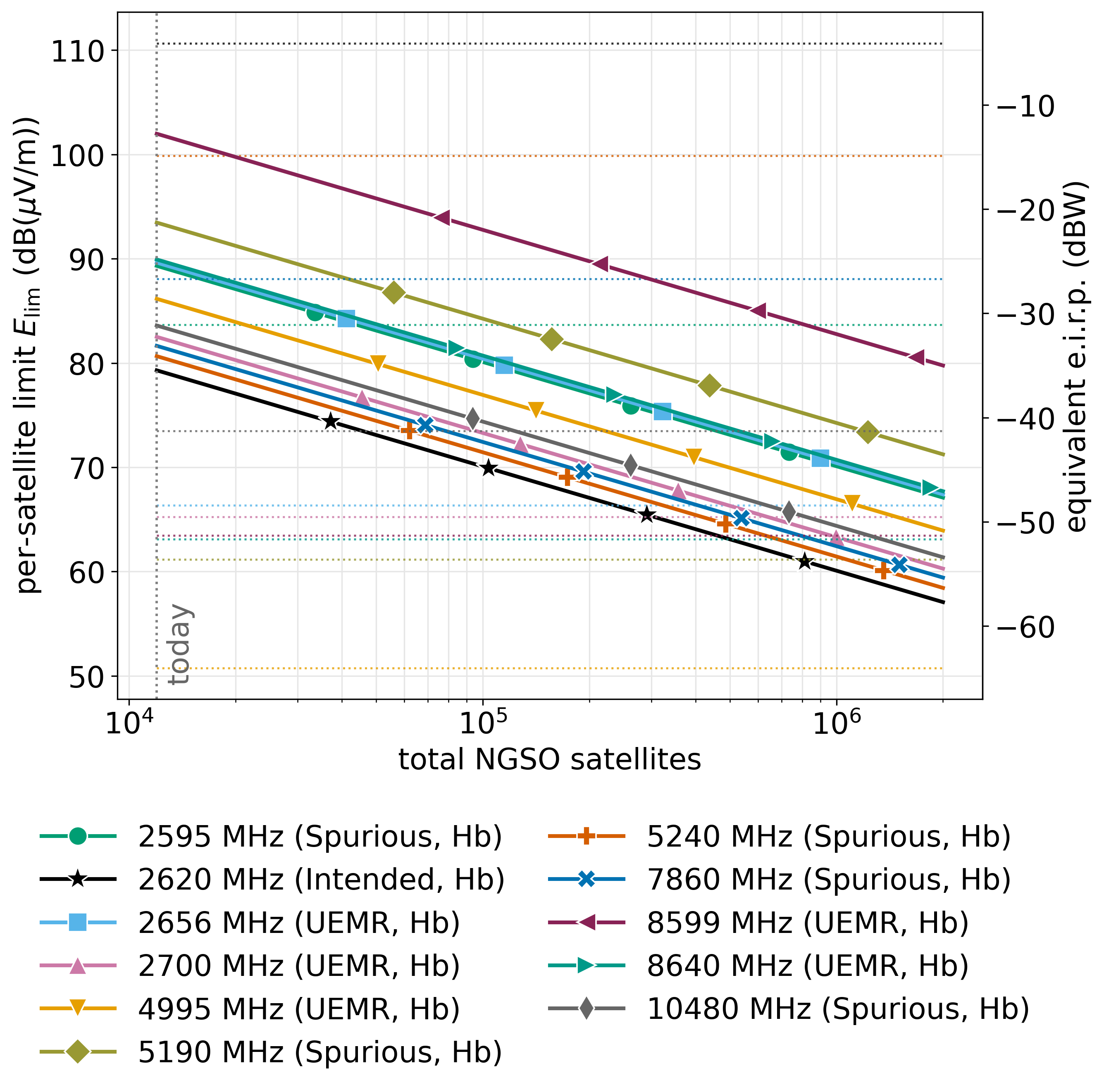}}
\caption{Per-satellite RA.769-protective field-strength limit $E_{\rm lim}$ for the \emph{VLBI} criterion (dB($\mu$V/m) at \SI{10}{m}, \SI{1}{MHz} reference bandwidth) versus total NGSO constellation size, for each SNIFFLES-I detection (all  detections ATCA-measured). $E_{\rm lim}$ falls by \SI{10}{dB} per decade of constellation growth because the aggregate is linear in $N$. Dotted horizontals mark the measured per-satellite level $E_{\rm meas}$; the vertical gap above each $E_{\rm lim}$ curve is the exceedance, which widens as the fleet grows. The right-hand axis gives the equivalent isotropically radiated power producing the same field at \SI{10}{m}, $\mathrm{e.i.r.p.} = E - \SI{114.8}{dB}$, the quantity conventionally used at ITU-R; both axes express the same physical level, so the conversion factor itself is bandwidth-free, but the level carries the reference bandwidth of its own detection: \SI{1}{MHz} for the broadband detections, whose e.i.r.p.\ can therefore be read directly as a spectral density in dBW/MHz, and the emission or radiation bandwidth for the narrowband ones, whose e.i.r.p.\ is the total power in the detection (Table~\ref{tab:cispr}). The limit depends only on the RA.769 threshold and the aggregate geometry, not on the (detection-limited) measured level. The L-band detections of Table~\ref{tab:cispr_cont} lie below AuScope's band, carry no VLBI tier, and do not appear here.}
\label{fig:elim}
\end{figure}

Inverting the aggregate requirement (Section~\ref{sec:thresholds}), we solve for the maximum tolerable \emph{per-satellite} level that keeps the 98th-percentile aggregate at the RA.769 VLBI threshold for a target constellation. For population $N$ this per-satellite \ac{e.i.r.p} spectral density is
\begin{equation}
\mathrm{EIRP}_{\rm lim}=\mathrm{EIRP}_{\rm meas}+M_0-10\log_{10}(N/N_0),
\end{equation}
where $M_0=S_{\rm lim}-\mathrm{pfd}_{98}$ is the present 98th-percentile margin. Because $\mathrm{EIRP}_{\rm meas}$ enters $M_0$ with the opposite sign it cancels, leaving
\begin{equation}
\mathrm{EIRP}_{\rm lim}=S_{\rm lim}-K-10\log_{10}(N/N_0),
\end{equation}
with $K$ the aggregate transfer factor. The limit is fixed by the RA.769 threshold and the aggregate geometry alone, \emph{independent} of the (detection-limited) measured level it is compared against. Table~\ref{tab:cispr} lists, per detection, the equivalent field strength in dB($\mu$V/m) at \SI{10}{m} in a \SI{1}{MHz} reference bandwidth (the CISPR measurement geometry above \SI{1}{GHz}) at the \num{50000}, \num{200000} and \num{300000}-satellite scenarios, the measured field strength $E_{\rm meas}$, and the decibels by which the measured level already exceeds the \num{300000}-satellite limit. Figure~\ref{fig:elim} traces $E_{\rm lim}$ continuously against constellation size, falling \SI{10}{dB} per decade of growth. For the unintended-radiation (UEMR) detections this is a CISPR-referenced radiated-disturbance limit; for the spurious-emission detections the same level expressed as an e.i.r.p. spectral density is the corresponding ITU-R SM quantity. The ceiling binds the \emph{unwanted} emission and the unintended radiation only. The intended \SI{2620}{MHz} downlink is a licensed in-band emission, and within its authorised band the operator may transmit at its licensed power; its row in Table~\ref{tab:cispr} is therefore diagnostic, a measure of the sharing problem in that band, not a compliance target. Among the detections the ceiling does bind, the \ac{DTD} harmonics are the binding ones for VLBI. At \num{300000} satellites the per-satellite ceiling for these broadband detections lies around 65--70\,dB($\mu$V/m), relaxing by $\sim$\SI{8}{dB} at \num{50000} and tightening by the same towards the million-satellite filings. Current spacecraft exceed this ceiling by $+20.4$\,dB at the \SI{7860}{MHz} third harmonic, just below the legacy X-band and the one bright detection whose flux is a clean, uncompressed measurement. The brighter detections exceed by more, but their fluxes are compression-limited lower bounds: $\geq+33.2$\,dB at the \SI{5240}{MHz} second harmonic and $\geq+45.4$\,dB at the \SI{2620}{MHz} intended downlink, the latter, as noted, not an emission-mask question. The narrowband \SI{2595}{MHz} LO fundamental exceeds its bandwidth-raised limit by $+8.3$\,dB; the \SI{10480}{MHz} fourth harmonic exceeds its limit by $+3.9$\,dB. The incidental platform UEMR, after the emission/radiation-bandwidth correction (Section~\ref{sec:bwcorr}), falls \emph{below} its VLBI-protective limit at \num{300000}, by $-3.3$ (\SI{2700}{MHz}), $-9.3$ (\SI{2656}{MHz}) and $-24.5$\,dB (\SI{8599}{MHz}), and so does not constrain geodetic VLBI. It remains bound by the single-dish criterion and by front-end saturation.

\begin{table}[t]
\caption{Per-satellite radiation/emission limits to protect VLBI as the constellation grows, and the measured exceedance. $E_{\rm meas}$ is the measured per-satellite field strength; $E_{\rm lim}^{N}$ is the RA.769-protective limit required at $N=$\,\num{50000}, \num{200000} and \num{300000} satellites; all in dB($\mu$V/m) at \SI{10}{m}, \SI{1}{MHz} reference bandwidth (the CISPR geometry above \SI{1}{GHz}). Exceed is the dB by which $E_{\rm meas}$ exceeds $E_{\rm lim}^{\rm 300k}$ (negative $=$ already below the limit). Exceed and the breach-onset population $N_{\rm breach}$ of Table~\ref{tab:key} are one statement read two ways, related by Exceed${}=10\log_{10}(\num{300000}/N_{\rm breach})$: it is negative precisely when breach onset lies \emph{beyond} the \num{300000}-satellite reference (\SI{2700}{MHz}: $N_{\rm breach}\approx\num{640000}$, hence $-3.3$\,dB, and a data loss of only $0.1\%$ at \num{300000}, well short of the $2\%$ that defines the 98th-percentile breach) and positive when it lies below it (\SI{2595}{MHz}: $N_{\rm breach}\approx\num{44000}$, $+8.3$\,dB). A detection can thus carry a finite sub-$2\%$ data loss at \num{300000} (Table~\ref{tab:key}) while sitting below its per-satellite limit there (negative Exceed); the two are consistent, not contradictory. The limit is RA.769-referenced and includes the emission/radiation-bandwidth correction of Section~\ref{sec:bwcorr}, which raises the threshold for the narrowband UEMR detections and leaves them below their VLBI-protective limit (negative Exceed). Field strengths use the CISPR measurement bandwidth (\SI{1}{MHz} above \SI{1}{GHz}), capped at the emission or radiation bandwidth for the narrowband detections so that the field reflects the detection's total power. For the spurious and intended emission detections the equivalent ITU-R SM e.i.r.p.\ spectral density is $E-\SI{174.8}{dB}$ for the detections that fill the \SI{1}{MHz} reference bandwidth; for the two narrowband ones it follows their own bandwidth, $E-\SI{114.8}{dB}-10\log_{10}(\Delta\nu_e/\SI{1}{Hz})$, that is $E-\SI{151.8}{dB}$ at \SI{2595}{MHz} (\SI{5}{kHz}) and $E-\SI{154.8}{dB}$ at \SI{5190}{MHz} (\SI{10}{kHz}). Class as in Table~\ref{tab:inputs}: UEMR $\to$ CISPR track, Spurious/Intended $\to$ ITU-R SM track. $^\dagger$The \SI{2620}{MHz} row is an intended, licensed in-band emission, listed for diagnosis only: within its authorised band an operator may transmit at its licensed power, so $E_{\rm lim}$ and Exceed there quantify the magnitude of the impact in that frequency and are not a compliance target. The ceiling binds the spurious-emission and UEMR detections. For the equivalent isotropically radiated power in dBW, as used at ITU-R, subtract \SI{114.8}{dB} from any field-strength entry ($\mathrm{e.i.r.p.} = E - \SI{114.8}{dB}$ at \SI{10}{m}); the factor itself is purely geometric, set by the \SI{10}{m} reference distance alone, and holds in any bandwidth, so the resulting e.i.r.p.\ carries the reference bandwidth of the entry it comes from: a power in \SI{1}{MHz} for the broadband detections, and in the emission or radiation bandwidth for the narrowband ones.}
\label{tab:cispr}
{\tablefont 
\begin{tabular}{@{\extracolsep{\fill}}rlrrrrr}
\toprule
Freq & Class & $E_{\rm meas}$ & $E_{\rm lim}^{50\rm k}$ & $E_{\rm lim}^{200\rm k}$ & $E_{\rm lim}^{300\rm k}$ & Exceed\\
(MHz) & & \multicolumn{4}{c}{dB($\mu$V/m)} & (dB)\\
\hline
2595 & Spurious & 83.7 & 83.1 & 77.1 & 75.3 & +8.3\\
2620$^\dagger$ & Intended & 110.7 & 73.1 & 67.1 & 65.3 & +45.4\\
2656 & UEMR & 66.4 & 83.4 & 77.4 & 75.6 & -9.3\\
2700 & UEMR & 65.3 & 76.3 & 70.3 & 68.5 & -3.3\\
4995 & UEMR & 50.8 & 80.0 & 73.9 & 72.2 & -21.4\\
5190 & Spurious & 61.2 & 87.3 & 81.2 & 79.5 & -18.3\\
5240 & Spurious & 99.9 & 74.5 & 68.4 & 66.7 & +33.2\\
7860 & Spurious & 88.1 & 75.4 & 69.4 & 67.7 & +20.4\\
8599 & UEMR & 63.5 & 95.8 & 89.8 & 88.0 & -24.5\\
8640 & UEMR & 63.1 & 83.7 & 77.7 & 75.9 & -12.8\\
10480 & Spurious & 73.5 & 77.4 & 71.4 & 69.6 & +3.9\\
\botrule
\end{tabular}
}
\end{table}

\begin{figure*}[t]
\centerline{\includegraphics[width=\textwidth]{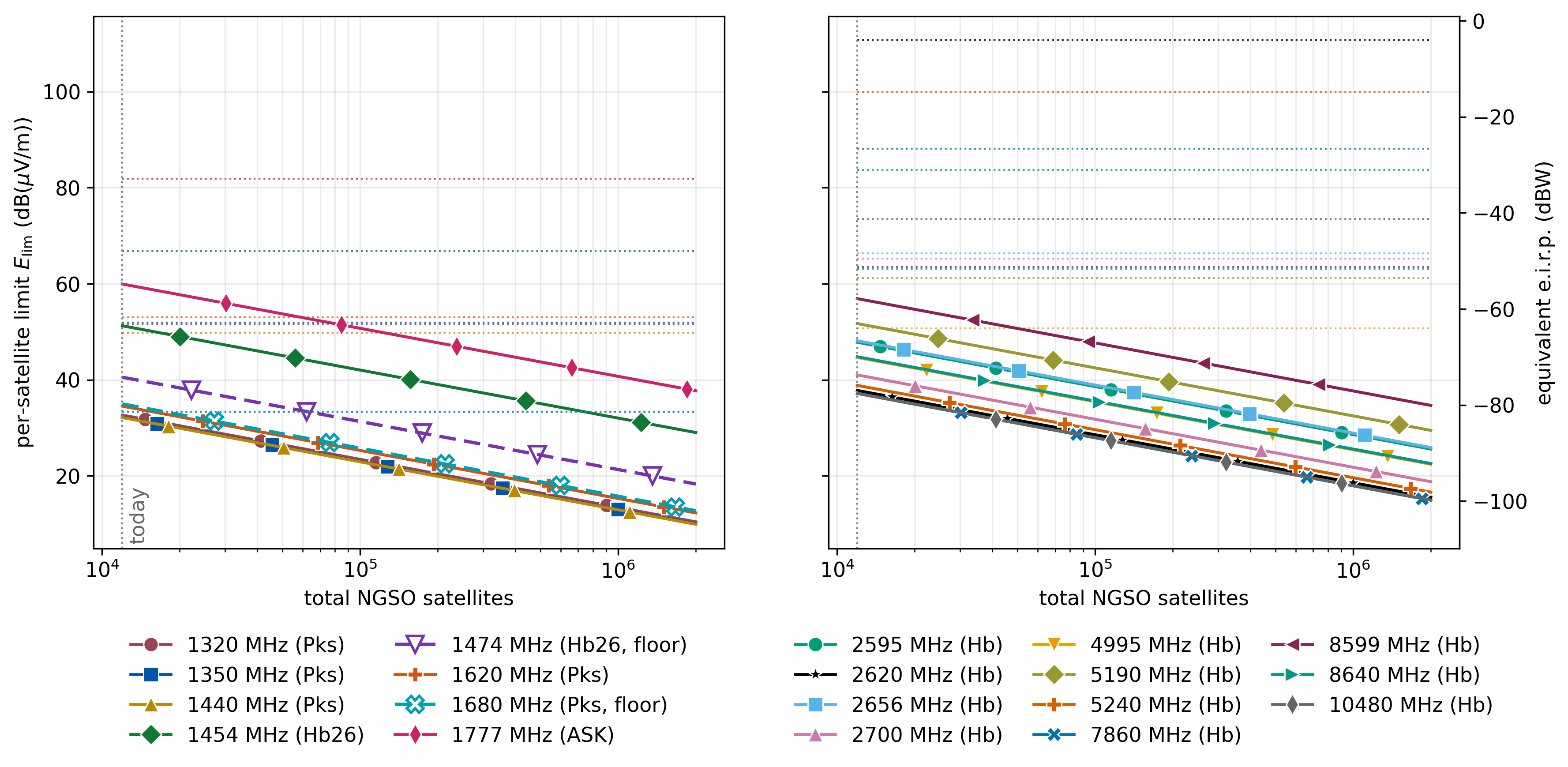}}
\caption{As Figure~\ref{fig:elim} but for \emph{single-dish radio astronomy}: the per-satellite field-strength limit set by the RA.769 continuum criterion versus total NGSO constellation size. \emph{Left:} the L-band detections, which lie below the AuScope band and are derived from the worst-affected facility of Table~\ref{tab:facilities-cont} (station in the legend); dashed curves with open markers mark the two floor-only detections, whose ceilings are provisional through their assumed active fractions. \emph{Right:} the detections above \SI{2.5}{GHz}, derived at the worst AuScope station as in Figure~\ref{fig:elim}; the right-hand axis gives the equivalent e.i.r.p. ($E-\SI{114.8}{dB}$), a power in each detection's reference bandwidth as in Table~\ref{tab:cispr}: \SI{1}{MHz} for the broadband detections, the radiation bandwidth (5--\SI{16}{kHz}) for the narrowband ones, so the L-band points are total power in the detection and not a per-MHz density. The continuum limits are some \SI{40}{dB} lower than the VLBI values, and every detection, at every frequency from L-band to X-band, lies above its single-dish limit at all constellation sizes (measured levels dotted).}
\label{fig:elim_cont}
\end{figure*}

The per-satellite limits above protect the interferometric observable of geodetic VLBI. The calibration tier of the geodetic chain, the single-dish-mode noise-diode (\ac{Tsys}) measurements on which the millimetre-level accuracy depends, is not covered by them. Under the conservative Report ITU-R RA.2507 treatment (the continuum methodology at the \SI{100}{s} calibration integration) its per-satellite ceilings are the continuum values of Table~\ref{tab:cispr_cont} raised by \SI{6.5}{dB}, of order 30--50\,dB($\mu$V/m) at \num{300000} satellites; the per-satellite limit is linear in the threshold, so that shift is exact. The permissive end of the range is set by the measurement physics: calibration against a strong continuum source is a differential measurement whose amplitude accuracy requires interference below the corresponding fraction of the system noise, and at \SI{1}{\percent} accuracy this coincides with the VLBI (1\%-of-noise) criterion itself. A dedicated derivation of the calibration threshold is left to future work; until then the RA.2507 treatment is the citable, conservative envelope. The single-dish radio astronomy service is protected by the \SI{40}{dB} more stringent continuum criterion, whose per-satellite limits are correspondingly lower (Table~\ref{tab:cispr_cont}; Figure~\ref{fig:elim_cont}), of order 23 to 43\,dB($\mu$V/m) at \num{300000} satellites. The same inversion applied at the worst-affected facility extends these ceilings to the L-band detections below AuScope's band: 18--46\,dB($\mu$V/m) at \num{300000} satellites, exceeded by the measured levels by 30--36\,dB at the calibrated detections. Radio astronomy in the L-band, home of the \SI{21}{cm} hydrogen line and the OH lines, thus requires per-satellite control of the same class as the higher bands. Every measured detection exceeds its single-dish limit today, including the narrowband UEMR that escapes the VLBI criterion: by $+75.0$\,dB at \SI{5240}{MHz}, $+64.9$ at \SI{7860}{MHz}, $+32.2$ at \SI{2656}{MHz}, $+38.2$ at \SI{2700}{MHz} and $+20.6$\,dB at the calibrated \SI{8599}{MHz} detection. The emission/radiation-bandwidth dilution that protects VLBI does not help single-dish radio astronomy, whose far higher sensitivity is breached even by kHz-wide radiation.

\begin{table}[t]
\caption{Per-satellite limits to protect single-dish radio astronomy, derived against the RA.769 continuum criterion exactly as Table~\ref{tab:cispr} is derived against the VLBI criterion. These limits are some \SI{40}{dB} lower than the VLBI values, and every detection, the narrowband UEMR included, exceeds them at the present fleet. Columns and units as in Table~\ref{tab:cispr}. The $^\dagger$ on the \SI{2620}{MHz} row carries the same meaning as in Table~\ref{tab:cispr}. The e.i.r.p. conversion of Table~\ref{tab:cispr} applies unchanged: subtract \SI{114.8}{dB} from any entry, which gives a power in that row's reference bandwidth, \SI{1}{MHz} for the broadband detections and the emission or radiation bandwidth for the narrowband ones. The L-band rows (\SIrange{1320}{1777}{MHz}) lie below the AuScope VGOS band and carry no VLBI tier; they are derived by the identical inversion from the worst-affected facility of Table~\ref{tab:facilities-cont} (chiefly Murriyang and ASKAP). The ceiling is independent of the measured level, which cancels in the inversion; the red rows are provisional through their assumed active fractions, not through the level.}
\label{tab:cispr_cont}
{\tablefont 
\begin{tabular}{@{\extracolsep{\fill}}rlrrrrr}
\toprule
Freq & Class & $E_{\rm meas}$ & $E_{\rm lim}^{50\rm k}$ & $E_{\rm lim}^{200\rm k}$ & $E_{\rm lim}^{300\rm k}$ & Exceed\\
(MHz) & & \multicolumn{4}{c}{dB($\mu$V/m)} & (dB)\\
\hline
1320 & UEMR & 51.6 & 26.5 & 20.4 & 18.7 & +32.9\\
1350 & UEMR & 52.0 & 26.0 & 20.0 & 18.2 & +33.7\\
1440 & UEMR & 49.8 & 25.9 & 19.9 & 18.2 & +31.6\\
1454 & UEMR & 66.8 & 45.0 & 39.0 & 37.3 & +29.6\\
1474 & UEMR & \textcolor{red}{33.4} & \textcolor{red}{34.3} & \textcolor{red}{28.3} & \textcolor{red}{26.6} & \textcolor{red}{+6.8}\\
1620 & UEMR & 53.1 & 28.3 & 22.3 & 20.5 & +32.5\\
1680 & UEMR & \textcolor{red}{33.4} & \textcolor{red}{28.8} & \textcolor{red}{22.8} & \textcolor{red}{21.0} & \textcolor{red}{+12.4}\\
1777 & UEMR & 81.8 & 53.7 & 47.7 & 46.0 & +35.9\\
2595 & Spurious & 83.7 & 41.6 & 35.6 & 33.8 & +49.8\\
2620$^\dagger$ & Intended & 110.7 & 31.6 & 25.6 & 23.8 & +86.9\\
2656 & UEMR & 66.4 & 41.9 & 35.9 & 34.1 & +32.2\\
2700 & UEMR & 65.3 & 34.8 & 28.8 & 27.0 & +38.2\\
4995 & UEMR & 50.8 & 38.5 & 32.4 & 30.7 & +20.1\\
5190 & Spurious & 61.2 & 45.5 & 39.5 & 37.7 & +23.4\\
5240 & Spurious & 99.9 & 32.6 & 26.6 & 24.9 & +75.0\\
7860 & Spurious & 88.1 & 30.9 & 24.9 & 23.2 & +64.9\\
8599 & UEMR & 63.5 & 50.7 & 44.7 & 42.9 & +20.6\\
8640 & UEMR & 63.1 & 38.6 & 32.6 & 30.8 & +32.3\\
10480 & Spurious & 73.5 & 31.0 & 25.0 & 23.2 & +50.3\\
\botrule
\end{tabular}
}
\end{table}

These limits are demanding but not without precedent in spacecraft engineering. A satellite is usually qualified for radiated emissions before flight: to be compatible with its launch vehicle it must meet radiated-emission masks set by \mbox{MIL-STD-461} (test method RE102), tailored per \mbox{MIL-STD-464} \citep{milstd461g, milstd464}, together with the launch provider's own payload limits \citep{spacex_rideshare}, expressly to protect the vehicle's GNSS, telemetry and flight-termination receivers. Nor is the practice confined to Western operators. China's national standard GJB 151B-2013 mirrors the MIL-STD-461 emission and susceptibility structure \citep{gjb151b}, \ac{JAXA} imposes its own \ac{EMC} design standard \citep{jaxa_jerg2241}, and European missions apply ECSS-E-ST-20-07C \citep{ecss_emc}; radiated-emission qualification against tailored masks is universal spacecraft engineering practice. 

\begin{figure}[t]
\centerline{\includegraphics[width=\columnwidth]{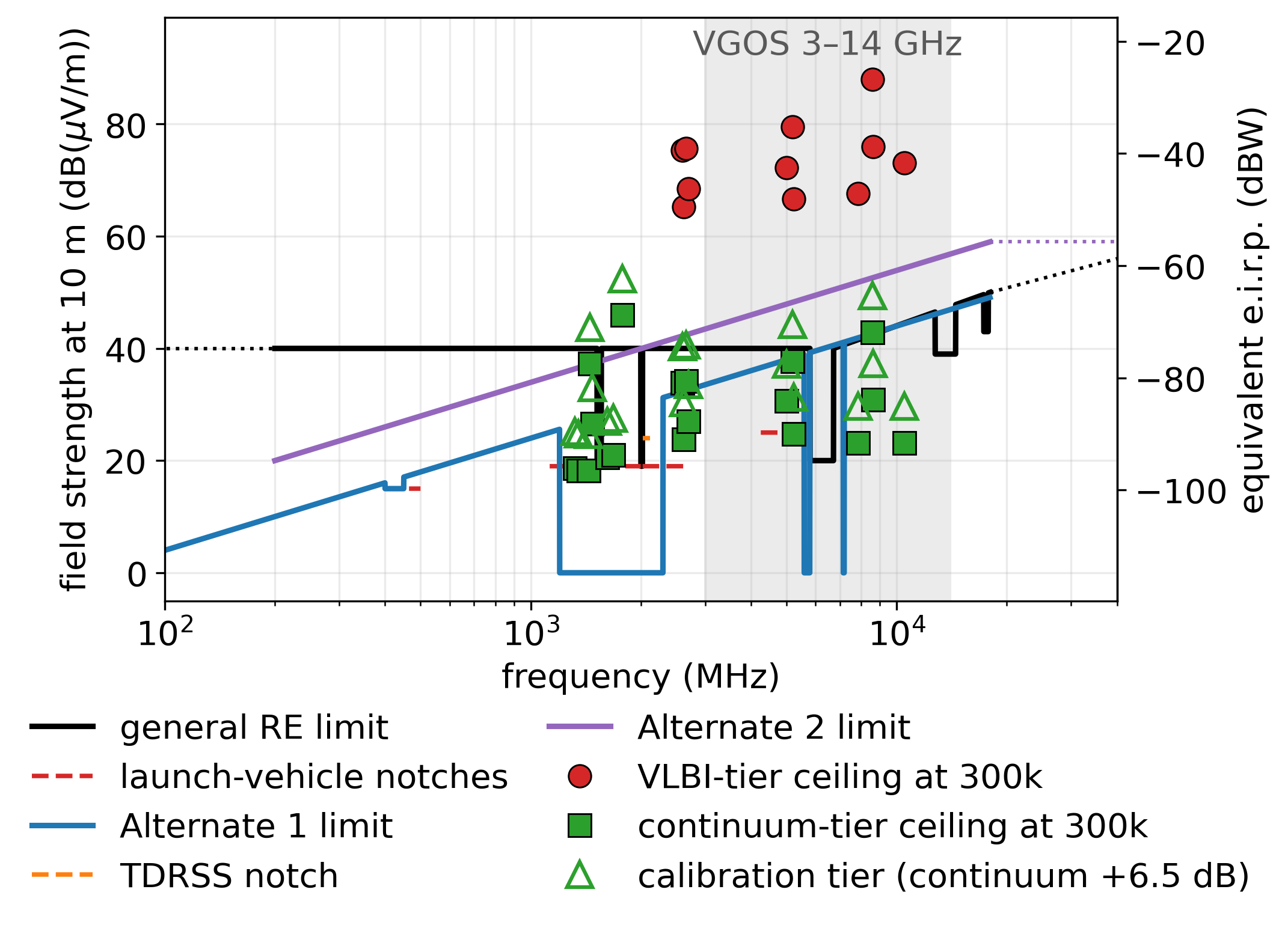}}
\caption{The spacecraft radiated-emissions masks of AIAA S-121A-2017 \citep{aiaa_s121a} compared with the per-satellite ceilings proposed here, all expressed at \SI{10}{m} in a \SI{1}{MHz} reference bandwidth (the standard follows MIL-STD-461G RE102, stated at \SI{1}{m}; far-field scaling applied; its resolution bandwidth above \SI{1}{GHz} equals ours, and the notches specified at narrower bandwidths are stricter still against broadband emission). Masks from the standard's Section~8.20: the general limit with its receive-band notches (Tables~8 and~10), the Alternate~1 low-emissions limit with its notches (Table~11), and the Alternate~2 limit for space science platforms, whose protected-band annex includes the \SIrange{2690}{2700}{} and \SIrange{4950}{5000}{MHz} passive bands. Circles and squares mark the proposed per-satellite ceilings at \num{300000} satellites under the VLBI and continuum criteria (Tables~\ref{tab:cispr} and~\ref{tab:cispr_cont}), and open triangles the geodetic calibration tier, the continuum ceilings raised by \SI{6.5}{dB} for the \SI{100}{s} calibration integration of Report ITU-R RA.2507 (Section~\ref{sec:limits}); the elevated values of the narrowband detections reflect the emission/radiation-bandwidth convention of Section~\ref{sec:bwcorr}. A spacecraft compliant with any of the three masks meets the VLBI-tier ceiling outright; the notch levels the industry already imposes on itself to protect its own receivers, 20--63\,dB($\mu$V/m) at \SI{1}{m}, or 0--43 in the \SI{10}{m} geometry, bracket the continuum-tier ceilings: per-satellite control at the levels proposed here is within existing aerospace EMC practice. The right-hand axis gives the equivalent isotropically radiated power producing the same field at \SI{10}{m} ($\mathrm{e.i.r.p.} = E - \SI{114.8}{dB}$). The continuum-tier points now extend to the L-band detections (worst-affected facility, as in Table~\ref{tab:cispr_cont}); their cluster at 18--21\,dB($\mu$V/m) coincides with the Alternate~1 mask level in the very band the standard's own \SIrange{1200}{2300}{MHz} receiver notch protects.}
\label{fig:aiaamasks}
\end{figure}

These masks shown in Figure~\ref{fig:aiaamasks} impose deep receive-band notches in the protected bands, of order 30--50\,dB($\mu$V/m) at \SI{1}{m}; the \ac{AIAA} masks of Figure~\ref{fig:aiaamasks} notch the bands of the space systems' own receivers, the \ac{GPS}, \ac{TDRSS}, and the launch vehicles', to 20--63\,dB($\mu$V/m) at \SI{1}{m}, equivalently 0--43\,dB($\mu$V/m) in the \SI{10}{m} geometry used here, and the standard's Alternate~2 limit extends the discipline to space science platforms, its protected-band annex naming the \SIrange{2690}{2700}{} and \SIrange{4950}{5000}{MHz} passive bands among those to be notched. These limits also live in the framework of the \ac{ISO}: ISO 14302 \citep{iso14302}, the space-systems EMC standard of ISO/TC~20/SC~14, sets out how equipment-level radiated-emission requirements are derived from system-level EMC requirements, and the general and low-emission limits of Figure~\ref{fig:aiaamasks} are currently under discussion in that committee. The far-field $1/r$ conversion is valid because we assume the radiation originates in the electrically small digital units and cabling rather than in the spacecraft structure as a coherent aperture; the far-field distance $2D^2/\lambda$ stays below \SI{10}{m} across \mbox{1--14\,GHz} even for the largest platforms. The SpaceX rideshare guide, for example, notches the \ac{GPS} band to $\sim$48\,dB($\mu$V/m) at \SI{1}{m}. These masks are expressed in the same field-strength units, and verified with the same instruments and anechoic chambers, as the per-satellite limits derived above. We understand from a satellite operator that a radiated field as low as 15\,dB($\mu$V/m) at \SI{10}{m} (35\,dB($\mu$V/m) at \SI{1}{m}) is achievable, within the notch depths routinely met on flight hardware. The per-satellite limits derived above, of order 23 to 70\,dB($\mu$V/m) at \SI{10}{m}, are no more stringent than that. The engineering capability and an established verification pathway therefore already exist. Verification across the full span is more laborious than a notch test, since measurement time scales with the frequency-by-aspect-angle matrix and the required sensitivity approaches chamber noise floors, but it is a scaling of established practice, not a new capability. What is new is extending that discipline to the radio-astronomy bands at the minimum, or across the radio spectrum ideally, and to the in-orbit operating state, since launch qualification tests the spacecraft on the ground rather than the constellation in service.

\subsection{Front-end saturation is a secondary constraint for VGOS apertures}
\label{sec:saturation}

For the \mbox{12m} VGOS apertures the strongest in-band emission we measure is the \SI{5240}{MHz} second harmonic of the \ac{DTD} downlink; the \SI{2620}{MHz} fundamental itself falls below the VGOS tuning range. Tracked through the ATCA main beam this detection reaches a compressed lower bound of $\geq$\SI{0.3}{MJy} over \SI{20}{MHz} (its brightest transits drive the calibrating back-end into compression in 21 of 36 measured DTD passes). It couples $\approx -83$\,dBm into a \mbox{12m} aperture on boresight, some \SI{43}{dB} below the \SI{-40}{dBm} linearity limit and far below the \SI{+10}{}--\SI{12}{dBm} damage level. The narrowband unintended-radiation detections couple far less despite their high spectral flux density: the front-end integrates total power, and their kHz bandwidths carry little of it (the \SI{2595}{MHz} fundamental reaches only $\approx -112$\,dBm). Front-end saturation is therefore \emph{not} the binding constraint for VGOS-class dishes at present emission levels; the dominant impact is the statistical sensitivity/correlation loss quantified in the \ac{EPFD} simulations above. For larger apertures the margin shrinks in proportion to collecting area, and is already exhausted in practice. The bright DTD carrier and its harmonics drive the \mbox{22m} signal chain used for the absolute-flux calibration into \emph{compression}, which is why those fluxes can only be quoted as lower bounds; the compression is in the digital back-end, not the antenna front-end.\footnote{The attribution to the digitiser rather than the receiver follows from the power budget. The brightest \SI{5240}{MHz} transits couple of order \SI{-77}{dBm} into the \mbox{22m} aperture, some \SI{40}{dB} below the input-referred \SI{1}{dB} compression point typical of a cryogenic low-noise amplifier, which is therefore comfortably linear; the compression also switches on and off within a single scan, tracking the detection rather than a thermal timescale. That same detection, however, carries power comparable to, and at the brightest transits above, the total system noise integrated over the \SI{4}{GHz} sampled band. Against the $\sim$\SI{50}{dB} usable dynamic range of \mbox{8-bit} sampling (with the sky noise set a few bits below full scale), its peaks then reach full scale and clip. A \mbox{16-bit} back-end ($\sim$\SI{98}{dB} range) would hold the detection within scale and recover the true, higher flux, so the levels we quote remain conservative lower bounds.} Saturation is thus a realised concern for the largest single dishes and for the brightest, closest transits.

Those couplings follow from the \emph{measured} fluxes, which for the bright DTD detections are compression-limited lower bounds and understate the beam-coupled case. SpaceX's \mbox{Gen2} DTD satellites carry a published peak e.i.r.p.\ of $+58$\,dBW (spectral density $-2.33$\,dB(W/Hz)) \citep{spacex_dtc_narrative}. They fly at a median \SI{358}{km} (over the \num{642} \mbox{STARLINK-11xxx} objects in our snapshot), within the \mbox{340--360\,km} shell the \mbox{FCC} authorised in March~2025, lowered from \mbox{525--535\,km} to reduce latency \citep{fcc_da25_197}. Should a beam dwell on a station, that peak e.i.r.p.\ puts an in-beam \ac{PFD} at the \SI{850}{km} transit range of $\approx -71.9$\,dB(W/m$^2$/MHz), some \SI{62}{dB} above the apparent measured level. It couples $\approx \SI{-14}{dBm}$ into a \SI{12}{m} aperture, \SI{26}{dB} \emph{past} its \SI{-40}{dBm} linearity limit (though still $>$\SI{20}{dB} below the \SI{+10}{}--\SI{12}{dBm} damage level), and $\approx \SI{-8}{dBm}$ into the \SI{22}{m} calibrator, whose consequent compression is exactly why these fluxes can only be quoted as lower bounds. The \SI{5240}{MHz} second harmonic, only $\sim$\SI{11}{dB} below the fundamental as measured and itself compressing, reaches an in-beam $\approx -83$\,dB(W/m$^2$/MHz) on the same basis. The operator's stated maximum in-service-area \ac{PFD} of $-80$\,dB(W/m$^2$/MHz), about \SI{16}{dB} below the antenna peak, is the typical service level rather than this worst case. This in-beam coupling is a single-satellite worst case, for a station inside a served cell with a beam centred on it. It is distinct from the aggregate \ac{EPFD} of Section~\ref{sec:results}, which uses the measured, compression-limited per-satellite levels. The implication is that front-end saturation, comfortable for a \SI{12}{m} VGOS dish at the apparent levels, vanishes once the true in-beam level is restored. Conversely, because saturation at the measured levels is confined to this beam-on-station geometry, operator-side exclusion zones that keep serving beams pointed away from radio astronomy and geodetic stations are likely sufficient to avoid it altogether; first trials of exactly this mitigation at the ATCA and Murriyang remove every bright near-boresight transit, and the receiver-compression episodes with them \citep{indermuehle2026odsba}. We phrase this as likely rather than certain: the margin rests on the beam counts, bandwidths and per-satellite levels of the currently observed DTD payloads, and a substantially expanded payload would warrant repeating the assessment.

\subsection{Apportionment among operators}

The per-satellite limit falls as the reciprocal of the radiating population (Figure~\ref{fig:elim}), because the aggregate is linear in $N$ while the protected total is fixed. Read naively, this appears to punish the small operator: an entity flying a few hundred satellites is asked to meet a ceiling set by a sky of order a million, most of which it did not launch. EMC standards already treat limits as a per-\emph{unit} emission mask rather than a per-operator budget; CISPR and \mbox{MIL-STD-461} impose a field strength in dB($\mu$V/m) on every emitting device at a reference distance. A uniform per-unit mask \emph{is} an apportionment, the equal-share-per-satellite one. If every satellite meets $E_{\rm lim}(N)$, an operator of $N_{\rm op}$ satellites contributes exactly $N_{\rm op}/N$ of the aggregate and the total equals the protected budget by construction; a per-constellation ceiling is simply $N_{\rm op}\,E_{\rm lim}$. The spread among the plotted ceilings at fixed $N$ originates from the fact that each point answers the question for the subfleet that radiates that detection, and so carries that detection's radiating share (source subpopulation times active fraction, spanning five to $\sim$\num{10000} present-day radiators), the transit-dominated tail statistics of the smallest subfleets, its emission/radiation bandwidth, and the frequency-dependent threshold; none of it derives from the measured levels, which cancel in the inversion. A standards body writing a single limit line would possibly base the calculation on a mixed fleet of satellites and assume that all of them, despite the different designs, would equally radiate at a certain frequency. That assumption yields a limit at or below the lower envelope of the plotted points: each point sits above it by its subfleet's radiating share, from under \SI{1}{dB} for the near-unity-share detections to $\sim$\SI{13}{dB} for the \ac{DTD} harmonics. Figure~\ref{fig:aiaamasks} sets the proposed ceilings against the spacecraft radiated-emissions masks of \ac{AIAA} S-121A-2017: the aerospace industry's own standard already contains masks of the class required here.

Among the workable architectures this is the least burdensome for the small operator, though not without cost. The recurring, per-unit cost of compliance (filtering, shielding, component selection) scales with fleet size: a few hundred clean satellites rather than a million. The fixed cost of a compliant design and its qualification campaign, however, is amortised over far fewer units, and weighs relatively more on an operator building from off-the-shelf components. That regressivity is not specific to this proposal; it is the cost structure of every type-approval regime, from CISPR conformity for consumer electronics to the MIL-STD-461 qualification the spacecraft already undergoes (Section~\ref{sec:limits}). The market such regimes create softens the impact: suppliers pushing towards compliant components make compliance purchasable off the shelf for everyone else. At today's population the mask is loose and small fleets sit well within it. The tightening over time is driven by, and falls hardest on, the operators filing for the hundreds of thousands to millions of satellites that set the aggregate. The alternative a small operator might prefer, an equal budget \emph{per operator}, is in fact worse for it: its share would shrink as $1/M$ with every new entrant. A uniform per-unit mask is moreover the only non-gameable form; a looser limit for ``small'' operators invites a large constellation to refile as many small ones. We therefore propose the per-satellite mask, set against the current and ITU-notified population and revised on the regulatory review cycle, as the primary instrument, complemented by a per-system aggregate cap so that no single operator consumes a disproportionate slice of the budget. This mirrors the single-entry and aggregate architecture the ITU already operates for non-geostationary equivalent power flux-density (Radio Regulations Article~22 and Resolution~76 \citep{ituRR_art22, ituRes76}), and the per-system aggregate is computable with the same \ac{EPFD} machinery used here. That architecture is itself in flux. The \ac{FCC} has resolved to replace the Article~22 \ac{EPFD} limits domestically as ``needlessly prescriptive, outdated, and overprotective'' \citep{fcc2026modernization}, and SpaceX's Gen3 application seeks waivers of both the FCC and ITU \ac{EPFD} compliance demonstrations \citep{spacex2026gen3}. The only satellite-aggregate instrument in the Radio Regulations is thus being eroded just as the aggregate problem reaches the passive services, which argues for a RAS-specific aggregate ceiling rather than reliance on the GSO-protection machinery.

\section{Global DTD deployment and the harmonic challenge}
\label{sec:global}

The threat the harmonics pose to geodetic VLBI is sharpened by two facts that compound each other: geodetic VLBI must observe the \emph{same} frequencies at every station in the network simultaneously (Section~\ref{sec:vgos}), and \ac{DTD} downlinks are being authorised across a wide and growing set of bands worldwide. A geodetic station cannot site-select or frequency-plan its way around an interferer, because the correlation that produces the geodetic observable requires all stations to record identical sky frequencies, so a band lost at one station is lost across every baseline to it. A DTD downlink deployed in \emph{any} country therefore places its harmonics into the geodetic bands observed \emph{everywhere}.

The first Australian DTD deployment is at \mbox{2620--2630\,MHz}, with \mbox{870--890} and \mbox{2670--2690\,MHz} to follow. New Zealand operates at \mbox{1875--1880} and \mbox{2620--2635\,MHz}, and the United States operates in \mbox{1990--1995\,MHz}, with \mbox{1990--2000} and \mbox{2180--2200\,MHz} authorised. Canada proposes a comparable set, across \mbox{617--652}, \mbox{728--756}, \mbox{869--894}, \mbox{1930--1995} and \mbox{2110--2180\,MHz} \citep{ised_smse006_24}. Europe is following. In the United Kingdom, \ac{DTD} operation is already authorised: licence-exemption regulations in force since February 2026 permit DTD handsets in all frequency-division-duplex and supplementary-downlink mobile bands below \SI{3}{GHz} (satellite downlinks at \mbox{758--788}, \mbox{791--821}, \mbox{925--960}, \mbox{1452--1492}, \mbox{1805--1880}, \mbox{2110--2170} and \mbox{2620--2690\,MHz}), with the first commercial services starting in 2026 \citep{ofcom_d2d_2026}. In the European Union, the Commission has mandated \ac{CEPT} to develop harmonised technical conditions for DTD in the same harmonised mobile bands, with final results due in late 2026 \citep{ec_d2d_mandate}, and separately proposes re-authorising the \SI{2}{GHz} \ac{MSS} band (downlink \mbox{2170--2200\,MHz}) beyond 2027 \citep{ec_ip26_1170}. The supplementary-downlink band at \mbox{1452--1492\,MHz} adds a path not present in the sets above: its second harmonic, \mbox{2904--2984\,MHz}, falls inside VGOS~A, and its fourth clips the upper edge of VGOS~B.
These national bands are subsets of the much broader space-to-Earth (downlink) sets already authorised for the two largest systems: SpaceX may transmit, outside the United States, in \mbox{1475--1518}, \mbox{1805--1880}, \mbox{1930--2000}, \mbox{2110--2180}, \mbox{2180--2200}, \mbox{2345--2360} and \mbox{2620--2690\,MHz} \citep{fcc_da24_1193}, and AST SpaceMobile in \mbox{617--652}, \mbox{728--821}, \mbox{852--894} and \mbox{902--960\,MHz} \citep{fcc_da26_391}; the FCC has no mandate to authorise DTD operation in other countries. We consider downlink (space-to-Earth) bands only, since it is the satellite that radiates towards the ground; the uplink is transmitted by handsets.

\begin{figure*}[t]
\centerline{\includegraphics[width=\textwidth]{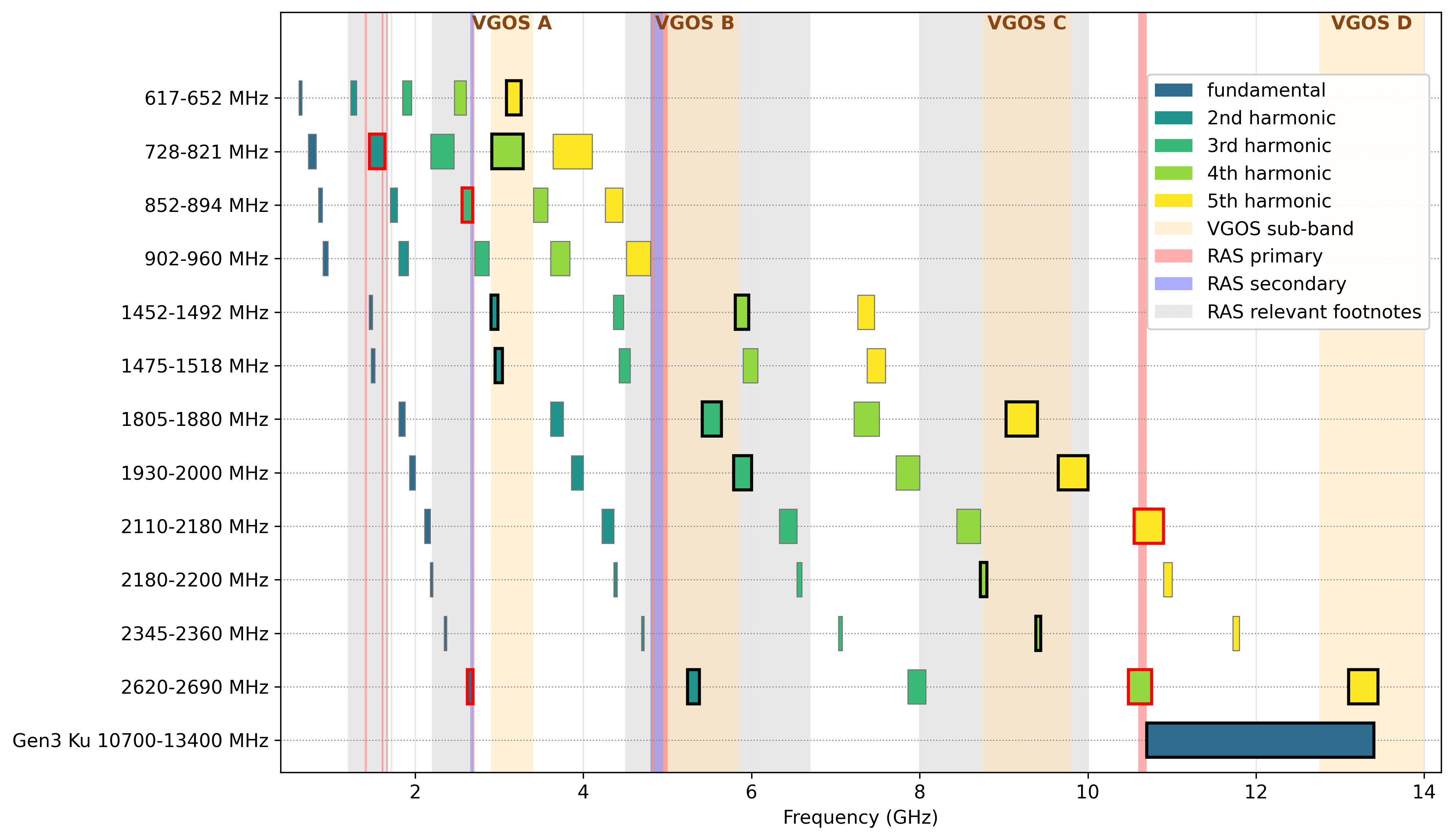}}
\caption{The fundamental and harmonics (2nd to 5th order) of the global DTD space-to-Earth (downlink) bands, against the four VGOS observing sub-bands (tan) and the Australian RAS bands of Table~\ref{tab:ras-alloc}: red for primary and blue for secondary RAS allocations, grey for the recognised but unallocated (footnoted) bands. Each row is a downlink band: the national deployments, the broader FCC-authorised sets for SpaceX \citep{fcc_da24_1193} and AST \citep{fcc_da26_391}, and, in the bottom row, the Ku-band downlink of the filed SpaceX Gen3 system \citep{spacex2026gen3}, whose \emph{fundamental} (not a harmonic) enters VGOS~D directly and abuts the \mbox{10.68--10.7\,GHz} band in which RR No.~5.340 prohibits all emissions. Coloured bars mark the harmonic orders; a black outline marks an emission that falls within a VGOS sub-band and a red outline one that falls within a RAS-allocated band. Every VGOS sub-band is reached by at least one DTD harmonic of order five or below, and several harmonics also land in the narrow RAS allocations (for example the \mbox{10.6--10.7\,GHz} primary band). Because geodetic VLBI observes these same frequencies at every station worldwide, a downlink deployed in any country places its harmonics into the geodetic bands observed everywhere.}
\label{fig:dtdharm}
\end{figure*}

Figure~\ref{fig:dtdharm} maps the harmonics of these downlink bands, to the fifth order, against the four VGOS sub-bands \citep{IVS-Res-2026-01}. \emph{Every} VGOS sub-band is reached by at least one DTD harmonic of order five or below:
\begin{itemize}
    \item \textbf{VGOS A} (\mbox{2.90--3.40\,GHz}): the 5th harmonic of \mbox{617--652}, the 4th of \mbox{728--821}, and the 2nd of \mbox{1452--1492} and \mbox{1475--1518\,MHz};
  \item \textbf{VGOS B} (\mbox{4.80--5.85\,GHz}): the 3rd harmonic of \mbox{1805--1880} and \mbox{1930--2000}, the 2nd of \mbox{2620--2690\,MHz} (the calibrated \SI{5240}{MHz} detection of this paper), and the 4th of \mbox{1452--1492\,MHz} at the sub-band's upper edge;
  \item \textbf{VGOS C} (\mbox{8.75--9.80\,GHz}): the 5th harmonic of \mbox{1805--1880} and \mbox{1930--2000}, and the 4th of \mbox{2180--2200} and \mbox{2345--2360\,MHz};
  \item \textbf{VGOS D} (\mbox{12.75--14.00\,GHz}): the 5th harmonic of \mbox{2620--2690\,MHz}, and, directly, the filed Gen3 Ku downlink (\mbox{10.7--13.4\,GHz}; discussed below).
\end{itemize}
The legacy X-band (\mbox{8.2--8.95\,GHz}), which connects VGOS to more than four decades of historical VLBI results, is affected by the 4th harmonic of the \mbox{2110--2200\,MHz} downlinks (\mbox{8440--8800\,MHz}); the 3rd harmonic of the \mbox{2620--2690\,MHz} downlink, the \SI{7860}{MHz} detection of this paper, lands just below it, in the VGOS receiver passband between sub-bands B and C.

The harmonic orders swept in Figure~\ref{fig:dtdharm} correspond to the scope of the Radio Regulations. RR Appendix~3 applies its spurious-domain limits to ``all emissions, including harmonic emissions'' with no cutoff in harmonic order (\S11), and sets the frequency range of measurement at \SI{9}{kHz} to \SI{110}{GHz}, or the second harmonic if higher (\S6). Recommendation ITU-R SM.329 \citep{ituSM329} narrows this for practical verification: for fundamentals between \SI{600}{MHz} and \SI{5.2}{GHz}, its Table~1 recommends measurement up to the \emph{fifth} harmonic. For the \mbox{2620--2690\,MHz} DTD downlink that scope runs to \mbox{13.10--13.45\,GHz}, so every harmonic we modelled, the 5th-order entry into VGOS~D included, lies within the range ITU-R itself recommends verifying. The applicable Appendix~3 limit for space stations, $43+10\log P$ or \SI{60}{dBc}, whichever is less stringent, in a \SI{4}{kHz} reference bandwidth, attaches to the station rather than to any band: it covers the space stations of every space service, the fixed- and mobile-satellite services included, at any spurious-domain frequency. The only outright exemption is for deep-space stations of the space research service (Appendix~3, Table~I, Note~17), which no NGSO constellation can claim. How loose that mask is follows from the units of Section~\ref{sec:limits}. Under the measured near-isotropic spurious radiation (Section~\ref{sec:vgos-methods-mc}), the allowance corresponds to $\approx$96--102\,dB($\mu$V/m) at \SI{10}{m} for transmitter powers up to a few hundred watts: some 26--37\,dB above the VLBI-protective per-satellite ceiling at \num{300000} satellites, and 58--79\,dB above the single-dish ceilings of the spurious-emission detections (Table~\ref{tab:cispr_cont}). The measured harmonics sit \emph{below} that allowance, so the constellation can be fully compliant with Appendix~3 while exceeding the RAS protection levels by the $+20$ to $+33$\,dB of Table~\ref{tab:cispr}: the mask carries no population term, and a single compliant emitter already approaches the aggregate budget. SM.329 (\S4.2) describes the space-service values as ``currently shown as design limits'', verified at design rather than in orbit, and Appendix~3 (\S4) reserves more stringent levels for the protection of passive services, subject to agreement at a world radiocommunication conference, the treaty pathway through which any per-satellite ceiling of the kind derived in Section~\ref{sec:limits} would be set. We can put this into perspective against terrestrial transmitters: for most digital terrestrial transmitters above \SI{1}{GHz} the Category~B limit of the same Recommendation (\S4.3, Table~3) is \SI{-30}{dBm} in a \SI{1}{MHz} reference bandwidth, stated as power supplied to the antenna transmission line, so in radiated terms a terrestrial antenna adds its gain on top of it. Placed in those units, the VLBI-protective ceiling for the broadband harmonics is $-12$ to $-18$\,dBm/MHz at \num{300000} satellites and $-17$ to $-23$\,dBm/MHz at a million: protecting geodetic VLBI asks of a satellite no deeper suppression than a terrestrial transmitter is already required to achieve. The single-dish continuum ceiling is the demanding one, $-60$ to $-62$\,dBm/MHz at \num{300000} and $-65$ to $-67$ at a million, some 30 to \SI{37}{dB} below the terrestrial limit. That is a genuine design challenge, but not one without precedent: it lies inside the $-85$ to $-42$\,dBm/MHz spanned by the receive-band notches that spacecraft manufacturers already can impose on themselves to protect their own receivers (Figure~\ref{fig:aiaamasks}).

We further note that harmonics are no longer the only path into a VGOS band. SpaceX's Gen3 application \citep{spacex2026gen3} requests \emph{intended} Ku-band downlink across \mbox{10.7--13.4\,GHz} from up to \num{100000} satellites. This places co-frequency intended emission directly inside VGOS~D (\mbox{12.75--13.4\,GHz}, the lower half of the sub-band), with its lower edge flush against the \mbox{10.68--10.7\,GHz} RAS band in which RR No.~5.340 prohibits all emissions\footnote{ECC Report 271 \citep{ecc271} concluded that already the Starlink Gen1 constellation was not compliant with RAS thresholds in \mbox{10.68--10.7\,GHz} unless the lowest carrier \mbox{10.7--10.95\,GHz} remained unused. An update of the report is due and comes to the same conclusion for Starlink Gen2.}. At the filing's own constant ground-level \ac{PFD} of $-116.1$\,dB(W\,m$^{-2}$\,MHz$^{-1}$), a single compliant satellite serving the telescope's cell delivers a band-integrated $\approx-88$\,dB(W\,m$^{-2}$) within the VGOS~D overlap alone, at the $-88.7$\,dB(W\,m$^{-2}$) front-end linearity bound of Section~\ref{sec:saturation}. Over the full \mbox{10.7--13.4\,GHz} occupancy of the \mbox{3--14\,GHz} feed passband it delivers $\approx-82$\,dB(W\,m$^{-2}$), some \SI{7}{dB} above that bound: saturation from a single serving beam, before any aggregate is counted. The filing names boresight avoidance as the mitigation and does not mention geodetic VLBI, nor does it acknowledge UEMR.

No VGOS sub-band can be sacrificed to escape the problem, since each is hit independently, and because the spanned bandwidth between the lowest and highest sub-band sets the geodetic precision (Section~\ref{sec:vgos}); losing any of them degrades the product.

This is a structural consequence of the deployment. As more administrations license more of the \mbox{0.6--2.7\,GHz} mobile spectrum for DTD, the integer multiples of those downlinks sweep across the entire \mbox{3--14\,GHz} VGOS span, and the global-frequency requirement of geodetic VLBI means the harmonics are seen at every station. The per-satellite limits of Section~\ref{sec:limits} must therefore hold for the worldwide aggregate of every DTD system, in every authorised band, simultaneously. As of our 2026 catalogue, roughly \num{640} Starlink \ac{DTD} satellites (the STARLINK-\num{11}xxx series) are already on orbit, alongside a handful of AST SpaceMobile spacecraft. SpaceX's Gen2 filing is for some \num{30000} satellites, a large and growing fraction \ac{DTD} capable, and its 2026 Gen3 application adds \num{100000} more \citep{spacex2026gen3}.

\section{Orbital data centres: a UEMR-dominated megaconstellation}
\label{sec:odc}

A qualitatively different deployment is also on file. In January 2026 SpaceX requested authorisation for up to one million ``AI1'' ODC satellites \citep{fcc_spacex_odc}: solar-powered compute platforms rather than communications relays, each a \mbox{$\sim$70\,m}-span spacecraft sustaining of order \SI{120}{kW} of processing. The constellation is anomalous in two ways that together reshape the interference picture. It flies in a tightly constrained dawn-dusk Sun-synchronous orbit. And it carries no radio communications payload other than relatively low data rate \ac{TTC} downlinks, whose details are unknown and could not be modelled, though they would be significant at the proposed scale. We therefore modelled only the unintended platform radiation (UEMR) of its digital hardware, assuming radiation signatures similar to those we found on other satellites. The absence of a broadband radio payload is credible because the broadband egress is optical: the filing states the system will primarily rely on optical inter-satellite links, both among the ODC platforms and to the Gen1 and Gen2 Starlink constellations \citep{fcc_spacex_odc}, so user traffic enters the radio domain only at the Starlink relays' existing Ku/Ka-band ground segment. Optical inter-satellite links are operationally mature in the Starlink fleet, with more than \num{9000} laser terminals at \SI{100}{Gbit/s} each, a fleet throughput above \SI{42}{PB} per day over link distances up to \SI{5400}{km} at \SI{99}{\percent} link uptime \citep{brashears2024}. Laser links radiate nothing in the radio spectrum, so the platform's radio signature is the low-rate TT\&C plus its \ac{UEMR}, the case modelled here; the corollary is that ODC traffic concentrates additional transmit activity onto the existing Starlink downlink bands used for communication with gateway stations.

While not relevant for the frequencies discussed in this paper, it should be mentioned that the increased traffic expected for both the Starlink network and the data centres will require more bandwidth for Starlink's gateway connections. SpaceX has requested substantial spectrum above \SI{71}{GHz}. The FCC granted access to \mbox{71--76\,GHz} (space-to-Earth) and \mbox{81--86\,GHz} (Earth-to-space) for Starlink's \num{7500} Gen2 satellites \citep{fcc_da24_222}. In addition, SpaceX can use \mbox{92--94}, \mbox{94.1--95}, \mbox{95--100}, \mbox{102--109.5} and \mbox{111.8--114.25\,GHz} for Earth-to-space gateway communication, but only on a \emph{non-protection, non-interference} basis \citep{fcc_da26_36}. These activities may affect millimetre and sub-millimetre radio astronomy even where the gateway stations are not in the immediate vicinity of observatories. Compatibility studies are under way within ITU-R but have not yet been published.

It should also be noted that the one million ODC satellites could have a severe impact on optical astronomy, too \citep[see e.g.][]{hainaut2026}.

\subsection{Deriving the constellation geometry}
\label{sec:odc-geom}

\begin{figure*}[t]
\centerline{\includegraphics[width=\textwidth]{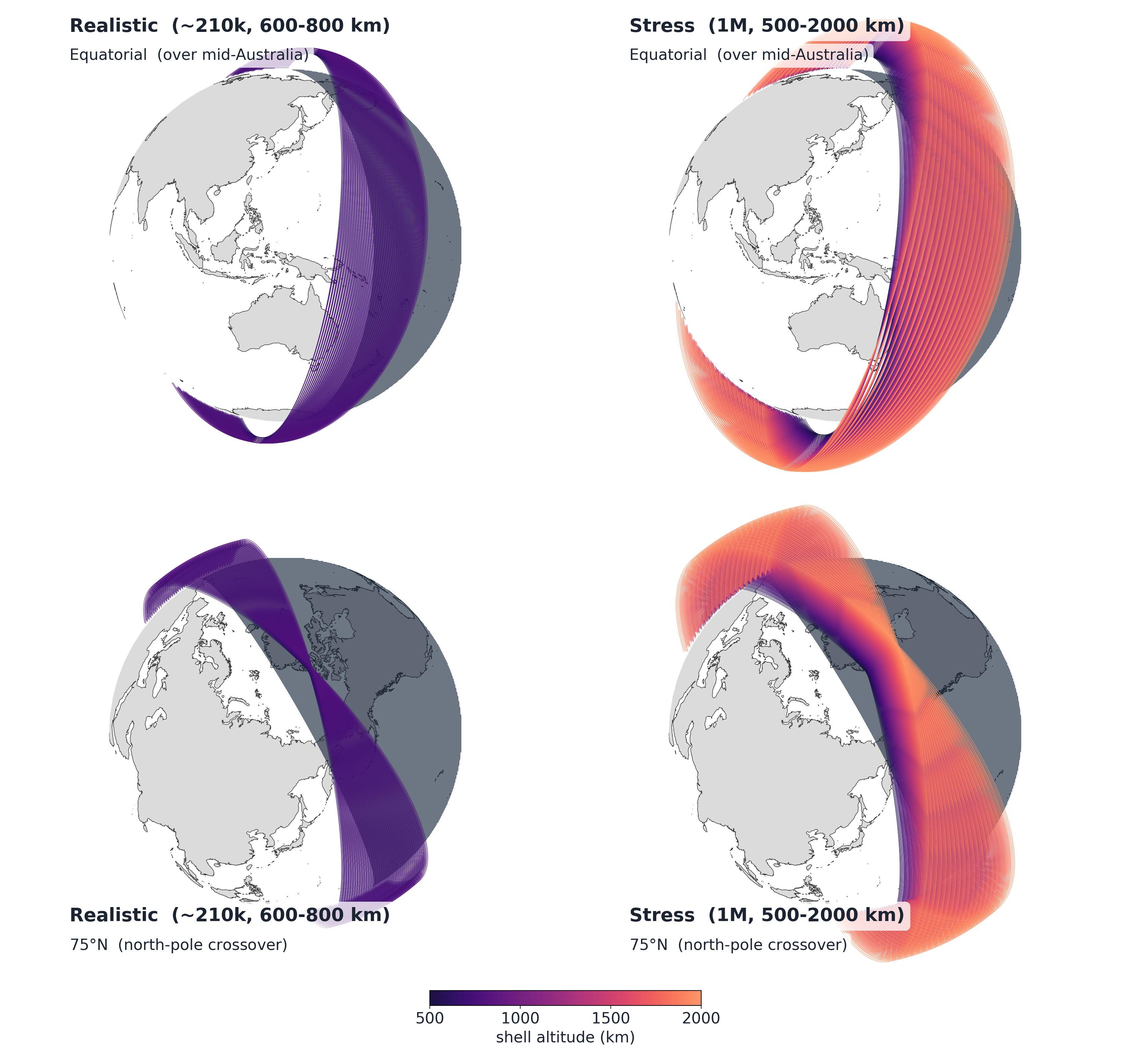}}
\caption{Snapshot of the modelled ODC constellation in dawn-dusk Sun-synchronous orbit, for the realistic (left, $\sim$\num{210000} satellites, \mbox{600--800\,km}) and stress (right, \num{1000000} satellites, \mbox{500--2000\,km}) presets, viewed over mid-Australia from the equatorial plane (top) and from \ang{75}N (bottom) to show the dawn and dusk streams crossing over the pole. Orbital tracks are coloured by shell altitude and the shaded hemisphere is the night side. The planes ride the day-night terminator and cluster near two meridians, so a ground station is swept at roughly fixed local times. Rendered tracks are a representative subsample of each fleet.}
\label{fig:odcsnap}
\end{figure*}

The orbital parameters follow from the platforms' need for continuous solar power and radiative cooling. SpaceX's filed envelope spans \mbox{500--2000\,km} altitude and inclinations from \ang{30} to Sun-synchronous. The operational design it has described, \mbox{600--800\,km} in a dawn-dusk Sun-synchronous orbit, is the configuration that keeps the solar panels in sunlight for \SIrange{90}{93}{\percent} of each orbit and the radiators in a stable thermal environment. We adopt that physics to fix the geometry. The Sun-synchronous condition, for which the $J_2$ nodal regression matches Earth's mean motion about the Sun (\ang{0.9856} per day), sets the inclination as a function of altitude: \ang{97.8} at \SI{600}{km}, \ang{98.6} at \SI{800}{km}, rising to \ang{104.9} at \SI{2000}{km}. All are near-polar and retrograde, so the shells cross every latitude and no ground station escapes them by siting. The dawn-dusk geometry fixes each orbital plane near the day-night terminator, so the ascending nodes cluster in a narrow band of right ascension rather than spreading uniformly as a communications shell would. The satellites concentrate near the dawn and dusk meridians and sweep a given site at roughly fixed local times, twice per day (Figure~\ref{fig:odcsnap}).

A million satellites cannot occupy a single altitude shell, so we stack shells at \SI{50}{km} intervals and cap each shell's occupancy by a collision-avoidance argument. Treating a shell as a spherical surface of area $4\pi r^2$ and requiring a minimum neighbour separation $s$ gives a capacity of order $4\pi r^2/s^2$. For the dawn-dusk case the planes are confined to a narrow ($\pm\ang{15}$) band of right ascension, so the usable area is two pole-to-pole lunes, about one sixth of the sphere. At a conservative $s=\SI{50}{km}$ (current Starlink shells are an order of magnitude sparser) one dawn-dusk shell holds $\sim$\num{41000} satellites. On this basis we model two presets: a \emph{realistic} deployment of five shells filling the \mbox{600--800\,km} design band to capacity at $\sim$\num{210000} satellites, and a \emph{stress} case of 31 shells spanning the full filed envelope, \mbox{500--2000\,km}, packed to the filed ceiling of \num{1000000}. The stress case fills the clumped packing capacity to roughly two thirds. A million \mbox{70\,m} platforms in dawn-dusk Sun-synchronous orbit therefore sit close to the geometric limit of what the orbital volume safely holds, and reaching the filed number requires abandoning the compact design band for the entire \mbox{500--2000\,km} envelope. For each shell we synthesise two-line elements (Sun-synchronous inclination, terminator-clustered nodes, Walker phasing within the band) and propagate them through the same SGP4 and Monte-Carlo \ac{EPFD} machinery used for the measured fleet (Section~\ref{sec:methods}).

\subsection{Emission/radiation model and projected impact}
\label{sec:odc-impact}

We assume that because an ODC carries no significant radio communications payload, its emission is predominantly UEMR: the broadband and clock-harmonic noise of dense digital hardware, with neither an intended downlink nor the downlink-derived harmonics that drive the geodetic threat of Sections~\ref{sec:limits} and~\ref{sec:global}. We anchor each satellite to the v2-Mini UEMR floor measured by SNIFFLES-I (of order $-119$\,dB(W/Hz) spectral e.i.r.p.) as a baseline parameter. The ODC's own UEMR is unmeasured, and \SI{120}{kW} of compute could plausibly exceed it, so we report the impact as a function of this level. The harmonic-order structure of the measured \ac{UEMR} makes this expectation plausible. The low-frequency detections are high-order harmonics of slow clocks: the \ac{LOFAR} detections at \SI{125}{}, \SI{150}{} and \SI{175}{MHz} are consistent with the 5th to 7th harmonics of a \SI{25}{MHz} clock \citep{divruno2023}. The GHz detections measured by SNIFFLES-I are instead consistent with fundamentals and low-order harmonics of GHz-class clocks (among the v2-Mini detections, \SI{2700}{MHz} is twice \SI{1350}{MHz} and \SI{8640}{MHz} is six times \SI{1440}{MHz}). AI compute hardware clocks its memory and serial interconnects at fundamentals of order \SIrange{1}{16}{GHz}, placing fundamentals and low-order harmonics, which carry far more power than high orders, directly across the \SIrange{1}{14}{GHz} range assessed here. The geodetic outcome then hinges on a second unmeasured property, the emission/radiation bandwidth, through the correction of Section~\ref{sec:bwcorr}. If the platform radiation is narrowband, like the measured v2-Mini UEMR forest, it is diluted by $\sim$\SI{38}{dB} against the \SI{32}{MHz} VGOS channel and geodetic VLBI is spared. If it is instead broadband, as \SI{120}{kW} of digital switching might well produce, it fills the channel and is not diluted. We therefore bracket the ODC between these two cases.

Running the aggregate \ac{EPFD} against both criteria, anchored to the floor level above, the single-dish continuum criterion is breached in either bandwidth case and at both fleet sizes. Across the Australian RAS telescopes (Section~\ref{sec:results}) the realistic \num{210000}-satellite deployment loses \mbox{40--60\%} of the sky, and the stress \num{1000000}-satellite case \mbox{65--80\%}, worst at Ceduna and Murriyang, driven by incidental platform noise alone. In the broadband case the million-satellite ODC breaches the VGOS array as well, most at the lowest frequencies: at Hobart the VLBI data loss reaches \SI{22}{\percent} at \SI{3.3}{GHz}, \SI{16}{\percent} at \SI{5.2}{GHz} and \SI{1}{\percent} at \SI{8.6}{GHz}, with Katherine and Yarragadee a few points lower. The realistic deployment stays near \SI{9}{\percent} and below. These broadband percentages are lower bounds anchored to the v2-Mini UEMR floor, not estimates of the ODC's own radiation, which is unmeasured and, at \SI{120}{kW} of compute per platform, could exceed it. The aggregate is linear in per-platform level, so the loss scales directly with any excess over that floor: every \SI{10}{dB} of additional per-platform UEMR raises the aggregate by \SI{10}{dB} and drives these fractions sharply higher, towards saturation at the lowest bands. The \SI{22}{\percent} at \SI{3.3}{GHz} is thus the optimistic end of the broadband case, not an estimate of it. In the narrowband case the same radiation is diluted away and the VGOS loss falls to essentially zero.

Direct simulations of the full \num{1000000}-satellite stress fleet (full fleet simulation, no scaling) at Hobart, on the actual clustered geometry, confirm these figures and that the linear extrapolation is conservative for VLBI at every band tested. The true VLBI loss is \SI{21.1}{\percent} against the \SI{22}{\percent} analytic estimate at \SI{3.3}{GHz}, \SI{16.3}{\percent} against \SI{16.4}{\percent} at \SI{5.2}{GHz}, and \SI{0.7}{\percent} against \SI{1.0}{\percent} at \SI{8.6}{GHz}. At the binding 98th percentile the analytic aggregate overestimates the direct one by $0.2$--$0.3$\,dB in each case. The single-dish continuum loss (\SI{85}{\percent} true versus \SI{83}{\percent} scaled at \SI{5.2}{GHz}) is saturated either way. This conservatism is not automatically inherited from the inclined-shell validation of Section~\ref{sec:vgos-methods-mc}: the mechanism behind it, the narrowing of the aggregate distribution as the sky fills, depends on the node distribution, and the dawn-dusk constellation clusters its planes into two terminator lunes rather than spreading them. The clustering sharpens the contrast between the occupied band and the empty sky, but the binding 98th-percentile statistic is set \emph{within} the occupied band. There, each cell still sums over more satellites as $N$ grows and narrows by the same central-limit argument, with the regular along-track phasing, if anything, reinforcing the uniformity. The \SI{3.3}{GHz} and \SI{5.2}{GHz} simulations above, run on the actual clustered geometry, reach directly into the higher-loss low-frequency bands and confirm this, each reproducing the analytic loss to within a point with the analytic side the conservative one. We therefore treat the broadband percentages as validated by direct simulation rather than as merely indicative. 

The conclusion is asymmetric but clear: a transmission-free compute megaconstellation is, through sheer multiplicity, a certain threat to single-dish radio astronomy, whereas whether it also corrupts geodetic VLBI hinges on a spacecraft radiation bandwidth that has not yet been measured.

\section{Discussion}
\label{sec:discussion}
This study provides the first quantitative assessment of the impact of directly measured intended emissions, spurious emissions, and unintended electromagnetic radiation from operational NGSO constellations on both single-dish radio astronomy and geodetic VLBI across the \mbox{1--14\,GHz} range. Three conclusions emerge.
\begin{enumerate}\itemsep2pt
\item The threat is \emph{already realised} for single-dish radio astronomy: at the measured emission and radiation levels the nominal RA.769 continuum criterion is exceeded across the whole sky at every frequency we tested, for the present fleet. Two allocated primary RAS bands at 2.7 and 5 GHz are affected, which must be protected from harmful interference. Our finding is consistent with, and extends to higher frequencies, the aggregate-interference picture drawn at \SI{150}{MHz} by \citet{divruno2023} and at \SI{1400}{MHz} by \citet{hellbourg2026}, who found the RA.769 power threshold levels overwhelmed by $\sim$4{,}400 and 30{,}000 satellites, respectively. At the unwanted-emission and unintended-radiation levels \emph{measured} in the SNIFFLES survey, the population required to exceed the limits is far smaller still. The $100\%$ data loss arises from each of the three classes independently. The intended \SI{2620}{MHz} downlink, the DTD harmonics, and the brightest platform UEMR detections each reach total loss on their own, and by wide margins: the present-fleet 98th-percentile margins at the worst-affected facility (Table~\ref{tab:facilities-cont-margin}) are $-$74.8\,dB for the downlink, $-$30.5 to $-$61.6\,dB for the harmonics, and $-$20.8 to $-$28.0\,dB for the bright platform UEMR; only the fainter incidental UEMR detections fall short (for instance the \SI{2656}{MHz} detection, at $44$--$72\%$ depending on the dish and its elevation limit, margin $-$20.1\,dB). In practice, $100\%$ loss means that at every frequency carrying one of these bright detections, no \SI{2000}{s} single-dish continuum or spectral-line integration meeting the RA.769 sensitivity standard can be made anywhere in the accessible sky while the constellation is overhead, which for a fully deployed fleet is at all times. 

High-sensitivity single-dish radio astronomy at these frequencies is therefore not ``degraded'' but foreclosed. What survives is bright-source work well above the protection level, and the interferometric and geodetic VLBI modes that the separate, $\sim$\SI{40}{dB} looser criterion of Section~\ref{sec:geovlbi} partially shields. However, foreclosed is not irreversible: the loss is set by today's per-satellite levels, and Table~\ref{tab:cispr_cont} states the levels at which single-dish RA.769-compliance can be achieved.

\item Geodetic VLBI is less affected owing to its less stringent 1\%-of-noise criterion, by the decorrelation of interference across long baselines, and by the narrowband nature of the unintended radiation, diluted against the broad \SI{32}{MHz} channel (Section~\ref{sec:bwcorr}). That protection holds against the platform UEMR but fails against the broadband downlink harmonics. The second harmonic at \SI{5240}{MHz} breaches the VLBI criterion today, and the third at \SI{7860}{MHz} degrades from $7.6\%$ to $71\%$ loss across the projected growth (the present-fleet $7.6\%$ is the \SI{2000}{s}-reference value and is window-dependent at the VGOS scan length, Section~\ref{sec:window}; the growth holds at every window). The data-loss percentage is the fraction of \SI{2000}{s} integrations and pointings in which the sidelobe aggregate exceeds the RA.769 VLBI protection level, the Table~3 (1\%-of-noise) values of RA.769-2 interpolated to the observing frequency (Section~\ref{sec:interp}); the IVS-computed VGOS levels of Report RA.2507 agree with these to better than \SI{1}{dB} (Section~\ref{sec:limits}). It counts scans in which the interference is detrimental by the RA.769 standard, not scans lost to front-end saturation, which is a separate and higher threshold (Section~\ref{sec:limits}). A \SI{71}{\percent} loss means roughly two scans in three exceed the protection criterion and are counted as lost in the RA.1513 accounting. The physical severity within those scans is graded, which is why the margins matter: interference at the threshold raises the system temperature by \SI{1}{\percent}, while at the \num{300000}-satellite 98th-percentile margin of $-$20\,dB it is of similar magnitude to \ac{Tsys} itself. The consequence carries over from single-dish radiometry: because the aggregate decorrelates between stations, it enters each baseline as added system noise, and a doubled \ac{Tsys} at both ends of a baseline doubles the fringe noise, so recovering the lost sensitivity would cost a fourfold integration time \citep{hase1999}. Scan lengths in VGOS are optimised to achieve the target accuracy. It is a compromise between short scans enabling a high number of observations to scan the atmosphere and a long enough scan duration to achieve sufficient delay precision in the individual measurement. Every VGOS sub-band contributes to the group-delay solution and to the ionospheric correction, so contamination anywhere in the span degrades the geodetic product: the reference frames, polar motion, and UT1 that underpin GNSS and timekeeping. The harmonics are especially pernicious: they are deterministic, broadband consequences of the \ac{DTD} downlink, they fall inside the protected-science span by construction, and they are not removable by siting or scheduling.

\item The binding constraint for VGOS-class antennas is statistical sensitivity loss (the graded, aggregate loss of sensitivity produced by many simultaneous emitters) rather than front-end saturation (Section~\ref{sec:limits}); saturation becomes the dominant failure mode only for larger apertures or the rare brightest transits. This argues that protection needs to be framed in aggregate, population-aware terms, as a per-satellite emission/radiation ceiling that scales with constellation size, rather than as a single-entry \ac{PFD} mask.
\end{enumerate}

That ceiling is the contribution of Table~\ref{tab:cispr}: a per-satellite limit, in the units the standards bodies use, below which unwanted emission and radiation (the harmonics; ITU-R SM, e.i.r.p. spectral density) and unintended radiation (CISPR, dB($\mu$V/m)) would need to fall to coexist with geodetic VLBI at \num{300000} satellites. The shortfall between the measured per-satellite levels and these ceilings, the compliance gap, is the quantitative statement of the coexistence problem: the number of decibels by which each unwanted-emission and unintended-radiation detection must be reduced. For VLBI the gap is set by the broadband harmonics, reaching $\sim$\SI{33}{dB} at \SI{5240}{MHz}, while the narrowband radiation already meets the VLBI ceiling but exceeds the single-dish one. Stated as requirements, the two ceilings separate the science cases. Geodetic VLBI coexists with \num{300000} satellites if every radiating satellite meets the 65--70\,dB($\mu$V/m) ceiling of Table~\ref{tab:cispr}, which the narrowband platform radiation already does; this protects the interferometric observable, while the calibration tier of the geodetic chain straddles the two cases, conservatively requiring continuum-class control \SI{6.5}{dB} above the single-dish ceilings (Section~\ref{sec:limits}). High-sensitivity single-dish radio astronomy at these frequencies recovers only when per-satellite levels meet the \SI{40}{dB} stricter continuum ceilings of Table~\ref{tab:cispr_cont}, 18--46\,dB($\mu$V/m) at that population, which every unwanted-emission and unintended-radiation detection currently exceeds: by 20--75\,dB at the calibrated detections, and by at least 7 and 12\,dB at the two floor-bounded L-band detections.

This aggregate framing is also reflected in a regulatory gap. Surveying the dark-and-quiet-skies protection frameworks of 77 jurisdictions, \citet{yakushina2026coherent} find that unintended radiation in particular escapes the standard licence-based interference controls, being unintended, poorly documented, and hard to test in advance. They argue that governance must operate at the aggregate scale of the interference rather than through single-entry allocation and coordination norms. A per-satellite ceiling that scales with constellation size is precisely such an instrument. Expressed in the units of the emission and radiation standards and verifiable before launch, it is directly usable within the governance channels those authors identify (e.g.\ national licensing and market-access conditions) as the near-term lever, and it gives the current ITU work on aggregate NGSO interference a concrete per-satellite target.

The same gap is visible in current filings: the 2026 application for \num{100000} satellites \citep{spacex2026gen3} contains no unintended-radiation disclosure, no spurious-emission figure beyond the default regulatory masks, and no consideration of geodetic VLBI; and requests waivers to transmit inside two passive bands.

Ultimately, the challenge identified here is not one of spectrum allocation alone, but of spacecraft engineering. Whether radio astronomy and future NGSO constellations can coexist will depend on how effectively unwanted and unintended emissions can be reduced before launch. The limits derived in this work provide a benchmark against which that coexistence can be assessed.

\section{Conclusions}
\label{sec:conclusions}

Extrapolating the directly measured emission and radiation of current NGSO constellations onto the Australian VGOS array and single-dish facilities, we find that the single-dish radio astronomy protection criterion of ITU-R RA.769 is already exceeded across the whole sky at the present fleet, by both the broadband harmonics and the narrowband unintended radiation. Among these frequencies, two allocated primary RAS bands at \SI{2.7}{} and \SI{5}{GHz} are affected, which must be protected from harmful interference. Recommendation ITU-R RA.2178 places geodetic VLBI stations under the radio astronomy service and invites administrations to register them as RAS stations, so that protection extends to the AuScope stations in those bands (Section~\ref{sec:thresholds}). While all other frequencies are not subject to protection within the ITU-R framework set out in the Radio Regulations, the results may still be considered useful by the astronomy community and administrations affording their facilities with radio quiet zones \citep[see e.g.][]{ituRA2259}. 

For geodetic VLBI, the threat is the bright DTD payload. The intended \SI{2620}{MHz} downlink already saturates the VLBI criterion as stated in our own interpolation and in ITU-R RA.2507 at the present fleet (its measured \SI{4.3}{MJy} being a compressed lower bound). The third harmonic at \SI{7860}{MHz}, just below the legacy X-band, degrades from $7.6\%$ to $71\%$ data loss as the fleet grows towards \num{300000} satellites; the present-fleet $7.6\%$ is a \SI{2000}{s}-reference value, lower at the \SI{10}{}--\SI{30}{s} VGOS scan length (Section~\ref{sec:window}). The incidental narrowband platform radiation, diluted against the \SI{32}{MHz} VGOS channel, does not breach the VLBI criterion at any plausible fleet size. Because the aggregate is linear in constellation size, we recast the result as a per-satellite ceiling: the \emph{unwanted} emission of each spacecraft, its harmonics and the products of its carrier-generating oscillator, must fall to order 65--70\,dB($\mu$V/m) at \SI{10}{m} in a \SI{1}{MHz} bandwidth for coexistence with geodetic VLBI at that scale. Current spacecraft exceed this level by $\sim$20\,dB at the cleanly-measured \SI{7860}{MHz} harmonic, and by more at the brighter, compression-limited unwanted-emission detections. The intended downlink is not subject to this ceiling: within its authorised band an operator may transmit at its licensed power, and its incompatibility with a co-frequency radio-astronomy or geodetic receiver is a question of spectrum allocation and of receiver linearity, not of an emission mask. 

The narrowband incidental radiation, while not a VLBI sensitivity threat, remains a single-dish radio astronomy problem and a front-end saturation risk. Meeting these limits will require emission/radiation control at the spacecraft, expressed in regulatory and standard-setting-body terms and verified before launch, alongside spectrum-management measures before the ITU-R. Such control is within reach: spacecraft already qualify for flight against comparable \mbox{MIL-STD-461} radiated-emission masks, with deep notches protecting the launch vehicle's navigation, telemetry, and flight-termination receivers, so both the engineering capability and a verification pathway already exist \citep{milstd461g}.

The two criteria set two distinct compliance targets. Coexistence with geodetic VLBI at \num{300000} satellites requires every radiating satellite, whatever its operator, to meet the 65--70\,dB($\mu$V/m) ceiling (Table~\ref{tab:cispr}; Figure~\ref{fig:elim}); the narrowband platform radiation already meets it. That ceiling protects the interferometric observable; the single-dish-mode calibration measurements on which geodetic VLBI equally depends are conservatively held to continuum-class levels \SI{6.5}{dB} above the single-dish ceilings (Section~\ref{sec:limits}), so the full geodetic chain has a direct stake in progress towards the stricter tier. Restoring high-sensitivity single-dish radio astronomy requires the \SI{40}{dB} more stringent continuum ceilings of 18--46\,dB($\mu$V/m) (Table~\ref{tab:cispr_cont}; Figure~\ref{fig:elim_cont}), which every unwanted-emission and unintended-radiation detection currently exceeds, by up to 75\,dB. Emission and radiation control to the continuum ceilings would return these bands to RA.769-compliant levels.

These results provide concrete, standards-ready technical input for standard setting bodies, the proposed WRC-31 agenda item on geodetic VLBI, and numbers against which compliance can be assessed.

\section*{Acknowledgement}

The \ac{ATCA} is part of the ATNF which is funded by the Australian Government for operation as a National Facility managed by CSIRO. We acknowledge the Gomeroi people as the Traditional Owners of the Paul Wild Observatory site. The AuScope VLBI project is managed by the University of Tasmania, contracted through Geoscience Australia. The VLBI array was established with funds provided by AuScope and the Australian Government via the National Collaborative Research Infrastructure Strategy (NCRIS): auscope.org.au. This work made use of the  \texttt{pycraf} \citep{winkel2018pycraf} and \texttt{cysgp4} \citep{cysgp4_software} packages, built on the scientific Python ecosystem: \texttt{astropy} \citep{astropy2022}, \texttt{numpy} \citep{harris2020numpy}, \texttt{scipy} \citep{virtanen2020scipy}, and \texttt{matplotlib} \citep{hunter2007matplotlib}.

We thank the anonymous referee for a thorough and constructive report that has substantially improved the manuscript.

\section*{Data Availability}

The SNIFFLES-I measurements underlying the per-satellite inputs are described in \citet{indermuehle2026sniffles}. Orbital elements (Orbit Mean-Elements Messages) were obtained from Space-Track via the CSIRO SNIFFLES-I archive.

\onecolumn  
\appendix
\makeatletter\@removefromreset{figure}{section}\@removefromreset{table}{section}\makeatother
\setcounter{figure}{0}
\setcounter{table}{0}
\renewcommand{\thefigure}{\alph{figure}}  
\renewcommand{\thetable}{\alph{table}}    
\section{Full EPFD results}
\label{app:full}

\begin{figure}[H]
\centering\includegraphics[width=0.6\textwidth]{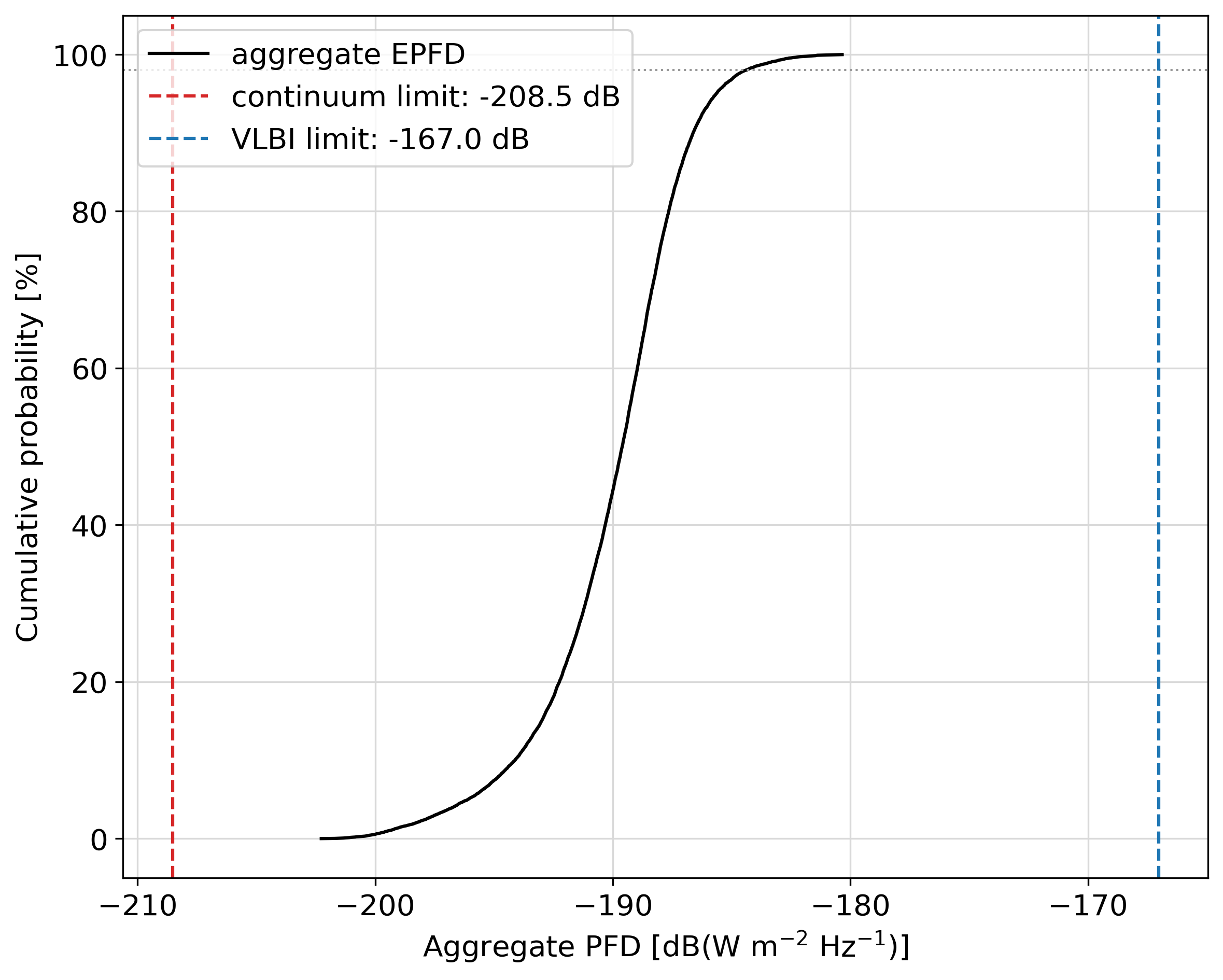}
\caption{Example appendix plot: aggregate-PFD CDF at Hobart for the \SI{2700}{MHz} unintended radiation, present fleet (ATCA), with the emission/radiation-bandwidth correction applied. The single-dish continuum limit is exceeded by the whole distribution ($100\%$ loss); the VLBI limit, raised \SI{38}{dB} by the \SI{5}{kHz} radiation bandwidth, is not reached ($\sim$$0\%$ VLBI loss).}
\label{fig:ex2700}
\end{figure}

\begin{figure}[H]
\centering\includegraphics[width=0.6\textwidth]{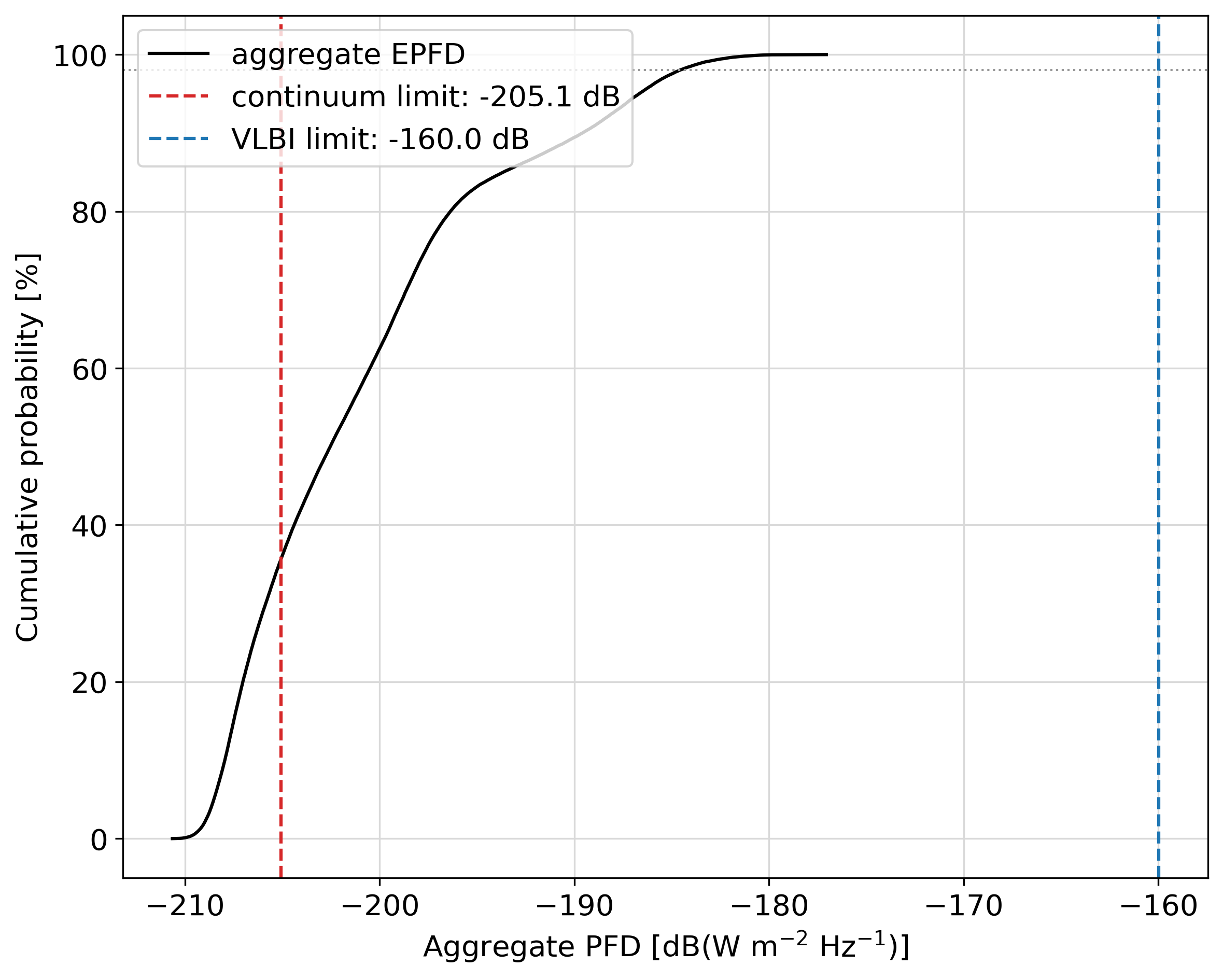}
\caption{Example appendix plot: aggregate-PFD CDF at Hobart for the OneWeb \SI{8599}{MHz} radiation at \num{300000} satellites, using the Y-factor calibrated peak \ac{PFD} of \SI{2786}{Jy}, with the emission/radiation-bandwidth correction applied. The single-dish continuum limit is exceeded over $\sim$64\% of the distribution; the VLBI limit, raised \SI{35}{dB} by the \SI{10}{kHz} radiation bandwidth, is not reached.}
\label{fig:ex8599}
\end{figure}

Table~\ref{tab:appendix} gives the complete grid of \ac{EPFD} data-loss percentages and 98th-percentile margins for every AuScope station (Hb $=$ Hobart, Ke $=$ Katherine, Yg $=$ Yarragadee), emission/radiation detection, level bracket (floor/ATCA), constellation scenario (now/100k/300k) and protection criterion (cont $=$ single-dish continuum, VLBI). To avoid reproducing all 144 plots, Figures~\ref{fig:cdf5240}--\ref{fig:loss7860} in the main text are representative. Two further examples are shown here, both illustrating the narrowband-UEMR case after the bandwidth correction (Section~\ref{sec:bwcorr}): Figure~\ref{fig:ex2700} (the \SI{2700}{MHz} radiation, present fleet) and Figure~\ref{fig:ex8599} (the OneWeb \SI{8599}{MHz} radiation at \num{300000} satellites), each exceeding the single-dish continuum limit while sitting below the VLBI limit. The complete plot set is available with the code.

\begin{figure}[H]
\centering\includegraphics[width=\textwidth]{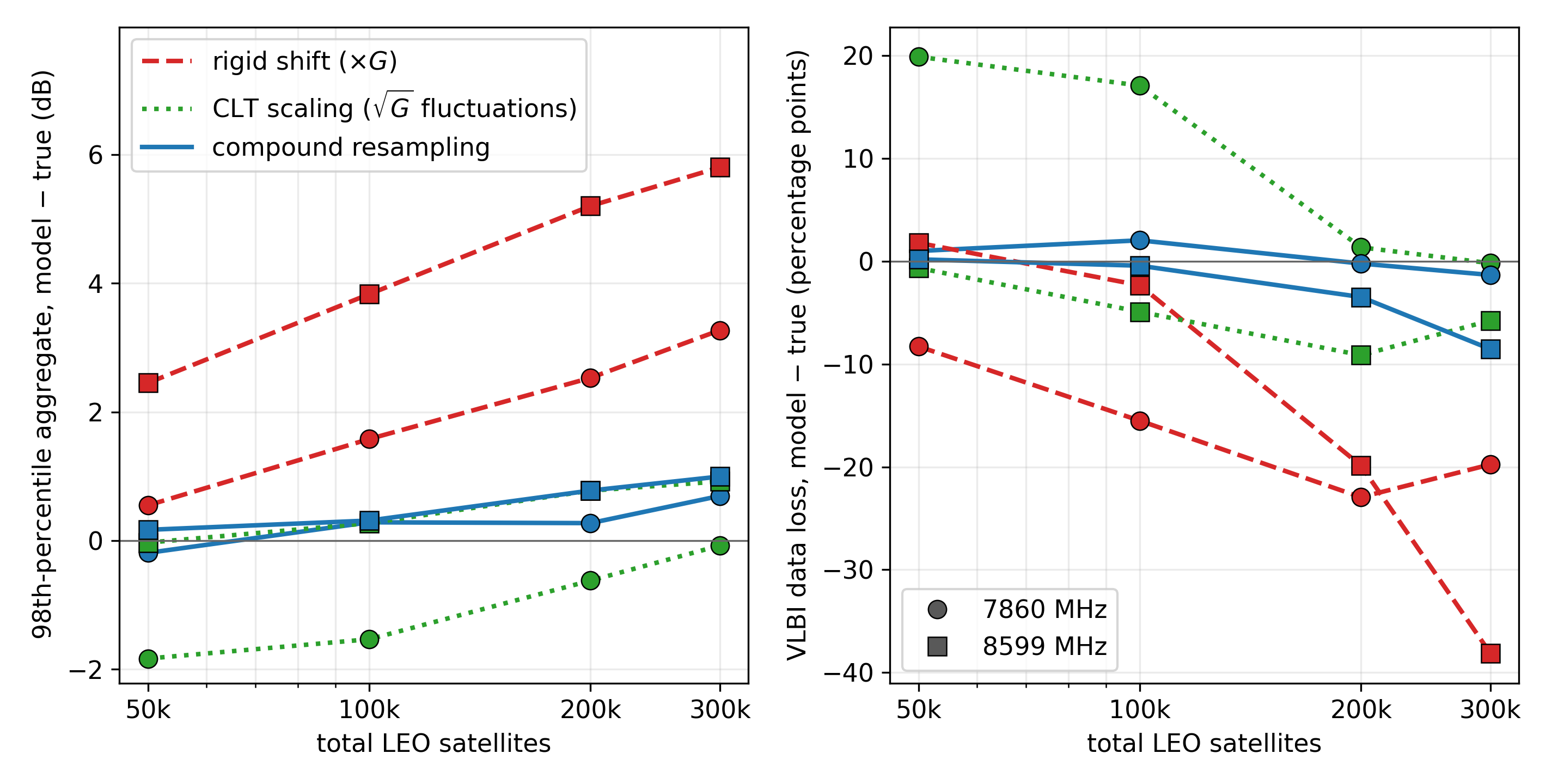}
\caption{Validation of the corrected $N$-scaling against direct Monte-Carlo of genuinely up-scaled fleets (Hobart; \SI{7860}{} and \SI{8599}{MHz}). \emph{Left:} error of each model's 98th-percentile aggregate relative to the true up-scaled simulation; \emph{right:} the same for the VLBI data loss. The rigid CDF shift ($\times G$) overstates the 98th percentile increasingly with population and understates the data loss, because the aggregate's fluctuations grow only as $\sqrt{N}$ while its mean grows as $N$. Pure $\sqrt{G}$ scaling of the fluctuations misrepresents the heavy transit tail at small $G$. Compound resampling, the sum of $G$ per-cell aggregates resampled from the Monte-Carlo ensemble, reproduces both statistics at every tested population.}
\label{fig:sqrtncorr}
\end{figure}

\clearpage
{\footnotesize
\begin{longtable}{@{}lll rrrrrr rr@{}}
\caption{Complete EPFD grid for every station, line and level bracket: data loss (\%) under the single-dish continuum ($C$) and VLBI ($V$) criteria at the present fleet, \num{100000} and \num{300000} satellites, and the present-fleet 98th-percentile margin per criterion, $M = S_{\rm lim}-\mathrm{pfd}_{98}$ (dB). Negative $M$ gives directly the aggregate attenuation required to meet the criterion; margins at \num{100000} and \num{300000} satellites follow by subtracting \SI{9.2}{} and \SI{14.0}{dB}. Floor-bracket rows bound the level from below: their losses are lower bounds and their margins upper bounds; for the floor-only (red) lines the margin entries carry an explicit $<$. The Monte-Carlo sampling uncertainty on the margins (bootstrap standard error of the pooled 98th percentile, resampled over iterations) has median \SI{0.1}{dB} and lies below \SI{0.3}{dB} for \SI{90}{\percent} of entries, reaching \SI{1.4}{dB} only for the sparsest radiating population (the 21-satellite GuoWang line at \SI{1777}{MHz}); the uncertainty budget is instead dominated by the input levels, bracketed by the floor and ATCA rows and by the compression-limited $\geq$ bounds. Freq in MHz.}\label{tab:appendix}\\
\toprule
Stn & Freq & Brkt & $C_{\rm now}$ & $V_{\rm now}$ & $C_{100k}$ & $V_{100k}$ & $C_{300k}$ & $V_{300k}$ & $M^{C}$ & $M^{V}$\\
\hline
\endfirsthead
\toprule
Stn & Freq & Brkt & $C_{\rm now}$ & $V_{\rm now}$ & $C_{100k}$ & $V_{100k}$ & $C_{300k}$ & $V_{300k}$ & $M^{C}$ & $M^{V}$\\
\hline
\endhead
ASKAP & 1320 & atca & 100.0 & 0.0 & 100.0 & 0.0 & 100.0 & 0.0 & $-$14.5 & 28.9\\
ASKAP & 1320 & floor & 0.0 & 0.0 & 82.6 & 0.0 & 99.2 & 0.0 & 3.8 & 47.1\\
ASKAP & 1350 & atca & 100.0 & 0.0 & 100.0 & 0.0 & 100.0 & 0.0 & $-$15.5 & 28.0\\
ASKAP & 1350 & floor & 0.0 & 0.0 & 89.2 & 0.0 & 99.7 & 0.0 & 3.1 & 46.6\\
ASKAP & 1440 & atca & 100.0 & 0.0 & 100.0 & 0.0 & 100.0 & 0.0 & $-$13.2 & 30.3\\
ASKAP & 1440 & floor & 0.0 & 0.0 & 88.3 & 0.0 & 99.6 & 0.0 & 3.2 & 46.7\\
ASKAP & 1454 & atca & 7.9 & 0.0 & 35.2 & 0.0 & 85.3 & 0.0 & $-$13.6 & 29.6\\
ASKAP & 1454 & floor & 0.0 & 0.0 & 0.0 & 0.0 & 1.0 & 0.0 & 15.1 & 58.4\\
ASKAP & 1474 & floor & \textcolor{red}{0.0} & \textcolor{red}{0.0} & \textcolor{red}{0.1} & \textcolor{red}{0.0} & \textcolor{red}{11.8} & \textcolor{red}{0.0} & \textcolor{red}{$<$11.3} & \textcolor{red}{$<$54.5}\\
ASKAP & 1620 & atca & 99.9 & 0.0 & 100.0 & 0.0 & 100.0 & 0.0 & $-$13.3 & 28.6\\
ASKAP & 1620 & floor & 0.0 & 0.0 & 37.3 & 0.0 & 93.4 & 0.0 & 6.4 & 48.3\\
ASKAP & 1680 & floor & \textcolor{red}{0.0} & \textcolor{red}{0.0} & \textcolor{red}{30.4} & \textcolor{red}{0.0} & \textcolor{red}{92.0} & \textcolor{red}{0.0} & \textcolor{red}{$<$6.8} & \textcolor{red}{$<$48.3}\\
ASKAP & 1777 & atca & 24.9 & 0.0 & 96.3 & 0.0 & 100.0 & 0.3 & $-$21.9 & 19.6\\
ASKAP & 1777 & floor & 0.0 & 0.0 & 0.0 & 0.0 & 0.0 & 0.0 & 21.8 & 63.3\\
ATCA & 1320 & atca & 100.0 & 0.0 & 100.0 & 0.0 & 100.0 & 0.0 & $-$16.0 & 27.4\\
ATCA & 1320 & floor & 0.1 & 0.0 & 82.3 & 0.0 & 97.4 & 0.0 & 2.2 & 45.6\\
ATCA & 1350 & atca & 100.0 & 0.0 & 100.0 & 0.0 & 100.0 & 0.0 & $-$17.0 & 26.4\\
ATCA & 1350 & floor & 0.3 & 0.0 & 87.6 & 0.0 & 98.8 & 0.0 & 1.6 & 45.0\\
ATCA & 1440 & atca & 100.0 & 0.0 & 100.0 & 0.0 & 100.0 & 0.0 & $-$14.7 & 28.7\\
ATCA & 1440 & floor & 0.2 & 0.0 & 86.5 & 0.0 & 98.6 & 0.0 & 1.7 & 45.1\\
ATCA & 1454 & atca & 9.5 & 0.0 & 40.6 & 0.0 & 90.8 & 0.0 & $-$15.1 & 28.2\\
ATCA & 1454 & floor & 0.0 & 0.0 & 0.0 & 0.0 & 2.1 & 0.0 & 13.6 & 56.9\\
ATCA & 1474 & floor & \textcolor{red}{0.0} & \textcolor{red}{0.0} & \textcolor{red}{2.0} & \textcolor{red}{0.0} & \textcolor{red}{13.6} & \textcolor{red}{0.0} & \textcolor{red}{$<$9.2} & \textcolor{red}{$<$52.4}\\
ATCA & 1620 & atca & 99.9 & 0.0 & 100.0 & 0.0 & 100.0 & 0.0 & $-$15.0 & 26.9\\
ATCA & 1620 & floor & 0.0 & 0.0 & 51.1 & 0.0 & 88.4 & 0.0 & 4.8 & 46.6\\
ATCA & 1680 & floor & \textcolor{red}{0.0} & \textcolor{red}{0.0} & \textcolor{red}{45.9} & \textcolor{red}{0.0} & \textcolor{red}{87.0} & \textcolor{red}{0.0} & \textcolor{red}{$<$5.1} & \textcolor{red}{$<$46.6}\\
ATCA & 1777 & atca & 36.0 & 0.0 & 99.7 & 0.0 & 100.0 & 1.0 & $-$21.9 & 19.6\\
ATCA & 1777 & floor & 0.0 & 0.0 & 0.0 & 0.0 & 0.4 & 0.0 & 21.8 & 63.3\\
ATCA & 2595 & atca & 100.0 & 0.1 & 100.0 & 6.3 & 100.0 & 11.4 & $-$36.0 & 5.5\\
ATCA & 2595 & floor & 0.0 & 0.0 & 0.2 & 0.0 & 1.8 & 0.0 & 14.2 & 55.8\\
ATCA & 2620 & atca & 100.0 & 100.0 & 100.0 & 100.0 & 100.0 & 100.0 & $-$73.2 & $-$31.7\\
ATCA & 2620 & floor & 57.1 & 0.0 & 100.0 & 0.0 & 100.0 & 0.0 & $-$18.9 & 22.6\\
ATCA & 2656 & atca & 51.5 & 0.0 & 100.0 & 0.0 & 100.0 & 0.0 & $-$18.5 & 23.0\\
ATCA & 2656 & floor & 0.0 & 0.0 & 0.2 & 0.0 & 1.6 & 0.0 & 14.5 & 56.0\\
ATCA & 2700 & atca & 100.0 & 0.0 & 100.0 & 0.0 & 100.0 & 0.0 & $-$23.2 & 18.3\\
ATCA & 2700 & floor & 0.0 & 0.0 & 3.4 & 0.0 & 44.6 & 0.0 & 8.7 & 50.2\\
ATCA & 4995 & atca & 32.0 & 0.0 & 96.2 & 0.0 & 100.0 & 0.0 & $-$6.1 & 35.4\\
ATCA & 4995 & floor & 0.0 & 0.0 & 0.0 & 0.0 & 3.3 & 0.0 & 13.4 & 54.9\\
ATCA & 5190 & atca & 4.7 & 0.0 & 44.1 & 0.0 & 83.7 & 0.0 & $-$8.3 & 33.5\\
ATCA & 5190 & floor & 0.0 & 0.0 & 0.1 & 0.0 & 0.7 & 0.0 & 18.6 & 60.4\\
ATCA & 5240 & atca & 100.0 & 48.1 & 100.0 & 100.0 & 100.0 & 100.0 & $-$59.9 & $-$18.1\\
ATCA & 5240 & floor & 24.1 & 0.0 & 94.2 & 0.0 & 100.0 & 0.0 & $-$14.4 & 27.4\\
ATCA & 7860 & atca & 100.0 & 2.0 & 100.0 & 21.8 & 100.0 & 51.1 & $-$44.8 & $-$0.2\\
ATCA & 7860 & floor & 32.5 & 0.0 & 99.7 & 0.0 & 100.0 & 0.0 & $-$11.5 & 33.0\\
ATCA & 8599 & atca & 4.7 & 0.0 & 33.6 & 0.0 & 55.3 & 0.0 & $-$6.8 & 38.2\\
ATCA & 8599 & floor & 0.0 & 0.0 & 0.0 & 0.0 & 0.1 & 0.0 & 21.9 & 67.0\\
ATCA & 8640 & atca & 99.7 & 0.0 & 100.0 & 0.0 & 100.0 & 0.0 & $-$19.2 & 25.9\\
ATCA & 8640 & floor & 0.0 & 0.0 & 5.5 & 0.0 & 17.3 & 0.0 & 7.1 & 52.2\\
ATCA & 10480 & atca & 100.0 & 0.1 & 100.0 & 0.8 & 100.0 & 1.1 & $-$27.3 & 19.1\\
ATCA & 10480 & floor & 36.6 & 0.0 & 100.0 & 0.0 & 100.0 & 0.0 & $-$9.6 & 36.8\\
Ceduna & 1620 & atca & 100.0 & 0.0 & 100.0 & 0.0 & 100.0 & 0.0 & $-$16.1 & 25.8\\
Ceduna & 1620 & floor & 0.1 & 0.0 & 56.3 & 0.0 & 85.2 & 0.0 & 3.7 & 45.6\\
Ceduna & 1680 & floor & \textcolor{red}{0.0} & \textcolor{red}{0.0} & \textcolor{red}{52.4} & \textcolor{red}{0.0} & \textcolor{red}{83.8} & \textcolor{red}{0.0} & \textcolor{red}{$<$4.0} & \textcolor{red}{$<$45.5}\\
Ceduna & 1777 & atca & 36.7 & 0.0 & 99.7 & 0.0 & 100.0 & 0.9 & $-$17.8 & 23.7\\
Ceduna & 1777 & floor & 0.0 & 0.0 & 0.0 & 0.0 & 0.4 & 0.0 & 25.9 & 67.4\\
Ceduna & 2595 & atca & 100.0 & 0.3 & 100.0 & 6.7 & 100.0 & 11.5 & $-$37.0 & 4.5\\
Ceduna & 2595 & floor & 0.0 & 0.0 & 0.4 & 0.0 & 2.6 & 0.0 & 13.3 & 54.8\\
Ceduna & 2620 & atca & 100.0 & 100.0 & 100.0 & 100.0 & 100.0 & 100.0 & $-$74.0 & $-$32.5\\
Ceduna & 2620 & floor & 62.5 & 0.0 & 100.0 & 0.0 & 100.0 & 0.0 & $-$19.7 & 21.8\\
Ceduna & 2656 & atca & 56.7 & 0.0 & 100.0 & 0.0 & 100.0 & 0.0 & $-$19.3 & 22.2\\
Ceduna & 2656 & floor & 0.0 & 0.0 & 0.4 & 0.0 & 2.2 & 0.0 & 13.7 & 55.2\\
Ceduna & 2700 & atca & 100.0 & 0.0 & 100.0 & 0.0 & 100.0 & 0.1 & $-$24.5 & 17.0\\
Ceduna & 2700 & floor & 0.0 & 0.0 & 8.1 & 0.0 & 45.9 & 0.0 & 7.4 & 48.9\\
Ceduna & 4995 & atca & 32.4 & 0.0 & 98.0 & 0.0 & 100.0 & 0.0 & $-$7.4 & 34.1\\
Ceduna & 4995 & floor & 0.0 & 0.0 & 0.2 & 0.0 & 6.2 & 0.0 & 12.1 & 53.6\\
Ceduna & 5190 & atca & 5.3 & 0.0 & 50.1 & 0.0 & 88.4 & 0.0 & $-$8.6 & 33.2\\
Ceduna & 5190 & floor & 0.0 & 0.0 & 0.2 & 0.0 & 0.8 & 0.0 & 18.4 & 60.1\\
Ceduna & 5240 & atca & 100.0 & 54.5 & 100.0 & 100.0 & 100.0 & 100.0 & $-$60.2 & $-$18.4\\
Ceduna & 5240 & floor & 29.7 & 0.0 & 97.5 & 0.0 & 100.0 & 0.0 & $-$14.6 & 27.2\\
Ceduna & 7860 & atca & 100.0 & 2.0 & 100.0 & 26.7 & 100.0 & 57.4 & $-$44.7 & $-$0.2\\
Ceduna & 7860 & floor & 38.0 & 0.0 & 99.9 & 0.0 & 100.0 & 0.0 & $-$11.4 & 33.1\\
Ceduna & 8599 & atca & 4.0 & 0.0 & 34.7 & 0.0 & 58.3 & 0.0 & $-$6.7 & 38.4\\
Ceduna & 8599 & floor & 0.0 & 0.0 & 0.0 & 0.0 & 0.2 & 0.0 & 22.1 & 67.2\\
Ceduna & 8640 & atca & 99.9 & 0.0 & 100.0 & 0.0 & 100.0 & 0.0 & $-$20.4 & 24.7\\
Ceduna & 8640 & floor & 0.0 & 0.0 & 6.9 & 0.0 & 17.2 & 0.0 & 5.9 & 51.0\\
Ceduna & 10480 & atca & 100.0 & 0.1 & 100.0 & 0.7 & 100.0 & 1.2 & $-$28.6 & 17.8\\
Ceduna & 10480 & floor & 43.3 & 0.0 & 100.0 & 0.0 & 100.0 & 0.0 & $-$10.9 & 35.5\\
Hb & 2595 & atca & 100.0 & 0.0 & 100.0 & 11.0 & 100.0 & 29.8 & $-$35.8 & 5.7\\
Hb & 2595 & floor & 0.0 & 0.0 & 0.0 & 0.0 & 1.5 & 0.0 & 14.4 & 56.0\\
Hb & 2620 & atca & 100.0 & 100.0 & 100.0 & 100.0 & 100.0 & 100.0 & $-$72.9 & $-$31.4\\
Hb & 2620 & floor & 75.9 & 0.0 & 100.0 & 0.0 & 100.0 & 0.0 & $-$18.6 & 22.9\\
Hb & 2656 & atca & 71.6 & 0.0 & 100.0 & 0.0 & 100.0 & 0.0 & $-$18.2 & 23.3\\
Hb & 2656 & floor & 0.0 & 0.0 & 0.0 & 0.0 & 1.2 & 0.0 & 14.8 & 56.3\\
Hb & 2700 & atca & 100.0 & 0.0 & 100.0 & 0.0 & 100.0 & 0.1 & $-$24.2 & 17.2\\
Hb & 2700 & floor & 0.0 & 0.0 & 6.3 & 0.0 & 66.0 & 0.0 & 7.7 & 49.2\\
Hb & 4995 & atca & 54.8 & 0.0 & 98.5 & 0.0 & 100.0 & 0.0 & $-$6.1 & 35.4\\
Hb & 4995 & floor & 0.0 & 0.0 & 0.0 & 0.0 & 3.1 & 0.0 & 13.4 & 54.9\\
Hb & 5190 & atca & 16.4 & 0.0 & 63.3 & 0.0 & 96.1 & 0.0 & $-$9.5 & 32.3\\
Hb & 5190 & floor & 0.0 & 0.0 & 0.0 & 0.0 & 0.4 & 0.0 & 17.5 & 59.2\\
Hb & 5240 & atca & 100.0 & 68.2 & 100.0 & 100.0 & 100.0 & 100.0 & $-$61.0 & $-$19.2\\
Hb & 5240 & floor & 38.9 & 0.0 & 99.8 & 0.0 & 100.0 & 0.0 & $-$15.5 & 26.4\\
Hb & 7860 & atca & 100.0 & 7.6 & 100.0 & 34.6 & 100.0 & 71.2 & $-$50.9 & $-$6.4\\
Hb & 7860 & floor & 49.8 & 0.0 & 100.0 & 0.0 & 100.0 & 0.0 & $-$17.6 & 26.9\\
Hb & 8599 & atca & 12.0 & 0.0 & 39.3 & 0.0 & 64.7 & 0.0 & $-$6.6 & 38.5\\
Hb & 8599 & floor & 0.0 & 0.0 & 0.0 & 0.0 & 0.0 & 0.0 & 22.2 & 67.2\\
Hb & 8640 & atca & 99.9 & 0.0 & 100.0 & 0.0 & 100.0 & 0.0 & $-$18.3 & 26.8\\
Hb & 8640 & floor & 0.0 & 0.0 & 4.6 & 0.0 & 36.3 & 0.0 & 8.0 & 53.1\\
Hb & 10480 & atca & 100.0 & 0.0 & 100.0 & 1.6 & 100.0 & 4.2 & $-$36.3 & 10.1\\
Hb & 10480 & floor & 55.8 & 0.0 & 100.0 & 0.0 & 100.0 & 0.0 & $-$18.6 & 27.8\\
Hb26 & 1320 & atca & 100.0 & 0.0 & 100.0 & 0.0 & 100.0 & 0.0 & $-$18.6 & 24.8\\
Hb26 & 1320 & floor & 2.6 & 0.0 & 89.4 & 0.0 & 98.5 & 0.0 & $-$0.3 & 43.1\\
Hb26 & 1350 & atca & 100.0 & 0.0 & 100.0 & 0.0 & 100.0 & 0.0 & $-$19.6 & 23.9\\
Hb26 & 1350 & floor & 4.2 & 0.0 & 92.3 & 0.0 & 99.2 & 0.0 & $-$1.0 & 42.5\\
Hb26 & 1440 & atca & 100.0 & 0.0 & 100.0 & 0.0 & 100.0 & 0.0 & $-$17.3 & 26.1\\
Hb26 & 1440 & floor & 3.9 & 0.0 & 91.5 & 0.0 & 99.1 & 0.0 & $-$0.9 & 42.5\\
Hb26 & 1454 & atca & 10.0 & 0.0 & 47.9 & 0.0 & 83.6 & 0.0 & $-$15.6 & 27.7\\
Hb26 & 1454 & floor & 0.0 & 0.0 & 0.1 & 0.0 & 2.4 & 0.0 & 13.2 & 56.4\\
Hb26 & 1474 & floor & \textcolor{red}{0.0} & \textcolor{red}{0.0} & \textcolor{red}{5.0} & \textcolor{red}{0.0} & \textcolor{red}{19.2} & \textcolor{red}{0.0} & \textcolor{red}{$<$7.2} & \textcolor{red}{$<$50.3}\\
Hb26 & 1620 & atca & 100.0 & 0.0 & 100.0 & 0.0 & 100.0 & 0.0 & $-$17.4 & 24.4\\
Hb26 & 1620 & floor & 0.3 & 0.0 & 68.3 & 0.0 & 92.7 & 0.0 & 2.3 & 44.2\\
Hb26 & 1680 & floor & \textcolor{red}{0.2} & \textcolor{red}{0.0} & \textcolor{red}{64.8} & \textcolor{red}{0.0} & \textcolor{red}{91.9} & \textcolor{red}{0.0} & \textcolor{red}{$<$2.7} & \textcolor{red}{$<$44.2}\\
Hb26 & 1777 & atca & 34.3 & 0.0 & 91.9 & 0.0 & 100.0 & 0.9 & $-$17.2 & 24.3\\
Hb26 & 1777 & floor & 0.0 & 0.0 & 0.0 & 0.0 & 0.4 & 0.0 & 26.5 & 68.0\\
Hb26 & 2595 & atca & 100.0 & 0.4 & 100.0 & 9.3 & 100.0 & 16.2 & $-$37.9 & 3.6\\
Hb26 & 2595 & floor & 0.0 & 0.0 & 0.5 & 0.0 & 3.4 & 0.0 & 12.4 & 53.9\\
Hb26 & 2620 & atca & 100.0 & 100.0 & 100.0 & 100.0 & 100.0 & 100.0 & $-$74.8 & $-$33.3\\
Hb26 & 2620 & floor & 75.7 & 0.0 & 100.0 & 0.0 & 100.0 & 0.0 & $-$20.5 & 21.0\\
Hb26 & 2656 & atca & 71.3 & 0.0 & 100.0 & 0.0 & 100.0 & 0.0 & $-$20.1 & 21.4\\
Hb26 & 2656 & floor & 0.0 & 0.0 & 0.4 & 0.0 & 2.8 & 0.0 & 12.9 & 54.4\\
Hb26 & 2700 & atca & 100.0 & 0.0 & 100.0 & 0.0 & 100.0 & 0.3 & $-$25.3 & 16.2\\
Hb26 & 2700 & floor & 0.0 & 0.0 & 12.1 & 0.0 & 57.4 & 0.0 & 6.6 & 48.1\\
Hb26 & 4995 & atca & 42.0 & 0.0 & 98.4 & 0.0 & 100.0 & 0.0 & $-$8.0 & 33.5\\
Hb26 & 4995 & floor & 0.0 & 0.0 & 0.3 & 0.0 & 8.4 & 0.0 & 11.5 & 53.0\\
Hb26 & 5190 & atca & 8.3 & 0.0 & 62.9 & 0.0 & 95.9 & 0.0 & $-$10.1 & 31.6\\
Hb26 & 5190 & floor & 0.0 & 0.0 & 0.2 & 0.0 & 1.0 & 0.0 & 16.8 & 58.6\\
Hb26 & 5240 & atca & 100.0 & 67.8 & 100.0 & 100.0 & 100.0 & 100.0 & $-$61.6 & $-$19.8\\
Hb26 & 5240 & floor & 39.6 & 0.0 & 99.7 & 0.0 & 100.0 & 0.0 & $-$16.1 & 25.8\\
Hb26 & 7860 & atca & 100.0 & 3.0 & 100.0 & 35.5 & 100.0 & 70.8 & $-$48.8 & $-$4.3\\
Hb26 & 7860 & floor & 49.5 & 0.0 & 100.0 & 0.0 & 100.0 & 0.0 & $-$15.5 & 29.0\\
Hb26 & 8599 & atca & 5.4 & 0.0 & 39.4 & 0.0 & 64.3 & 0.0 & $-$7.7 & 37.4\\
Hb26 & 8599 & floor & 0.0 & 0.0 & 0.0 & 0.0 & 0.1 & 0.0 & 21.0 & 66.1\\
Hb26 & 8640 & atca & 99.9 & 0.0 & 100.0 & 0.0 & 100.0 & 0.0 & $-$20.8 & 24.4\\
Hb26 & 8640 & floor & 0.0 & 0.0 & 8.7 & 0.0 & 23.3 & 0.0 & 5.5 & 50.6\\
Hb26 & 10480 & atca & 100.0 & 0.1 & 100.0 & 1.0 & 100.0 & 1.7 & $-$30.5 & 15.9\\
Hb26 & 10480 & floor & 55.3 & 0.0 & 100.0 & 0.0 & 100.0 & 0.0 & $-$12.8 & 33.5\\
Ke & 2595 & atca & 100.0 & 0.0 & 100.0 & 4.1 & 100.0 & 16.9 & $-$33.5 & 8.0\\
Ke & 2595 & floor & 0.0 & 0.0 & 0.0 & 0.0 & 0.3 & 0.0 & 16.7 & 58.2\\
Ke & 2620 & atca & 100.0 & 100.0 & 100.0 & 100.0 & 100.0 & 100.0 & $-$70.6 & $-$29.1\\
Ke & 2620 & floor & 45.3 & 0.0 & 100.0 & 0.0 & 100.0 & 0.0 & $-$16.3 & 25.2\\
Ke & 2656 & atca & 40.9 & 0.0 & 100.0 & 0.0 & 100.0 & 0.0 & $-$16.0 & 25.5\\
Ke & 2656 & floor & 0.0 & 0.0 & 0.0 & 0.0 & 0.3 & 0.0 & 17.0 & 58.5\\
Ke & 2700 & atca & 100.0 & 0.0 & 100.0 & 0.0 & 100.0 & 0.0 & $-$20.8 & 20.7\\
Ke & 2700 & floor & 0.0 & 0.0 & 0.0 & 0.0 & 32.1 & 0.0 & 11.1 & 52.6\\
Ke & 4995 & atca & 26.6 & 0.0 & 87.4 & 0.0 & 100.0 & 0.0 & $-$3.1 & 38.4\\
Ke & 4995 & floor & 0.0 & 0.0 & 0.0 & 0.0 & 0.1 & 0.0 & 16.4 & 57.9\\
Ke & 5190 & atca & 9.3 & 0.0 & 35.2 & 0.0 & 69.8 & 0.0 & $-$7.2 & 34.6\\
Ke & 5190 & floor & 0.0 & 0.0 & 0.0 & 0.0 & 0.2 & 0.0 & 19.8 & 61.5\\
Ke & 5240 & atca & 100.0 & 38.5 & 100.0 & 99.9 & 100.0 & 100.0 & $-$58.7 & $-$16.9\\
Ke & 5240 & floor & 18.5 & 0.0 & 80.2 & 0.0 & 100.0 & 0.0 & $-$13.2 & 28.7\\
Ke & 7860 & atca & 100.0 & 4.0 & 100.0 & 15.7 & 100.0 & 41.9 & $-$48.2 & $-$3.7\\
Ke & 7860 & floor & 25.8 & 0.0 & 92.3 & 0.0 & 100.0 & 0.0 & $-$14.9 & 29.6\\
Ke & 8599 & atca & 9.0 & 0.0 & 30.6 & 0.0 & 53.1 & 0.0 & $-$5.8 & 39.3\\
Ke & 8599 & floor & 0.0 & 0.0 & 0.0 & 0.0 & 0.0 & 0.0 & 22.9 & 68.0\\
Ke & 8640 & atca & 96.2 & 0.0 & 100.0 & 0.0 & 100.0 & 0.0 & $-$16.1 & 29.0\\
Ke & 8640 & floor & 0.0 & 0.0 & 1.0 & 0.0 & 17.5 & 0.0 & 10.2 & 55.3\\
Ke & 10480 & atca & 100.0 & 0.0 & 100.0 & 0.7 & 100.0 & 2.1 & $-$32.7 & 13.7\\
Ke & 10480 & floor & 29.3 & 0.0 & 95.5 & 0.0 & 100.0 & 0.0 & $-$15.1 & 31.3\\
Mopra & 1320 & atca & 100.0 & 0.0 & 100.0 & 0.0 & 100.0 & 0.0 & $-$16.3 & 27.1\\
Mopra & 1320 & floor & 0.3 & 0.0 & 83.1 & 0.0 & 97.8 & 0.0 & 1.9 & 45.4\\
Mopra & 1350 & atca & 100.0 & 0.0 & 100.0 & 0.0 & 100.0 & 0.0 & $-$17.3 & 26.2\\
Mopra & 1350 & floor & 0.5 & 0.0 & 88.3 & 0.0 & 99.0 & 0.0 & 1.3 & 44.8\\
Mopra & 1440 & atca & 100.0 & 0.0 & 100.0 & 0.0 & 100.0 & 0.0 & $-$15.0 & 28.4\\
Mopra & 1440 & floor & 0.5 & 0.0 & 87.2 & 0.0 & 98.7 & 0.0 & 1.4 & 44.8\\
Mopra & 1454 & atca & 9.5 & 0.0 & 41.4 & 0.0 & 91.0 & 0.0 & $-$15.1 & 28.2\\
Mopra & 1454 & floor & 0.0 & 0.0 & 0.0 & 0.0 & 2.2 & 0.0 & 13.7 & 57.0\\
Mopra & 1474 & floor & \textcolor{red}{0.0} & \textcolor{red}{0.0} & \textcolor{red}{2.1} & \textcolor{red}{0.0} & \textcolor{red}{14.3} & \textcolor{red}{0.0} & \textcolor{red}{$<$9.2} & \textcolor{red}{$<$52.3}\\
Mopra & 1620 & atca & 99.9 & 0.0 & 100.0 & 0.0 & 100.0 & 0.0 & $-$15.2 & 26.7\\
Mopra & 1620 & floor & 0.0 & 0.0 & 52.7 & 0.0 & 88.9 & 0.0 & 4.6 & 46.4\\
Mopra & 1680 & floor & \textcolor{red}{0.0} & \textcolor{red}{0.0} & \textcolor{red}{47.3} & \textcolor{red}{0.0} & \textcolor{red}{87.6} & \textcolor{red}{0.0} & \textcolor{red}{$<$4.9} & \textcolor{red}{$<$46.4}\\
Mopra & 1777 & atca & 37.2 & 0.0 & 99.7 & 0.0 & 100.0 & 1.0 & $-$21.3 & 20.2\\
Mopra & 1777 & floor & 0.0 & 0.0 & 0.0 & 0.0 & 0.3 & 0.0 & 22.4 & 63.9\\
Mopra & 2595 & atca & 100.0 & 0.1 & 100.0 & 6.7 & 100.0 & 12.2 & $-$36.4 & 5.2\\
Mopra & 2595 & floor & 0.0 & 0.0 & 0.2 & 0.0 & 2.0 & 0.0 & 13.9 & 55.4\\
Mopra & 2620 & atca & 100.0 & 100.0 & 100.0 & 100.0 & 100.0 & 100.0 & $-$73.4 & $-$31.9\\
Mopra & 2620 & floor & 58.7 & 0.0 & 100.0 & 0.0 & 100.0 & 0.0 & $-$19.1 & 22.4\\
Mopra & 2656 & atca & 53.0 & 0.0 & 100.0 & 0.0 & 100.0 & 0.0 & $-$18.8 & 22.7\\
Mopra & 2656 & floor & 0.0 & 0.0 & 0.2 & 0.0 & 1.8 & 0.0 & 14.2 & 55.7\\
Mopra & 2700 & atca & 100.0 & 0.0 & 100.0 & 0.0 & 100.0 & 0.0 & $-$23.4 & 18.1\\
Mopra & 2700 & floor & 0.0 & 0.0 & 4.0 & 0.0 & 45.8 & 0.0 & 8.5 & 50.0\\
Mopra & 4995 & atca & 33.2 & 0.0 & 96.8 & 0.0 & 100.0 & 0.0 & $-$6.4 & 35.1\\
Mopra & 4995 & floor & 0.0 & 0.0 & 0.0 & 0.0 & 3.7 & 0.0 & 13.1 & 54.6\\
Mopra & 5190 & atca & 5.2 & 0.0 & 45.9 & 0.0 & 84.9 & 0.0 & $-$8.8 & 33.0\\
Mopra & 5190 & floor & 0.0 & 0.0 & 0.1 & 0.0 & 0.7 & 0.0 & 18.2 & 59.9\\
Mopra & 5240 & atca & 100.0 & 50.0 & 100.0 & 100.0 & 100.0 & 100.0 & $-$60.4 & $-$18.6\\
Mopra & 5240 & floor & 25.5 & 0.0 & 95.1 & 0.0 & 100.0 & 0.0 & $-$14.8 & 27.0\\
Mopra & 7860 & atca & 100.0 & 2.2 & 100.0 & 22.8 & 100.0 & 52.9 & $-$45.9 & $-$1.4\\
Mopra & 7860 & floor & 33.8 & 0.0 & 99.8 & 0.0 & 100.0 & 0.0 & $-$12.6 & 31.9\\
Mopra & 8599 & atca & 4.8 & 0.0 & 33.6 & 0.0 & 55.8 & 0.0 & $-$7.1 & 38.0\\
Mopra & 8599 & floor & 0.0 & 0.0 & 0.0 & 0.0 & 0.1 & 0.0 & 21.7 & 66.8\\
Mopra & 8640 & atca & 99.8 & 0.0 & 100.0 & 0.0 & 100.0 & 0.0 & $-$19.4 & 25.7\\
Mopra & 8640 & floor & 0.0 & 0.0 & 5.9 & 0.0 & 18.2 & 0.0 & 6.8 & 52.0\\
Mopra & 10480 & atca & 100.0 & 0.1 & 100.0 & 0.8 & 100.0 & 1.2 & $-$27.8 & 18.6\\
Mopra & 10480 & floor & 38.5 & 0.0 & 100.0 & 0.0 & 100.0 & 0.0 & $-$10.1 & 36.3\\
Murriyang & 1320 & atca & 100.0 & 0.0 & 100.0 & 0.0 & 100.0 & 0.0 & $-$18.9 & 24.5\\
Murriyang & 1320 & floor & 3.3 & 0.0 & 44.0 & 0.0 & 91.4 & 0.0 & $-$0.7 & 42.7\\
Murriyang & 1350 & atca & 100.0 & 0.0 & 100.0 & 0.0 & 100.0 & 0.0 & $-$19.7 & 23.8\\
Murriyang & 1350 & floor & 4.8 & 0.0 & 49.9 & 0.0 & 97.3 & 0.0 & $-$1.1 & 42.4\\
Murriyang & 1440 & atca & 100.0 & 0.0 & 100.0 & 0.0 & 100.0 & 0.0 & $-$17.6 & 25.8\\
Murriyang & 1440 & floor & 5.0 & 0.0 & 47.6 & 0.0 & 97.0 & 0.0 & $-$1.2 & 42.2\\
Murriyang & 1454 & atca & 7.8 & 0.0 & 35.8 & 0.0 & 89.1 & 0.0 & $-$7.5 & 35.8\\
Murriyang & 1454 & floor & 0.0 & 0.0 & 0.4 & 0.0 & 0.7 & 0.0 & 21.2 & 64.5\\
Murriyang & 1474 & floor & \textcolor{red}{0.2} & \textcolor{red}{0.0} & \textcolor{red}{2.0} & \textcolor{red}{0.0} & \textcolor{red}{3.0} & \textcolor{red}{0.0} & \textcolor{red}{$<$9.4} & \textcolor{red}{$<$52.5}\\
Murriyang & 1620 & atca & 100.0 & 0.0 & 100.0 & 0.0 & 100.0 & 0.0 & $-$18.5 & 23.3\\
Murriyang & 1620 & floor & 0.7 & 0.0 & 26.1 & 0.0 & 51.3 & 0.0 & 1.2 & 43.1\\
Murriyang & 1680 & floor & \textcolor{red}{0.5} & \textcolor{red}{0.0} & \textcolor{red}{24.7} & \textcolor{red}{0.0} & \textcolor{red}{47.2} & \textcolor{red}{0.0} & \textcolor{red}{$<$1.6} & \textcolor{red}{$<$43.1}\\
Murriyang & 1777 & atca & 32.6 & 0.0 & 99.7 & 0.2 & 100.0 & 0.4 & $-$17.1 & 24.4\\
Murriyang & 1777 & floor & 0.0 & 0.0 & 0.1 & 0.0 & 0.3 & 0.0 & 26.6 & 68.1\\
Murriyang & 2595 & atca & 100.0 & 0.5 & 100.0 & 1.0 & 100.0 & 2.9 & $-$28.3 & 13.2\\
Murriyang & 2595 & floor & 0.0 & 0.0 & 0.5 & 0.0 & 0.8 & 0.0 & 22.0 & 63.5\\
Murriyang & 2620 & atca & 100.0 & 100.0 & 100.0 & 100.0 & 100.0 & 100.0 & $-$65.4 & $-$23.9\\
Murriyang & 2620 & floor & 49.4 & 0.0 & 100.0 & 0.0 & 100.0 & 0.3 & $-$11.1 & 30.5\\
Murriyang & 2656 & atca & 43.5 & 0.0 & 100.0 & 0.0 & 100.0 & 0.2 & $-$10.7 & 30.8\\
Murriyang & 2656 & floor & 0.0 & 0.0 & 0.4 & 0.0 & 0.7 & 0.0 & 22.3 & 63.9\\
Murriyang & 2700 & atca & 100.0 & 0.0 & 100.0 & 0.1 & 100.0 & 2.5 & $-$28.0 & 13.5\\
Murriyang & 2700 & floor & 0.1 & 0.0 & 8.2 & 0.0 & 14.3 & 0.0 & 3.9 & 45.4\\
Murriyang & 4995 & atca & 6.9 & 0.0 & 97.0 & 0.0 & 100.0 & 0.0 & $-$10.6 & 30.9\\
Murriyang & 4995 & floor & 0.0 & 0.0 & 2.1 & 0.0 & 4.2 & 0.0 & 8.9 & 50.4\\
Murriyang & 5190 & atca & 4.4 & 0.0 & 36.4 & 0.0 & 80.9 & 0.0 & $-$2.0 & 39.7\\
Murriyang & 5190 & floor & 0.0 & 0.0 & 0.1 & 0.0 & 0.2 & 0.0 & 24.9 & 66.7\\
Murriyang & 5240 & atca & 100.0 & 39.7 & 100.0 & 100.0 & 100.0 & 100.0 & $-$53.6 & $-$11.8\\
Murriyang & 5240 & floor & 18.4 & 0.0 & 95.1 & 0.0 & 100.0 & 0.1 & $-$8.1 & 33.8\\
Murriyang & 7860 & atca & 100.0 & 1.6 & 100.0 & 16.0 & 100.0 & 43.4 & $-$44.0 & 0.5\\
Murriyang & 7860 & floor & 25.5 & 0.0 & 99.8 & 0.0 & 100.0 & 0.0 & $-$10.7 & 33.8\\
Murriyang & 8599 & atca & 1.8 & 0.0 & 29.9 & 0.0 & 52.0 & 0.0 & 0.1 & 45.2\\
Murriyang & 8599 & floor & 0.0 & 0.0 & 0.1 & 0.0 & 0.2 & 0.0 & 28.9 & 74.0\\
Murriyang & 8640 & atca & 99.8 & 0.0 & 100.0 & 0.0 & 100.0 & 0.1 & $-$16.2 & 28.9\\
Murriyang & 8640 & floor & 0.6 & 0.0 & 1.9 & 0.0 & 2.7 & 0.0 & 10.1 & 55.2\\
Murriyang & 10480 & atca & 100.0 & 0.0 & 100.0 & 0.1 & 100.0 & 0.6 & $-$29.3 & 17.1\\
Murriyang & 10480 & floor & 29.2 & 0.0 & 100.0 & 0.0 & 100.0 & 0.0 & $-$11.7 & 34.7\\
Yg & 2595 & atca & 100.0 & 0.0 & 100.0 & 6.4 & 100.0 & 22.8 & $-$34.2 & 7.3\\
Yg & 2595 & floor & 0.0 & 0.0 & 0.0 & 0.0 & 0.4 & 0.0 & 16.1 & 57.6\\
Yg & 2620 & atca & 100.0 & 100.0 & 100.0 & 100.0 & 100.0 & 100.0 & $-$71.2 & $-$29.7\\
Yg & 2620 & floor & 59.2 & 0.0 & 100.0 & 0.0 & 100.0 & 0.0 & $-$16.9 & 24.6\\
Yg & 2656 & atca & 55.4 & 0.0 & 100.0 & 0.0 & 100.0 & 0.0 & $-$16.7 & 24.8\\
Yg & 2656 & floor & 0.0 & 0.0 & 0.0 & 0.0 & 0.3 & 0.0 & 16.3 & 57.8\\
Yg & 2700 & atca & 100.0 & 0.0 & 100.0 & 0.0 & 100.0 & 0.0 & $-$22.3 & 19.2\\
Yg & 2700 & floor & 0.0 & 0.0 & 1.3 & 0.0 & 47.8 & 0.0 & 9.6 & 51.1\\
Yg & 4995 & atca & 40.3 & 0.0 & 95.3 & 0.0 & 100.0 & 0.0 & $-$4.3 & 37.2\\
Yg & 4995 & floor & 0.0 & 0.0 & 0.0 & 0.0 & 0.6 & 0.0 & 15.2 & 56.7\\
Yg & 5190 & atca & 13.1 & 0.0 & 48.6 & 0.0 & 86.8 & 0.0 & $-$8.1 & 33.7\\
Yg & 5190 & floor & 0.0 & 0.0 & 0.0 & 0.0 & 0.2 & 0.0 & 18.9 & 60.6\\
Yg & 5240 & atca & 100.0 & 52.1 & 100.0 & 100.0 & 100.0 & 100.0 & $-$59.6 & $-$17.8\\
Yg & 5240 & floor & 26.7 & 0.0 & 95.2 & 0.0 & 100.0 & 0.0 & $-$14.1 & 27.8\\
Yg & 7860 & atca & 100.0 & 5.8 & 100.0 & 23.1 & 100.0 & 54.5 & $-$49.5 & $-$5.0\\
Yg & 7860 & floor & 35.8 & 0.0 & 99.4 & 0.0 & 100.0 & 0.0 & $-$16.2 & 28.3\\
Yg & 8599 & atca & 10.0 & 0.0 & 34.3 & 0.0 & 57.3 & 0.0 & $-$6.0 & 39.1\\
Yg & 8599 & floor & 0.0 & 0.0 & 0.0 & 0.0 & 0.0 & 0.0 & 22.7 & 67.8\\
Yg & 8640 & atca & 99.6 & 0.0 & 100.0 & 0.0 & 100.0 & 0.0 & $-$16.9 & 28.2\\
Yg & 8640 & floor & 0.0 & 0.0 & 1.7 & 0.0 & 26.2 & 0.0 & 9.4 & 54.5\\
Yg & 10480 & atca & 100.0 & 0.0 & 100.0 & 1.0 & 100.0 & 3.1 & $-$34.7 & 11.7\\
Yg & 10480 & floor & 41.2 & 0.0 & 100.0 & 0.0 & 100.0 & 0.0 & $-$17.1 & 29.3\\
\botrule
\end{longtable}}

\clearpage
\section{Per-facility EPFD sky coverage}
\label{app:facplots}

\begin{figure}[H]\centering\includegraphics[width=\textwidth]{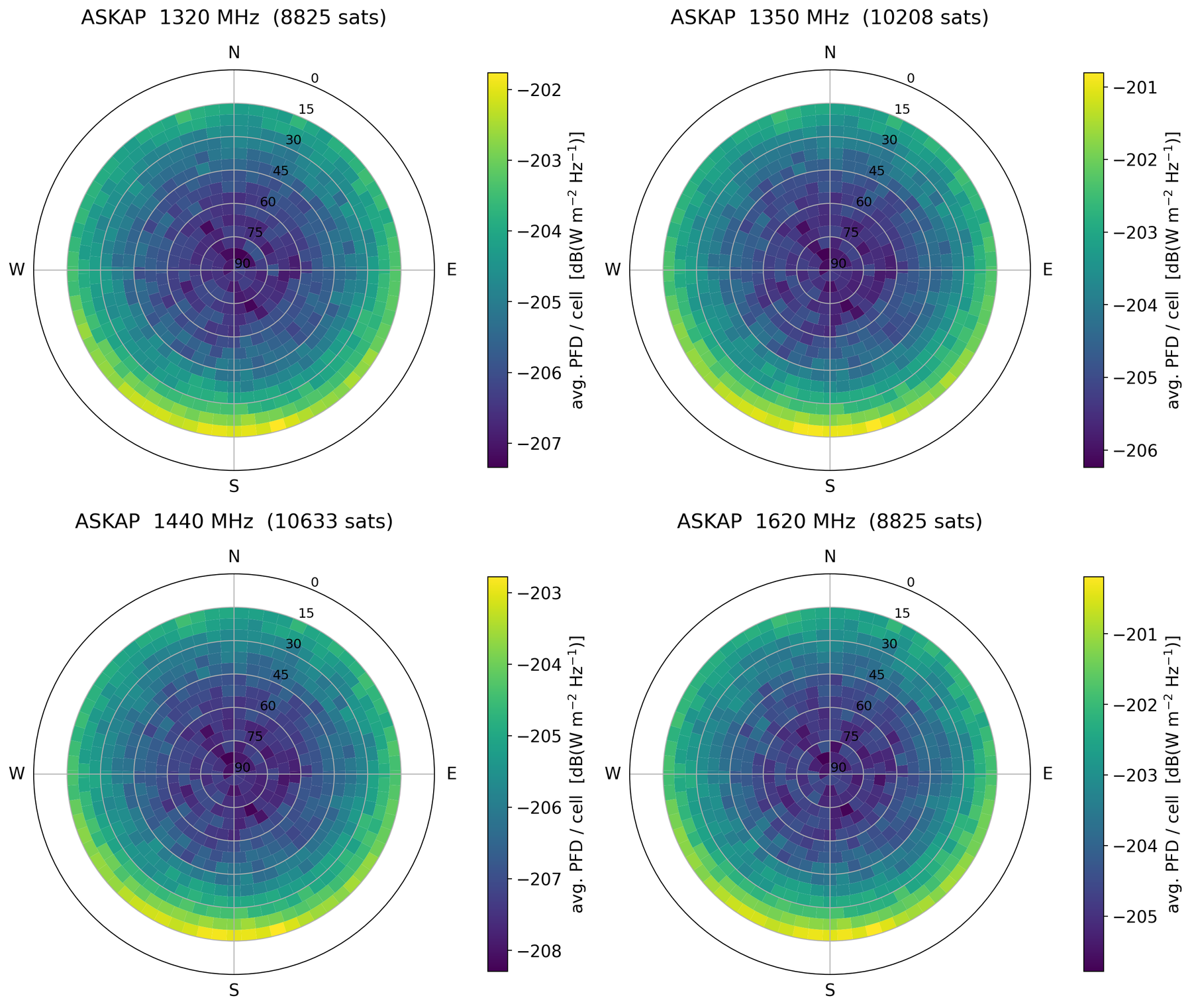}
\caption{ASKAP: aggregate-\ac{EPFD} sky coverage for four v2-Mini L-band unintended-radiation detections (\SI{1320}{}, \SI{1350}{}, \SI{1440}{} and \SI{1620}{MHz}; present fleet, ATCA level). ASKAP observes only \mbox{0.7--1.8\,GHz}; with $\sim$\num{10000} radiating satellites in band its sky is saturated against the single-dish continuum criterion (Table~\ref{tab:facilities-cont}).}
\label{fig:appASK}\end{figure}

For each tested facility we show the aggregate-\ac{EPFD} sky coverage for the three non-saturated detections that carry a per-cell map for the present fleet: the \SI{2656}{MHz} Starlink unintended-radiation forest detection, the \SI{5190}{MHz} second-harmonic UEMR and the OneWeb \SI{8599}{MHz} detection. Each panel is labelled with its facility and detection. The single-dish continuum data loss is not mapped: for these and the brighter detections it saturates to near-total across the accessible sky and carries little spatial detail (Table~\ref{tab:facilities-cont}; the one illustrative continuum map is Figure~\ref{fig:loss7860}). ASKAP observes only \mbox{0.7--1.8\,GHz}, so it sees none of these detections and is shown separately (Figure~\ref{fig:appASK}) for four of its now-calibrated L-band detections.

\begin{figure}[H]\centering\includegraphics[width=\textwidth]{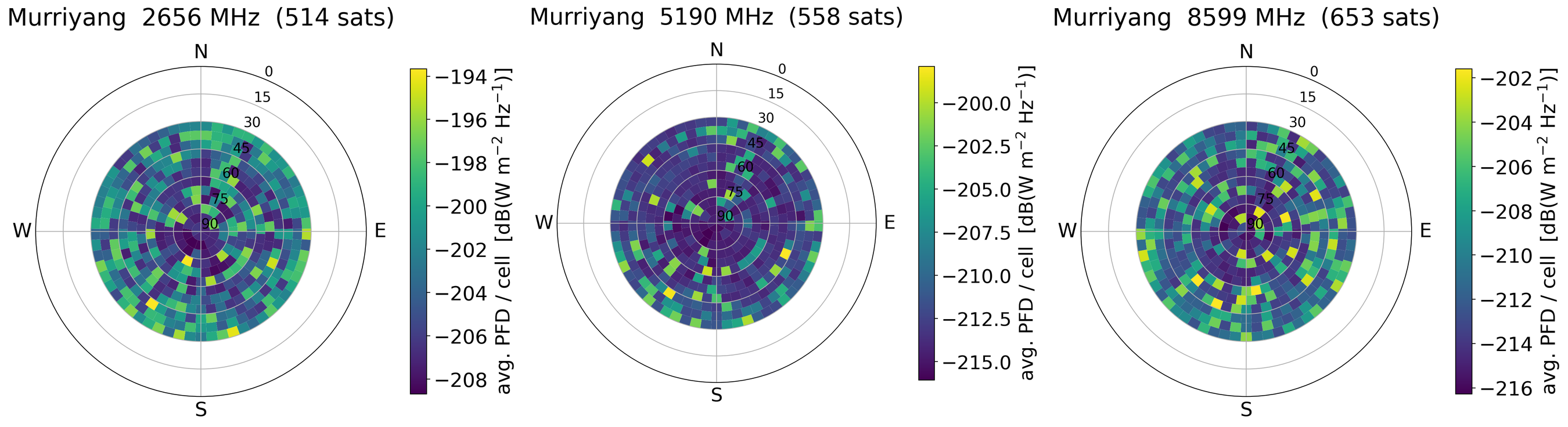}
\caption{Murriyang: aggregate-\ac{EPFD} sky coverage for the \SI{2656}{}, \SI{5190}{} and \SI{8599}{MHz} detections (present fleet, ATCA level).}
\label{fig:appPks}\end{figure}
\begin{figure}[H]\centering\includegraphics[width=\textwidth]{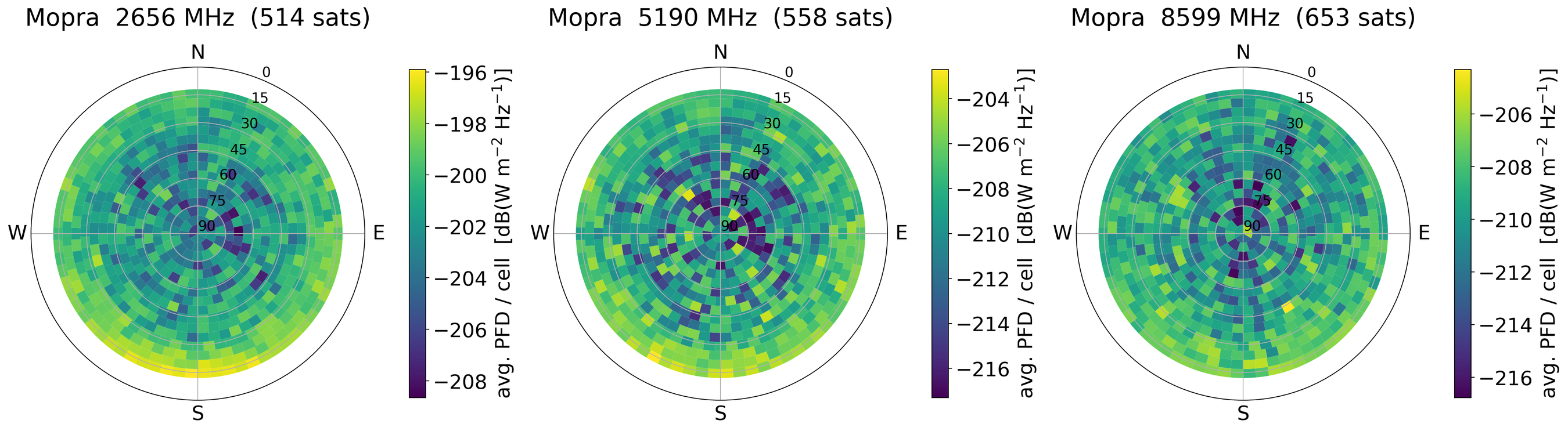}
\caption{Mopra: as Figure~\ref{fig:appPks}.}\label{fig:appMop}\end{figure}
\begin{figure}[H]\centering\includegraphics[width=\textwidth]{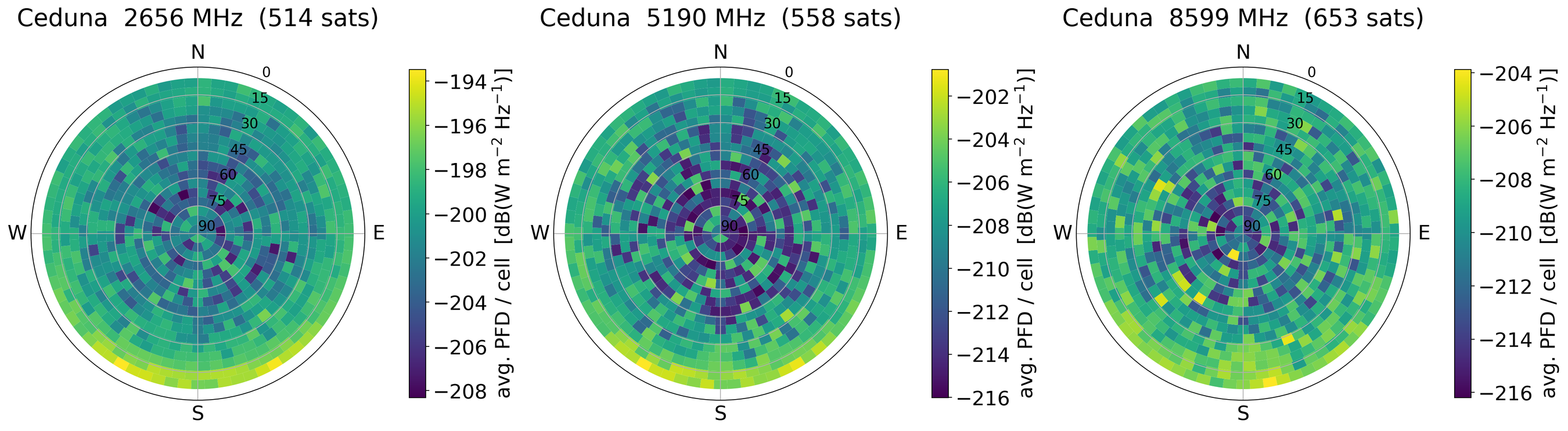}
\caption{Ceduna: as Figure~\ref{fig:appPks}.}\label{fig:appCed}\end{figure}
\begin{figure}[H]\centering\includegraphics[width=\textwidth]{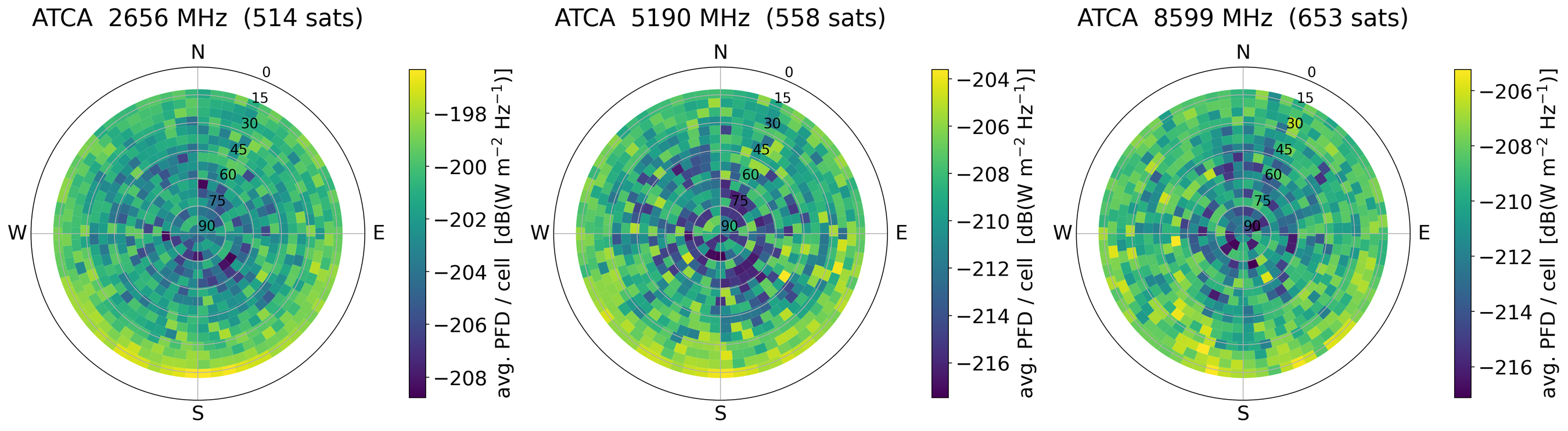}
\caption{ATCA (single-dish mode): as Figure~\ref{fig:appPks}.}\label{fig:appATCA}\end{figure}
\begin{figure}[H]\centering\includegraphics[width=\textwidth]{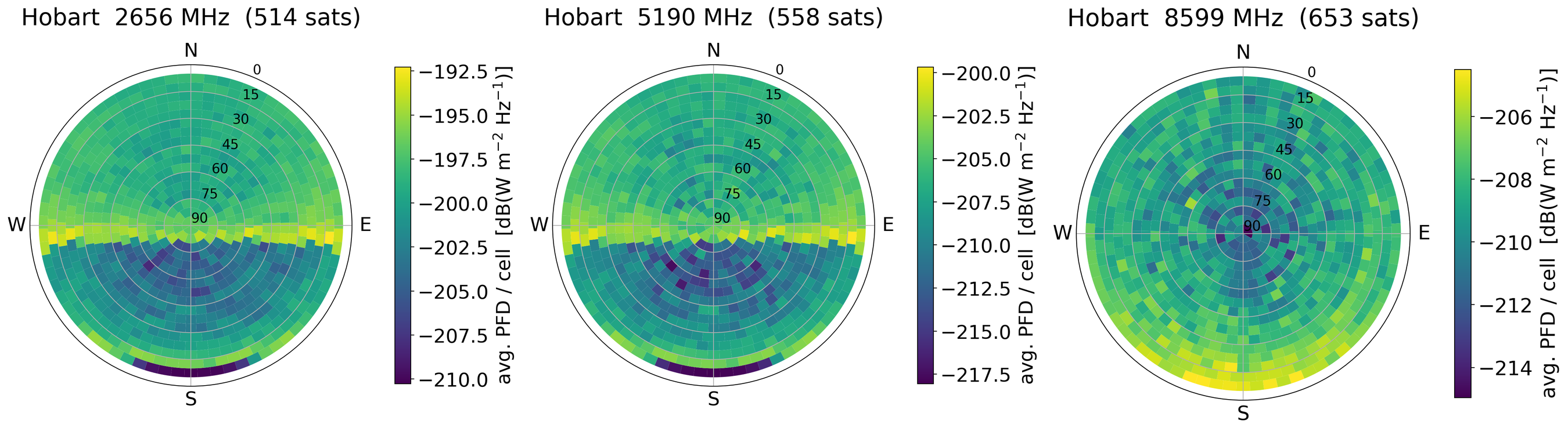}
\caption{Hobart: as Figure~\ref{fig:appPks}.}\label{fig:appHb}\end{figure}
\begin{figure}[H]\centering\includegraphics[width=\textwidth]{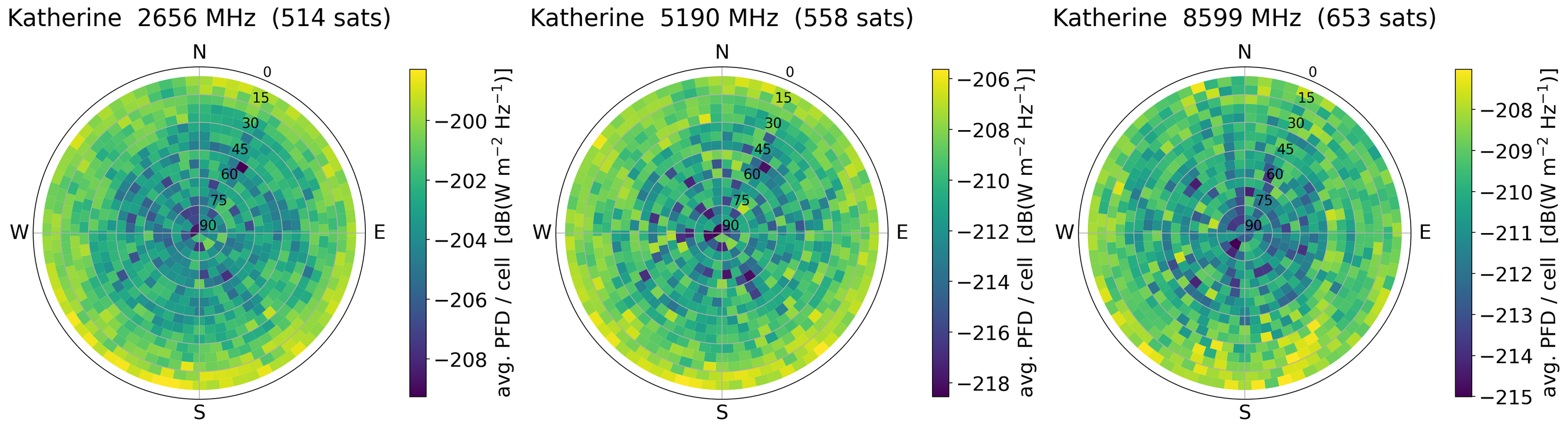}
\caption{Katherine: as Figure~\ref{fig:appPks}.}\label{fig:appKe}\end{figure}
\begin{figure}[H]\centering\includegraphics[width=\textwidth]{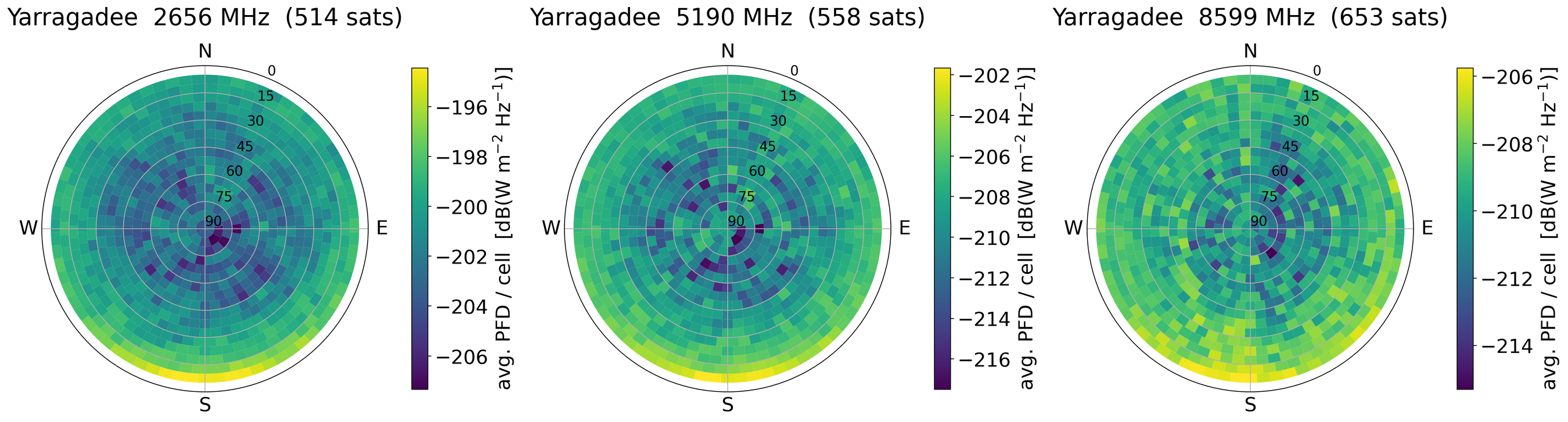}
\caption{Yarragadee: as Figure~\ref{fig:appPks}.}\label{fig:appYg}\end{figure}

\clearpage
\twocolumn
\section{Acronyms and abbreviations}

\begin{acronym}
    \acro{AIAA}{American Institute of Aeronautics and Astronautics}
    \acro{ASKAP}{Australian SKA Pathfinder}
    \acro{ATCA}{Australia Telescope Compact Array}
    \acro{ATNF}{Australia Telescope National Facility}
    \acro{BIGCAT}{Broadband Integrated GPU Correlator for ATCA}
    \acro{CABB}{Compact Array Broadband Backend}
    \acro{CDF}{cumulative distribution function}
    \acrodef{CEPT}{European Conference of Postal and Telecommunications Administrations}
    \acro{CISPR}{Comité international spécial des perturbations radioélectriques}
    \acro{CSIRO}{Commonwealth Scientific and Industrial Research Organisation}
    \acro{DORIS}{Doppler Orbitography and Radiopositioning Integrated by Satellite}
    \acro{DTD}{Direct-to-device}
    \acro{DTC}{Direct-to-Cell}
    \acro{DTH}{Direct-to-Handset}
    \acro{DTM}{Direct-to-mobile}
    \acro{ECC}{Electronic Communications Committee}
    \acro{ECSS}{European Cooperation for Space Standardization}
    \acro{e.i.r.p}{effective isotropic radiated power}
    \acro{EMC}{electromagnetic compatibility}
    \acro{EMI}{electromagnetic interference}
    \acro{EOP}{Earth Orientation Parameters}
    \acro{EPFD}{equivalent power flux density}
    \acro{FCC}{Federal Communications Commission}
    \acro{FSS}{Fixed Satellite Service}
    \acro{GGOS}{Global Geodetic Observing System}
    \acro{GNSS}{Global Navigation Satellite Systems}
    \acro{GPS}{Global Positioning System}
    \acro{GSO}{geostationary satellite orbit}
    \acro{ICRF3}{International Celestial Reference Frame 3}
    \acro{ISO}{International Organization for Standardization}
    \acro{ITRF2020}{International Terrestrial Reference Frame 2020}
    \acro{ITU-R}{International Telecommunication Union Radiocommunication Sector}
    \acro{IVS}{International VLBI Service for Geodesy and Astrometry}
    \acro{JAXA}{Japan Aerospace Exploration Agency}
    \acro{LEO}{low-Earth orbit}
    \acro{LNA}{low-noise amplifier}
    \acro{LO}{local oscillator}
    \acro{LOFAR}{LOw-Frequency ARray}
    \acro{MEO}{medium-Earth orbit}
    \acro{MSS}{Mobile-satellite service}
    \acro{NCRIS}{National Collaborative Research Infrastructure Strategy}
    \acro{NGSO}{non-geostationary satellite orbit}
    \acro{NRAO}{National Radio Astronomy Observatory}
    \acro{ODC}{orbital data centre}
    \acro{OMM}{Orbit Mean-Elements Message}
    \acro{OOBE}{out-of-band emission}
    \acro{PFD}{power flux density}
    \acro{RAS}{radio astronomy service}
    \acro{RFI}{radio-frequency interference}
    \acro{RR}{Radio Regulations}
    \acro{SKA}{Square Kilometre Array}
    \acro{SLR}{satellite laser ranging}
    \acro{SNIFFLES}{Satellite measuremeNt of Intended emission, unwanted emission, and radio radiation to develop, Follow up, and veriFy mitigation measures, regulatory compliance, and Lawful usE of the radio Spectrum}
    \acro{TDRSS}{Tracking and Data Relay Satellite System}
    \acro{Tsys}{system temperature}
    \acro{TTC}[TT\&C]{telemetry, tracking, and command}
    \acro{UEMR}{unintended electromagnetic radiation}
    \acro{UT1}{Universal Time 1}
    \acro{UTC}{Coordinated Universal Time}
    \acro{VGOS}{VLBI Global Observing System}
    \acro{VLBI}{very long baseline interferometry}
    \acro{WRC}{World Radiocommunication Conference}
\end{acronym}
\twocolumn
\bibliographystyle{paslike}
\bibliography{emlimits}

\end{document}